\documentclass[a4paper]{article}

\AtBeginDocument{%
  \providecommand\BibTeX{{%
    \normalfont B\kern-0.5em{\scshape i\kern-0.25em b}\kern-0.8em\TeX}}}

\usepackage{amsmath}
\usepackage{scalerel,amssymb}
\usepackage{mathtools}

\DeclarePairedDelimiter{\floor}{\lfloor}{\rfloor}
\usepackage{svg}
\usepackage{wrapfig}
\usepackage{tikz}
\usepackage{pgfplots}
\usepackage{booktabs}
\usepackage{float}
\usepackage{subfig}
\usepackage{hyperref}

\usepackage{enumitem}

\setenumerate[1]{label={(\arabic*)}}

\usepackage{tikz-uml}
\usepackage{pgf-pie}
\usetikzlibrary{backgrounds}

\usepackage{setspace}

\usepackage[normalem]{ulem}
\usepgfplotslibrary{fillbetween}

\usepackage[framemethod=tikz]{mdframed}
\usetikzlibrary{scopes, backgrounds}

\usepackage{listings}
\usepackage[ruled,vlined,linesnumbered]{algorithm2e}

\usepackage{textcomp}
\definecolor{lbcolor}{rgb}{0.93,0.93,0.93}
\definecolor{gnuplot@orange}{RGB}{229,158,0}
\definecolor{gnuplot@purple}{RGB}{148,0,212}
\definecolor{gnuplot@red}{RGB}{200,0,0}
\definecolor{gnuplot@lightblue}{RGB}{87,181,232}
\definecolor{gnuplot@green}{RGB}{0,158,65}
\definecolor{gnuplot@darkblue}{RGB}{0,115,179}
\definecolor{gnuplot@yellow}{RGB}{240,227,66}
\pgfplotscreateplotcyclelist{colorGPL}{%
gnuplot@darkblue,every mark/.append style={fill=gnuplot@darkblue!80!black},mark=o\\%
gnuplot@red,every mark/.append style={fill=gnuplot@red!50!black},mark=square\\%
gnuplot@green,every mark/.append style={fill=gnuplot@green!50!black},mark=triangle*\\%
gnuplot@orange,every mark/.append style={fill=gnuplot@orange!80!black,mark size=2.5pt},mark=x\\%
gnuplot@purple,mark=diamond\\%
black,densely dashed,every mark/.append style={solid,fill=gnuplot@darkblue!80!black},mark=square*\\%
gnuplot@lightblue,densely dashed,every mark/.append style={solid,fill=gnuplot@lightblue!30!black},mark=otimes*\\%
red!80!white,densely dashed,every mark/.append style={solid,fill=red!40!white},mark=oplus*\\%
}
\makeatother

\graphicspath{{svg_with_latex/}}

\usepackage{amsopn}

\usepackage{cite}
\usepackage{natbib}
\setcitestyle{numbers}
\setcitestyle{square}

\newif\ifwithacks

\begin{document}

\title{hyper.deal: An efficient, matrix-free finite-element library for high-dimensional partial differential equations\thanks{This work was supported by the German Research Foundation (DFG) under
    the project ``High-order discontinuous Galerkin for the exa-scale''
    (ExaDG) within the priority program ``Software for Exascale Computing''
    (SPPEXA), grant agreement no.~KO5206/1-1 and KR4661/2-1.
The authors gratefully acknowledge the Gauss Centre for Supercomputing e.V.~(\texttt{www.gauss-centre.eu}) for funding this project
by providing computing time on the GCS Supercomputer SuperMUC-NG at Leibniz Supercomputing Centre (LRZ, \texttt{www.lrz.de})
through project id pr83te. }}

\author{Peter Munch\thanks{Institute for Computational Mechanics,
    Technical University of Munich, Boltzmannstr.~15, 85748 Garching, Germany (\texttt{\{munch,kronbichler\}@lnm.mw.tum.de}).}~\thanks{Institute of Materials Research, Materials Mechanics, Helmholtz-Zentrum Geesthacht, Max-Planck-Str.~1, 21502 Geesthacht, Germany (\texttt{peter.muench@hzg.de}).}
\and
%
Katharina Kormann\thanks{Max Planck Institute for Plasma Physics,
    Boltzmannstr.~2, 85748 Garching, Germany, and Department of Mathematics, Technical University of Munich, Boltzmannstr.~3, 85748 Garching, Germany,
    (\texttt{katharina.kormann@ipp.mpg.de}).}
\and
%
Martin Kronbichler\footnotemark[2]}


\maketitle
  
\begin{abstract}
This work presents the efficient, matrix-free finite-element library \texttt{hyper.deal} for solving partial differential equations in two to six dimensions with high-order  discontinuous Galerkin methods. 
It builds upon the low-dimensional finite-element library \texttt{deal.II} to create complex low-dimensional meshes and to operate on them individually. These meshes are combined via a tensor product on the fly and the library provides new special-purpose highly optimized matrix-free functions exploiting domain decomposition as well as shared memory via \texttt{MPI-3.0} features.
Both node-level performance analyses and strong/weak-scaling studies on up to 147,456 CPU cores confirm the efficiency of the implementation.
Results of the library \texttt{hyper.deal} are reported for high-dimensional advection problems and for the solution of the Vlasov--Poisson equation in up to 6D phase space.
\end{abstract}

\noindent \textbf{Key words.}
  matrix-free operator evaluation, discontinuous Galerkin methods, high-dimensional, high-order,  Vlasov--Poisson equation, MPI-3.0 shared memory.

\section{Introduction}\label{sec:into}

Three-dimensional problems are today simulated in great detail based on domain 
decomposition codes up to supercomputer scale. With the increase in 
computational power, it becomes feasible to simulate also higher-dimensional 
problems. With this contribution, we target the case of moderately 
high-dimensional ($\le$6D) problems with complex geometry requirements in some 
of the dimensions. Our primary target are kinetic equations that describe the 
evolution of a distribution function in phase space. Such Boltzmann-type 
equations are for instance used in magnetic confinement fusion where the 
evolution of a plasma is described by a distribution function that evolves 
according to the Vlasov equation coupled to a system of Maxwell's equations for 
its self-consistent fields. Other areas of application of phase space are, e.g., 
cosmic microwave background radiation or magnetic reconnection in the earth's 
magnetosphere.

In these applications, there are two distinct sets of variables, configuration 
and velocity space. Especially in configuration space, the domain can be 
geometrically complex (e.g., torus-like shapes as in the case of tokamak and 
stellarator fusion reactors), necessitating a flexible description of the mesh as 
is commonly provided by finite-element method (FEM) libraries for grids in up to three 
dimensions. Another typical feature of kinetic equations is the coupling of 
high-dimensional problems in phase space to lower-dimensional problems in the 
spatial variables only.  With the library \texttt{hyper.deal}, we are targeting 
these applications: For a phase-space simulation, two 
separate---possibly complex---meshes describing configuration and velocity 
space are defined based on the capabilities of a base low-dimensional 
finite-element library---in our case the library 
\texttt{deal.II}~\cite{Alzetta2018,Arndt2020}---and combined by taking their 
tensor product on the fly. The equations that we have in mind are 
advection-dominated, which is why we mostly focus on the discretization of the 
advection equation in this paper.

The general concept presented is not limited to the description of the phase 
space. Possible applications of a high-dimensional FEM library could for 
instance include three-dimensional problems that involve a low-dimensional 
parameter space or low-dimensional Fokker--Planck-type equations.

\subsection{Previous work}

While the decomposition of complex domains in up to three dimensions is a 
well-studied problem, partial differential equations on complex domains in 
dimensions higher than three are not tackled by generic FEM libraries to date. 
Some libraries have started to extend their capabilities to higher dimensions 
on structured grids. An example is the \texttt{YaspGrid} module of the 
finite-element library \texttt{DUNE}, which implements structured grids in 
arbitrary dimensions \cite{Exadune16}. Also, higher-dimensional problems are
solved based on sparse grids \cite{Guo2016}, reduced basis methods, or
low-rank tensors \cite{Bachmayr2016}. On the other hand, in the plasma
community some specialized codes exist that target gyrokinetic or kinetic
equations. The \texttt{Gekyll} code \cite{Hakim2019} is a discontinuous
Galerkin-method solver for plasma physics applications, and a fully kinetic
version for Cartesian grids was presented in \cite{Juno2018}. The authors use
Serendipity elements to reduce the number of degrees of freedom. On the other
hand, such a reduced basis does not form a tensor product and is therefore not
amenable to the algorithms presented in this work. The specifics of domain 
decomposition in higher dimensions have only been studied recently. In 
\cite{Kormann2019}, a six-dimensional domain decomposition for a 
semi-Lagrangian solver on a Cartesian grid of the Vlasov--Poisson system was 
studied, highlighting the high demands on memory transfer between neighboring 
processes due to the increased surface-to-volume ratio with increasing 
dimensionality. The parallelization of a similar algorithm has also been 
addressed in \cite{Umeda2012}. However, the domain decomposition is limited to 
configuration space in that work, which poses a strong limit to the 
scalability of the implementation. 

\subsection{Model problem}

The Vlasov equation is given as 
\begin{align}
\frac{\partial\,}{\partial\, t}f_s(t,\vec{x},\vec{v})  
+\vec{v} \cdot \nabla_{\vec{x}} f_s(t,\vec{x},\vec{v})  
+\frac{q_s}{m_s}\left( \vec{E}(t,\vec{x})   + \vec{v} \times \vec{B}(t,\vec{x})   \right) \cdot \nabla_{\vec{v}} f_s(t,\vec{x},\vec{v})  
= 0, \label{equ:intro:vp:full}
\end{align}
where $f_s(t,\vec{x},\vec{v})$ denotes the probability density of a particle 
species $s$ with charge $q_s$ and mass $m_s$ as a function of the phase space,
 $\vec{E}$ the electric field, 
and $\vec{B}$ the magnetic field.
The phase space  is defined as the tensor product of configuration space,
$\vec{x}\in \Omega_{\vec{x}}  \subset\mathbb{R}^{d_{\vec{x}}}$, and 
velocity space,  $\vec{v}\in \mathbb{R}^{d_{\vec{v}}}$.
This equation is coupled either to the Maxwell or to the Poisson equation 
(see Section~\ref{sec:vp}) for the self-consistent fields.
If we define the gradient operator as 
$\nabla^\top := (\nabla_{\vec{x}}^\top,\;\nabla_{\vec{v}}^\top) $, 
Equation~\eqref{equ:intro:vp:full} can be rewritten as
\begin{align}
\frac{\partial }{\partial t}f_s(t,\vec{x},\vec{v})   + 
\nabla \cdot \left(\left( 
\begin{array}{c}
\vec{v} \\
\frac{q_s}{m_s}\left(\vec{E}(t,\vec{x}) + \vec{v} \times \vec{B}(t,\vec{x}) \right)
\end{array} 
\right)f_s(t,\vec{x},\vec{v}) \right)  = 0. \label{equ:intro:vp:simple}
\end{align}
This is a non-linear, high-dimensional ($d=d_{\vec{x}} + d_{\vec{v}}$), 
hyperbolic partial differential equation. To simplify the presentation in the 
following, we will consider the advection equation defined on the arbitrary 
$d$-dimensional domain $\Omega$, denoting the independent variable again by 
$\vec{x}$,
\begin{align}
\frac{\partial f}{\partial t} + \nabla \cdot (\vec{a}(f,\, t,\,\vec{x})  f) = 0 \qquad \text{on}
\qquad \Omega\times [0,t_{\text{final}}], \label{equ:intro:advection}
\end{align}
with $\vec{a}(f,\,t,\,\vec{x})$ being the solution-, time-, and space-dependent 
advection coefficient. The system is closed by an initial condition 
$f(0, \vec{x})$ and suitable boundary conditions. In the following, we 
concentrate on periodic boundary conditions.
We return to the Vlasov--Poisson equation in Section~\ref{sec:vp} where we 
combine an advection solver from the library \texttt{hyper.deal} and a Poisson 
solver based on \texttt{deal.II}.

\subsection{Discontinuous Galerkin discretization of the advection equation}

High-order discontinuous Galerkin methods are attractive methods for solving 
hyperbolic partial differential equations like 
Equation~\eqref{equ:intro:advection} due to their high accuracy in terms of 
dispersion and dissipation, while maintaining geometric flexibility through 
unstructured grids. The discontinuous Galerkin (DG) discretization of these 
equations can be derived by the following steps: The partial differential 
equation is multiplied with the test function $g$, integrated over the element
 domain $\Omega^{(e)}$ with $\displaystyle{\biguplus_e \Omega^{(e)}=\Omega}$,
 and the divergence theorem is applied:
\begin{align}
\left(g, \frac{\partial f}{\partial\, t}\right)_{\Omega^{(e)}} - \bigg(\nabla g, \vec{a} f \bigg)_{\Omega^{(e)}} + \bigg< g,\, \vec{n} \cdot (\vec{a} f)^* \bigg>_{\Gamma^{(e)}} = 0,
\end{align}
with $(\vec{a}f)^*$ being the numerical flux, like a central or an upwind flux. 
Integration $\int_\Omega  \, \mathrm{d}\Omega$ and derivation $\nabla $ are
 not performed in the real space but in the reference space $\Omega_0^{(e)}$
 and  $\Gamma_0^{(e)}$ and require a mapping to the reference coordinates:
\begin{align}\label{eq:into:adv:ref}
\left(g, |\mathcal{J}| \frac{\partial f}{\partial\, t}\right)_{\Omega^{(e)}_0} - \bigg( \mathcal{J}^{-T} \nabla_{\vec{\xi}} g,\; |\mathcal{J}|  \vec{a} f \bigg)_{\Omega^{(e)}_0} + \bigg< g,\, |\mathcal{J}| \; \vec{n} \cdot (\vec{a} f)^* \bigg>_{\Gamma^{(e)}_0} = 0,
\end{align}
where $\mathcal{J}$ is the Jacobian matrix of the mapping from reference to
real space and $|\mathcal{J}|$ its determinant.

To discretize this equation in space, we consider nodal polynomials with 
nodes in the Gauss--Lobatto points. 
The integrals are evaluated numerically by weighted sums. We consider both the 
usual Gauss(--Legendre) quadrature rules and the integration directly in the 
Gauss--Lobatto points without the need of interpolation (collocation 
setup~\cite{Deville2003}).
The resulting semi-discrete system has the following form:
\begin{align}
\mathcal{M} \frac{\partial \vec{f}}{\partial t} = \mathcal{A}(\vec{f},\,t)
\qquad\leftrightarrow\qquad
\frac{\partial \vec{f}}{\partial t} = \mathcal{M}^{-1} \mathcal{A}(\vec{f},\,t),
\end{align}
where $\vec{f}$ is the vector containing the coefficients for the polynomial
 approximation of $f$, $\mathcal{M}$ the mass matrix, and $\mathcal{A}$ the
 discrete advection operator. This system of ordinary differential equations
 can be solved with classical time integration schemes such as explicit
 Runge--Kutta methods. They require the right-hand side
 $\mathcal{M}^{-1} \mathcal{A}(\vec{f},\,t)$ to be evaluated efficiently. 
The particular structure of the mass matrix $\mathcal{M}$ should be noted: 
It is diagonal in the collocation case and block-diagonal in the case of the 
normal Gauss quadrature, which leads to a simple inversion of the mass
 matrix \cite{2016_73}. 

For 2D and 3D high-order DG methods, efficient matrix-free operator evaluations
 \cite{Kronbichler2012,Kronbichler2017a} for individual operators
 $\mathcal{M}^{-1}$, $\mathcal{A}$ as well as for the merged operator
 $\mathcal{M}^{-1}\mathcal{A}$ are common in the context of fluid
 mechanics~\cite{Krank2017a}, structural mechanics~\cite{Davydov2019},
 and acoustic wave propagation~\cite{Schoeder2018}. These methods do not
 assemble matrices, but evaluate the effect of the (linear or non-linear)
 operator on element vectors on the fly, often exploiting the tensor-product
 structure of the shape functions on hexahedron meshes. Since generally the
 access to main memory is the major limiting factor on modern CPU hardware,
 the fact that no matrix has to be loaded for matrix-free operator evaluation
 directly translates into a significant performance improvement. 
Discontinuous Galerkin methods and matrix-free operator evaluation kernels are
 part of many low-dimensional general-purpose FEM libraries nowadays.

\subsection{Our contribution}

This work shows that an extension of low-dimensional general-purpose high-order
 matrix-free FEM libraries to high dimensions is possible. We discuss how to
 cope with possible performance deteriorations due to the
 ``curse of dimensionality'' and present special-purpose concepts exploiting
the structure of the phase space, using domain decomposition and shared
 memory, all taking hardware characteristics into account.

All concepts that we describe in this work have been implemented and are 
available under the LGPL~3.0 license as the library \texttt{hyper.deal} hosted 
at \url{https://github.com/hyperdeal/hyperdeal}. It 
extends---as an example---the open-source FEM library 
\texttt{deal.II}~\cite{Alzetta2018} to high dimensions. 
Both node-level performance and strong/weak-scaling analyses, conducted for
 this library, confirm the suitability of the described concepts for solving
 partial differential equations in high dimensions.

The remainder of this work is organized as follows. In 
Section~\ref{sec:dealii},  we present the concepts of a generic modern 
low-dimensional finite-element library; special focus 
is put on mesh generation, parallelization, and efficient matrix-free operator 
evaluation as well as on why these concepts are only partly applicable for
higher-dimensional problems. 
Sections~\ref{sec:tensorproduct}-\ref{sec:concept:sm} give an introduction
into the concept of a tensor product of partitioned low-dimensional meshes
and details its implementation for phase space. Section~\ref{sec:performance}
presents performance results for the advection equation, confirming the
efficiency of the design decisions made during the implementation. 
Section~\ref{sec:vp} explains how \texttt{hyper.deal} can efficiently be 
combined with a \texttt{deal.II}-based Poisson solver and shows scaling results 
for a benchmark problem from plasma physics. 
Finally, Section~\ref{sec:outlook} summarizes our conclusions and points to 
further research directions.

\section{Matrix-free finite-element algorithms}\label{sec:dealii}

In the following, we describe building blocks commonly used in modern 
general-purpose FEM libraries with a special focus on triangulations,
parallelization, and efficient matrix-free methods. 

The concept of this work is generic and could in principle be built upon any
general-purpose FEM library such as MFEM~\cite{mfem2020},
DUNE~\cite{dune2010}, FEniCS~\cite{AlnaesBlechta2015a}, or
Firedrake~\cite{bercea2016structure}. Our description is, however,
specialized to the implementation in \texttt{hyper.deal} that is constructed
on top of the  \texttt{deal.II} library and uses some of naming conventions
in that project.

\subsection{Triangulation}

A triangulation $\mathcal{T}$ covers the computational domain $\Omega$ and is
defined by a set of point coordinates $\mathcal{P}$ as well as by a set of
cells $\mathcal{C}$. Each cell $c \in \mathcal{C}$ consists of a fixed number
of vertices. Since we consider unstructured quad/hex-only meshes, cells may
only consist of 2, 4, or 8 vertices.

\begin{figure}

\def\svgwidth{0.16\textwidth}{\footnotesize 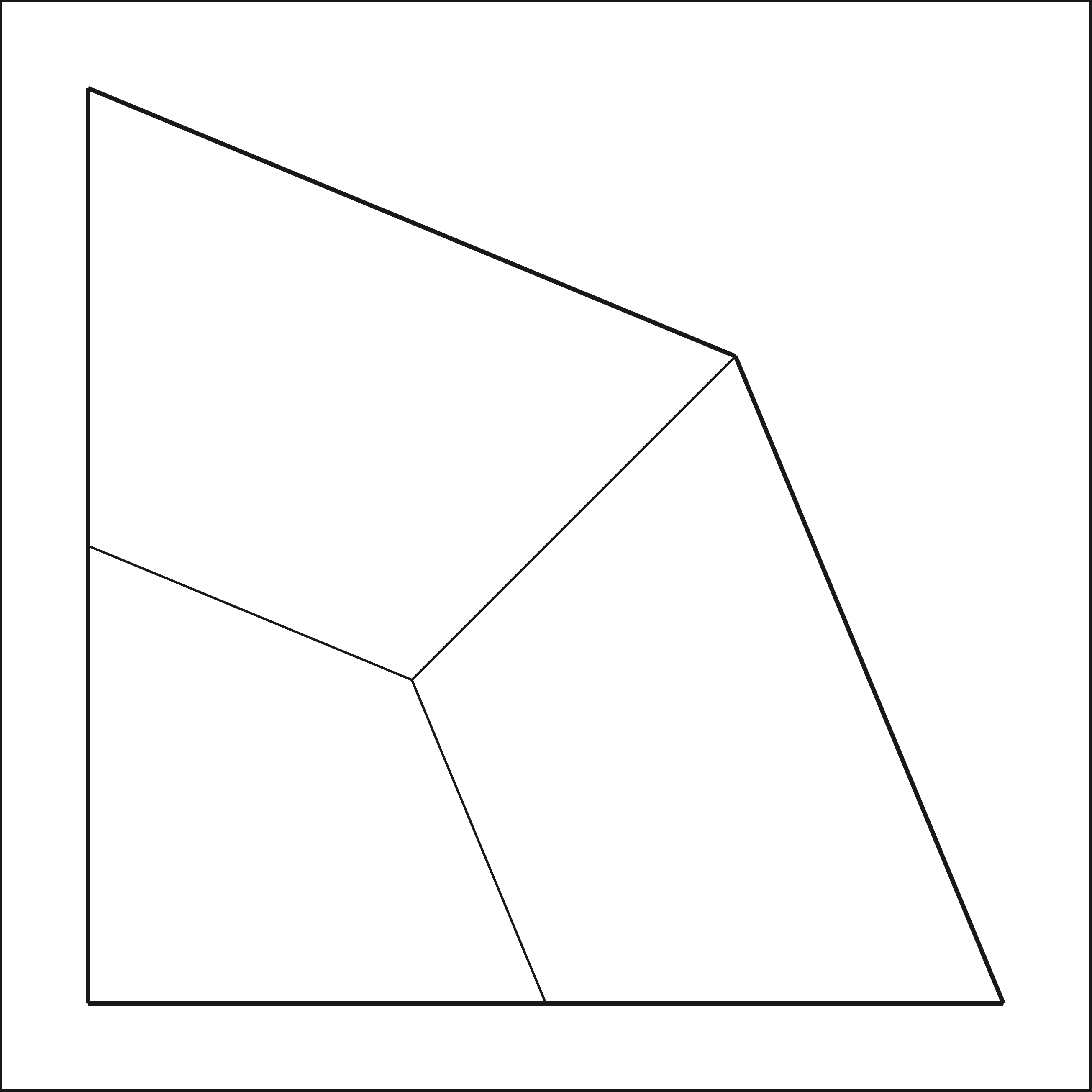}
\def\svgwidth{0.16\textwidth}{\footnotesize 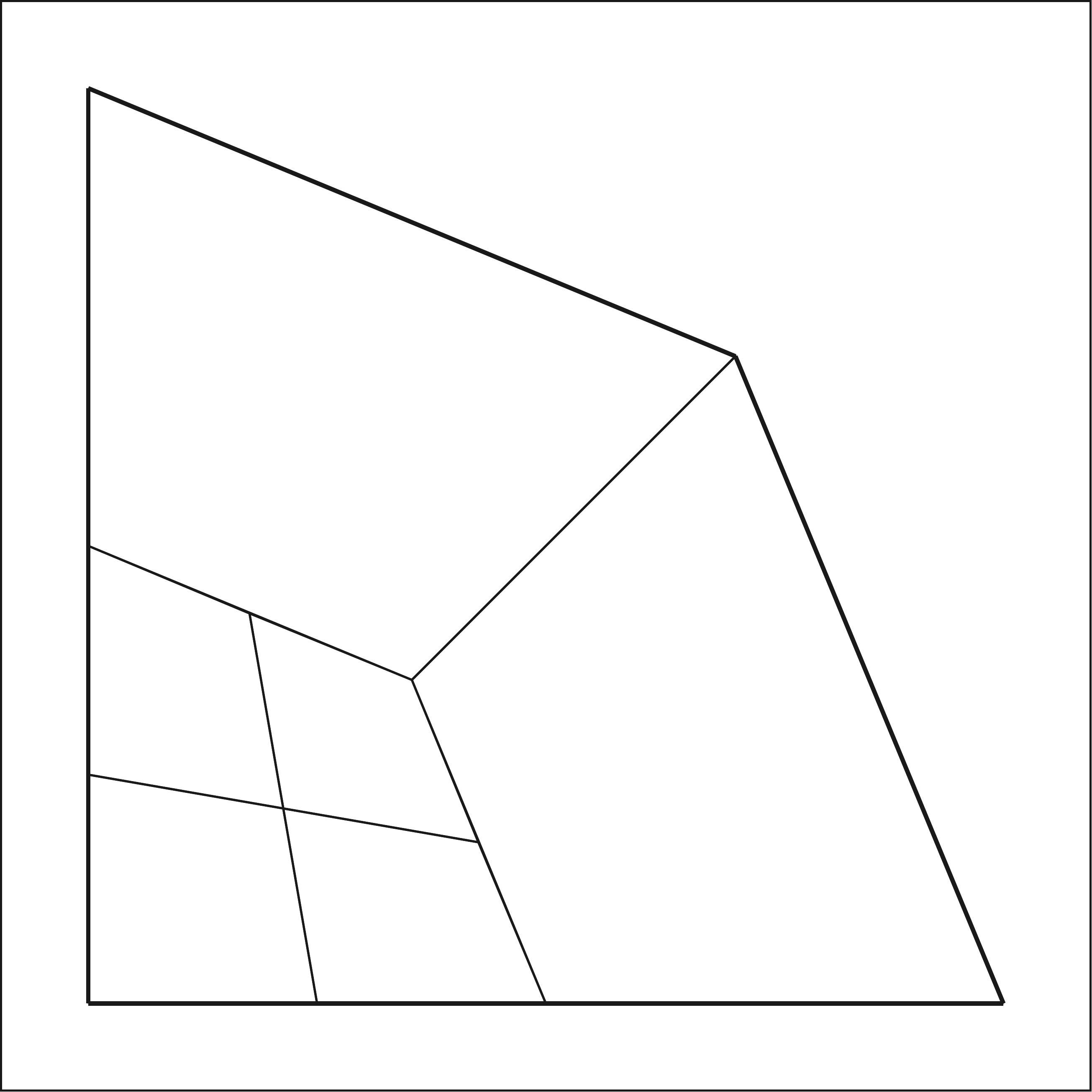}
\def\svgwidth{0.16\textwidth}{\footnotesize 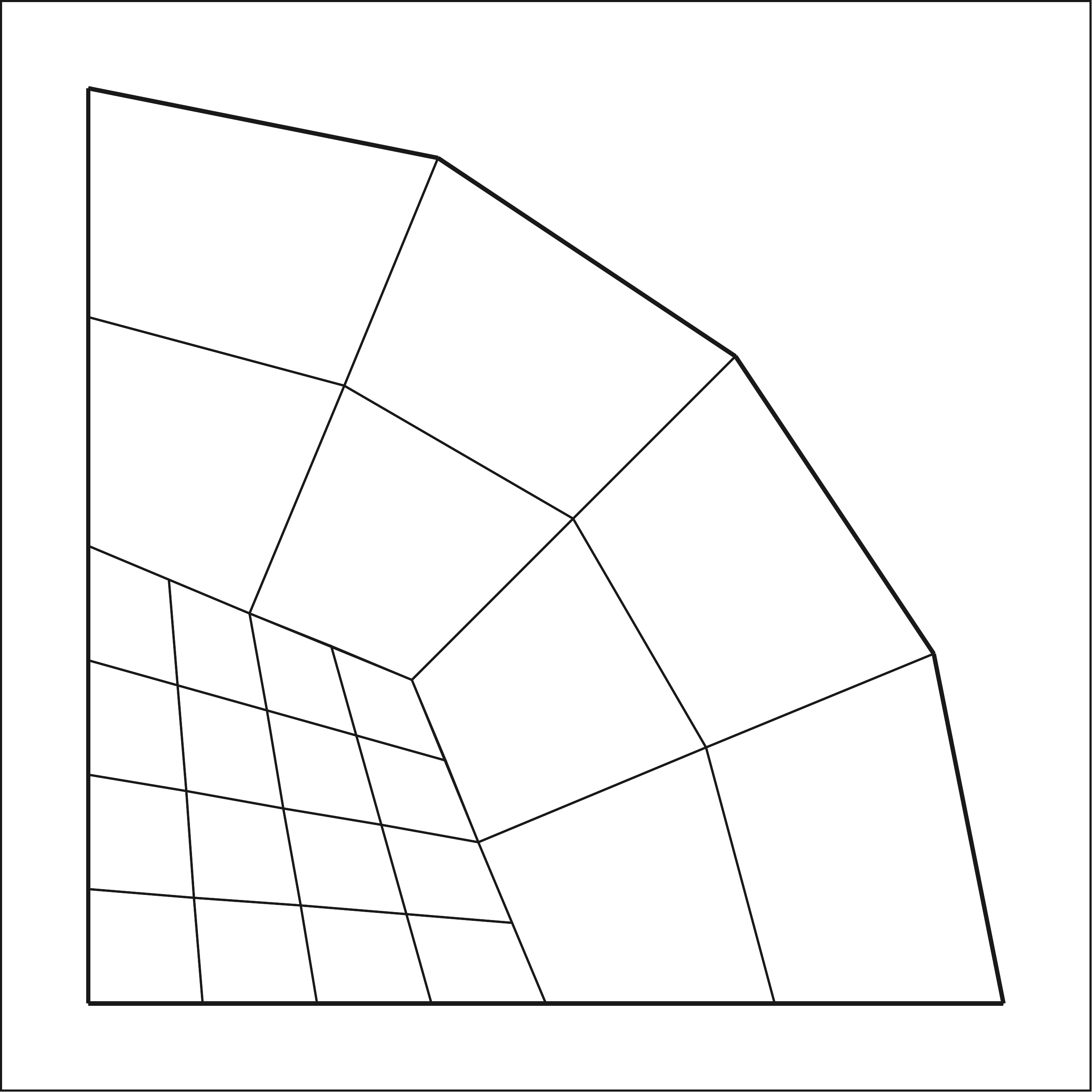}
\def\svgwidth{0.16\textwidth}{\footnotesize 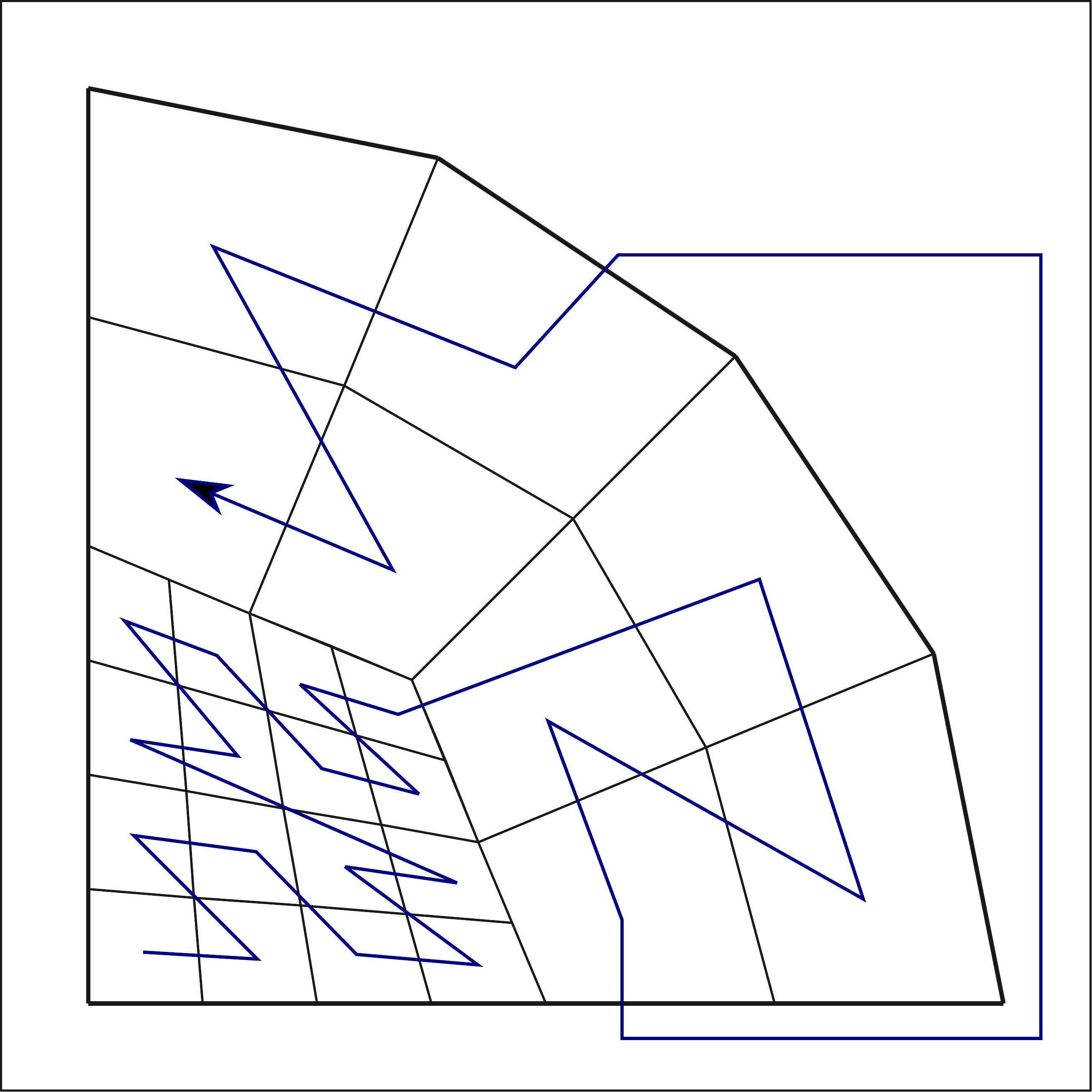}
\def\svgwidth{0.16\textwidth}{\footnotesize 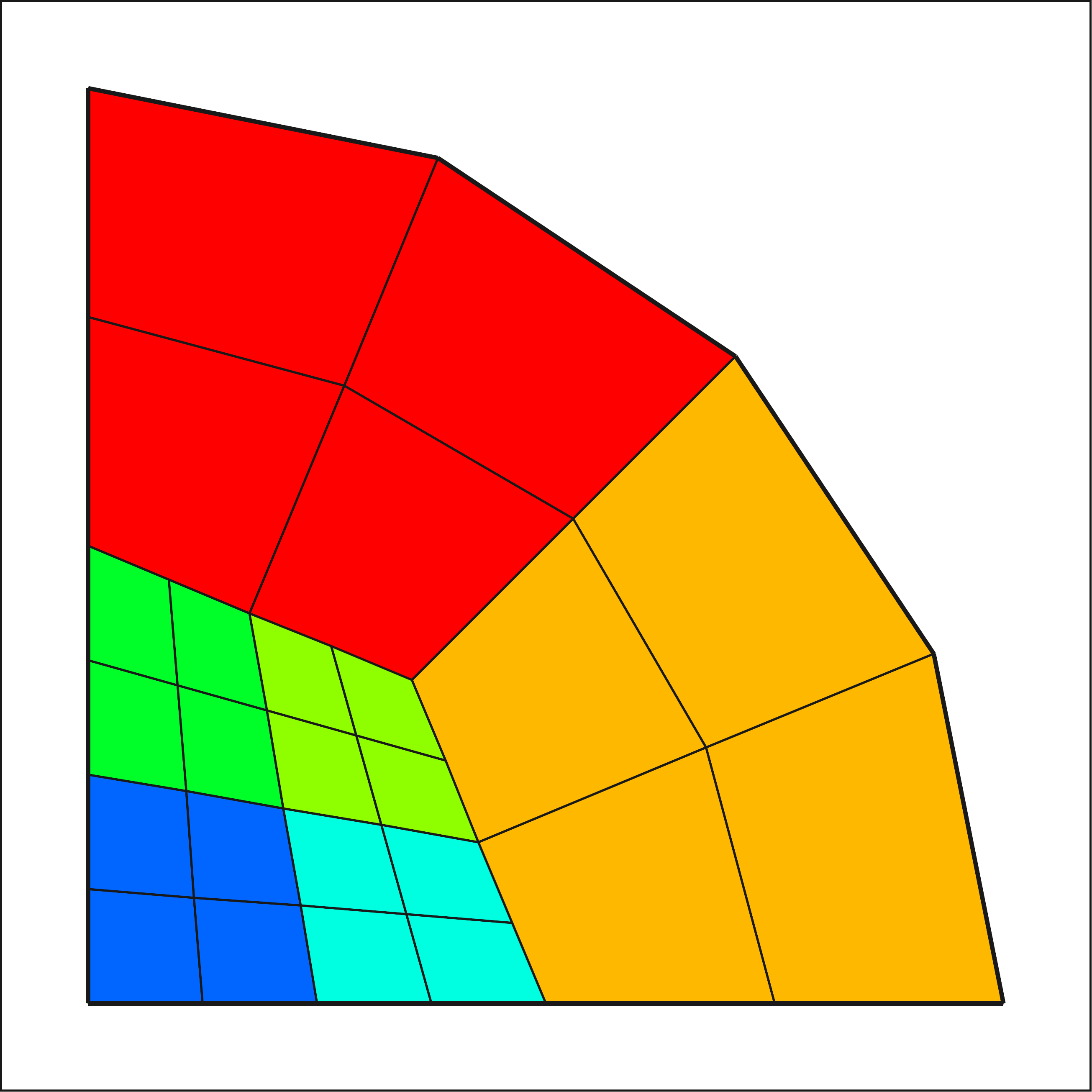}
\def\svgwidth{0.16\textwidth}{\footnotesize 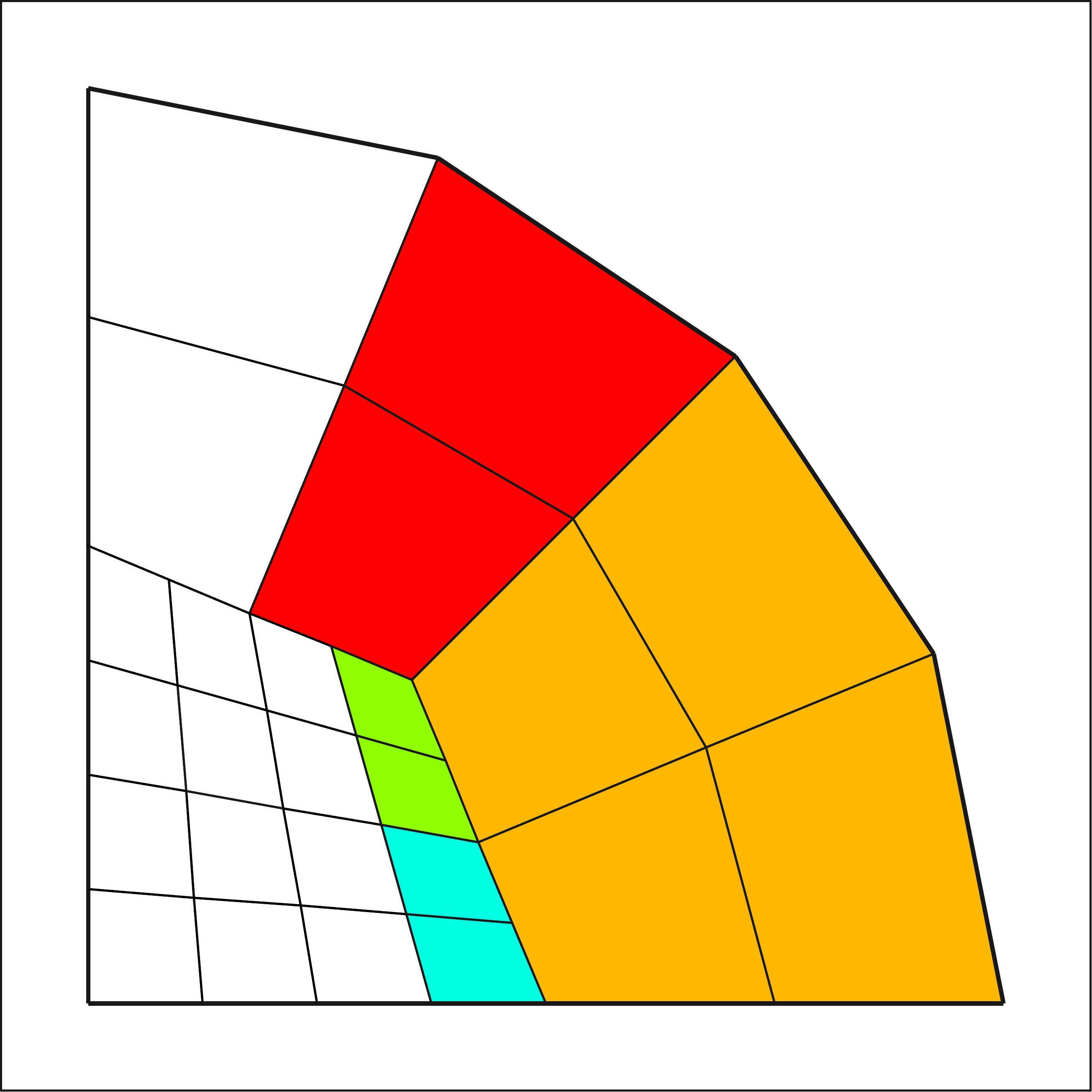}

\caption{Generation of a low-dimensional mesh in the base library
\texttt{deal.II}: a) coarse grid; b-c) refinement of a coarse grid; d)
enumeration of cells along a space-filling curve (sfc); e) partitioning of the
space-filling curve; f) local view of process 4 (showing only active local
cells and active ghost cells)}\label{fig:dealii:tria}

\end{figure}

The triangulation $\mathcal{T}$ can be the result of an external mesh generator
or of the repeated refinement of cells in a coarse mesh $\mathcal{T}_0$ in an
octree manner (see Figures~\ref{fig:dealii:tria}a-c). The latter way to create
the actual triangulation fits into the concept  of adaptive mesh refinement and
geometric multigrid methods 
(used in Section~\ref{sec:vp} for the Poisson problem), since the refinement
levels can also be used as multigrid levels. 
The cells in the mesh are continuously enumerated, which leads to a
space-filling curve (see also Figure~\ref{fig:dealii:tria}d).

\subsection{Parallelization: partitioning}

Distributed memory parallelization of finite-element codes is generally
implemented based on a domain decomposition via the Message Passing Interface 
(MPI). 
For this purpose, the space-filling curve of cells mentioned above can be 
partitioned---uniformly---among all processes within an MPI communicator 
(generally MPI\_COMM\_WORLD) so that a process $i$ is only responsible for a 
subset of (locally owned) cells 
$ \displaystyle{\biguplus_i \mathcal{L}_i = \mathcal{C}}$.

For the management of large-scale distributed meshes, libraries based on space-filling 
curves like \texttt{p4est} \cite{Bangerth2011,Burstedde2011}, which uses the Morton order/z-curves, or \texttt{Peano} \cite{weinzierl2019peano}, which uses Peano curves,
 and graph-partitioning algorithms as 
implemented in \texttt{Metis}~\cite{Karypis_1998} and \texttt{Zoltan} 
\cite{Heroux2003} are utilized.

Besides locally owned cells $\mathcal{L}_i$, each process has  a halo of ghost 
cells $\mathcal{G}_i \subseteq \mathcal{C}$ with 
$\forall i:\, \mathcal{L}_i \cap \mathcal{G}_i = \emptyset$. Regularly, the 
degrees of freedom in these ghost cells have to be exchanged between neighbors,
e.g., to evaluate fluxes in DG methods, requiring communication (see also 
Subsection~\ref{sec:dealii:fe}).

Cells owned by a process can be processed in parallel by threads. They can 
exploit shared memory, however, they have to be synchronized  to prevent race
conditions. For this purpose, the libraries \texttt{TBB} and \texttt{OpenMP} are 
integrated in the \texttt{deal.II} library. In Subsection~\ref{sec:concept:sm}, 
we present a novel hybrid parallelization scheme, which exploits the 
shared-memory capabilities of MPI-3.0 with the advantage that no threads have to be 
created explicitly.

%
%
%

As the last type of parallelization, explicit vectorization for SIMD (single 
instruction stream, multiple data streams) can be used. A detailed explanation of 
SIMD in context of matrix-free methods follows in Subsection~
\ref{sec:dealii:matrixfree}.

%

\subsection{Elements and degrees of freedom}\label{sec:dealii:fe}

Unknowns are assigned for all cells $c \in \mathcal{C}$. 
We use $d$-dimensional scalar, discontinuous tensor-product shape functions of 
polynomial degree $k$, based on Gauss-Lobatto  support points (see also 
Figure~\ref{fig:dealii:matrixfree:steps}):
\begin{align}
\mathcal{P}_k^d = \underbrace{\mathcal{P}_k^1 \otimes ... \otimes \mathcal{P}_k^1 }_{\times d}.
\end{align}
In DG, unknowns are not shared between cells so that each cell holds 
$(k+1)^d$ unknowns. The total number of degrees of freedom (DoF) is:
\begin{align}\label{eq:dealii:totalunkonwns}
N = |\mathcal{C}| \cdot (k+1)^d .
\end{align}
The unknowns are coupled via fluxes of DG schemes (see also 
Section~\ref{sec:into}). To be able to compute the fluxes, the degrees of 
freedom of neighboring cells have to be accessed. The dependency region 
for computing all contributions of a cell is in the case of advection: 
\begin{align}
\underbrace{(k+1)^d}_{\text{cell}} + \underbrace{2\cdot d \cdot (k+1)^{d-1}}_{\text{faces}},
\end{align}
i.e., the union of all unknowns of the cell and the unknowns residing on faces 
of the $2 \cdot d$ neighboring cells due to the nodal polynomials, which 
are combined with nodes at the boundary (see 
Figure~\ref{fig:dealii:matrixfree:steps}).

\subsection{Parallelization: communication}\label{sec:dealii:comm}
Due to the selection of nodal polynomials with nodes at the element faces, 
the amount of data (in doubles) to be transferred during ghost-value 
exchanges can be estimated as follows
\begin{align}\label{eq:dealii:comm:amount}
| \mathcal{L}_i' |^{\frac{d-1}{d}} \le \frac{(\texttt{data amount})_i}{2 \cdot d \cdot  (k+1)^{d-1}} \le |\mathcal{L}_i|, \quad i=0,\ldots, p-1,
\end{align}
where $p$ is the number of partitions. The lower bound is given by the data to 
be transferred on an idealized hypercube partition with 
$|\mathcal{L}_i'| \approx \hat{N} /(k+1)^d$ cells, where $\hat{N} \approx N/p$ 
is the averaged number of locally owned degrees of freedom of process $i$,  
and the upper bound is defined by the case that all locally owned cells are unconnected.
The total amount of data to be transferred,
\begin{align}\label{eq:dealii:comm:tamount}
2 \cdot N^{\frac{d-1}{d}} \cdot p^{\frac{1}{d}}
\le
\sum_i(\texttt{data amount})_i ,
\end{align}
can be minimized by well-shaped---hypercube-like---partitions and by
subdomains containing a large amount of cells. The latter approach competes 
with the domain-decomposition parallelization approach of a given mesh 
size. However, the amount of data exchange becomes smaller when using 
shared memory (see also Section~\ref{sec:concept:sm}).



The partitioning does not only influence the data volume to be transferred 
but also the memory overhead resulting from the need to allocate memory for 
at least two buffers of a size bounded by Equation~\eqref{eq:dealii:comm:tamount} 
for sending and receiving the data. Generally, one buffer is part of the actual 
distributed solution vector, which makes the data access to the degrees of 
freedom of neighboring cells simpler during operator evaluations. At least on one 
side of the transfer, data has to be packed and unpacked before or after each 
communication step, leading to the requirement of the second 
buffer~\cite{Kronbichler2017a}.

\subsection{Quadrature}

%
%
%

Similar to the shape functions, arbitrary dimensional quadrature rules are 
expressed as a tensor product of 1D quadrature rules (with $n_q$ points). For the 
cell integral, we get
\begin{align}
\mathcal{Q}^d_{n_q}
=
\underbrace{
\mathcal{Q}^1_{n_q}
\otimes
...
\otimes
\mathcal{Q}^1_{n_q}
}_{\times d}
\end{align}
with the evaluation points given as 
$\vec{x}^d_{n_q}= x^1_{n_q} \otimes ... \otimes x^1_{n_q}$ and the quadrature 
weights as  $w^d_{n_q}= w^1_{n_q} \cdot ... \cdot w^1_{n_q}$.  For the face 
integrals of the faces $2f$ and $2f+1$, we have
\begin{align}
\mathcal{Q}^{d-1}_{n_q}
=
\underbrace{
\mathcal{Q}^1_{n_q}
\otimes
\dots
\otimes
\mathcal{Q}^1_{n_q}
}_{\times (f-1)}
\otimes
\mathcal{Q}^1_f
\otimes
\underbrace{
\mathcal{Q}^1_{n_q}
\otimes
\dots
\otimes
\mathcal{Q}^1_{n_q}
}_{\times (d-f)}
\quad
\texttt{with}
\quad
\mathcal{Q}^1_f \in \{0,1\}.
\end{align}
A widespread approach is the Gauss--Legendre family of quadrature rules (see also 
Figure~\ref{fig:dealii:matrixfree:steps}), which are 
exact for polynomials of degree $2n_q-1$. Most efficient implementations of these 
rules perform a non-trivial interpolation operation from the Gauss--Lobatto to 
the Gauss--Legendre points. Subsection~\ref{sec:dealii:matrixfree} discusses an 
efficient approach to perform this basis change and to compute gradients at the 
quadrature points.

An alternative is the Gauss--Lobatto family of quadrature rules, which do not 
require a basis change, however, are only exact for polynomials of degree 
$2n_q-3$. We will consider the Gauss--Legendre family of quadrature rules to 
quantify the overhead resulting from the basis change.

To be able to evaluate Equation~\eqref{eq:into:adv:ref}, the Jacobian matrix $
\mathcal{J}_q \in \mathbb{R}^{d\times d}$, which can be derived from the element 
geometry, and its determinant $|\mathcal{J}_q| \in \mathbb{R}$ are needed at each 
cell quadrature point. At face quadrature points, the  Jacobian determinant and 
the face normal $\vec{n}_q \in \mathbb{R}^d$ are needed. These quantities are 
generally precomputed once during initialization. This leads to an additional 
memory consumption per final unknown:
\begin{align}
\frac{n^d \cdot (d^2+1) + 2 \cdot d \cdot n^{d-1}(d+1)}{(k+1)^d} = \mathcal{O} (d^2).
\end{align}

For affine meshes, only one set of mapping quantities has to be precomputed and 
cached, since they are the same for all quadrature points. 
As we are considering complex non--affine meshes in this paper, we will use 
simulations with these optimization techniques specific for affine or Cartesian 
grid only to quantify the quality of our implementation.

\subsection{Matrix-free operator evaluation}\label{sec:dealii:matrixfree}

\begin{figure}[t]
\centering

    \hspace{-0.9cm}\def\svgwidth{0.78\columnwidth}
    {\footnotesize 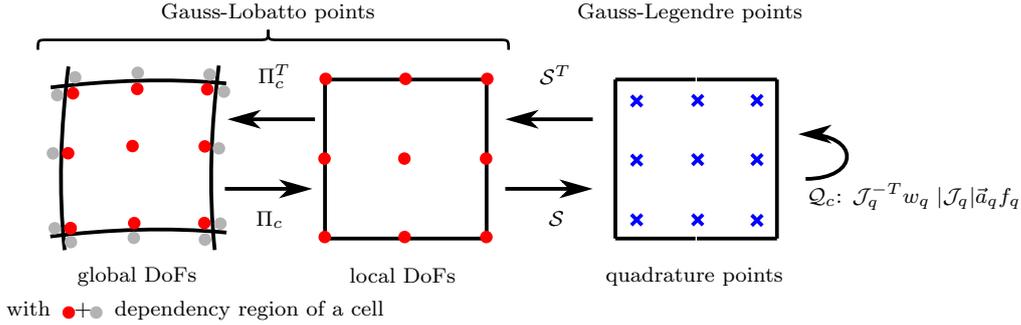}
    
    \vspace{0.3cm}

\caption{Visualization of the 5 steps of a matrix-free cell-integral evaluation for polynomial degree $k=2$ and number of quadrature points $n_q=3$.}\label{fig:dealii:matrixfree:steps}

\end{figure}


Let us now turn to the cell integral of the advection operator $\mathcal{A}$ (see 
Equation~\ref{eq:into:adv:ref}) as an example:
\begin{align}
- \bigg( \mathcal{J}^{-T} \nabla_{\vec{\xi}} v,\; |\mathcal{J}|  \vec{a} f \bigg)_{\Omega^{(e)}_0} \approx
 -\sum_q \bigg( \nabla_{\vec{\xi}} v_q,\; {\mathcal{J}_q^{-1} w_q \;  |\mathcal{J}_q|  \vec{a}_q } f_q \bigg)_{\Omega^{(e)}_0}.
\end{align}
As visualized in Figure~\ref{fig:dealii:matrixfree:steps}, the steps of the 
evaluation of cell integrals are as follows: 1) gather $(k+1)^d$ cell-local 
values $f_i$,
2) interpolate values to quadrature points $f_q$, 
3) perform the operations ${\mathcal{J}_q^{-1} w_q \;  |\mathcal{J}_q|  \vec{a}_q \, f_q}$ at each quadrature point, 
4) test with the gradient of the shape functions, and
5) write back the local contributions into the global vector.

The most efficient implementations of the basis change (from Gauss--Lobatto to 
Gauss--Legendre points) perform a sequence of $d$ 1D interpolation sweeps, 
utilizing the tensor-product form of the shape functions in an even-odd 
decomposition fashion with 
$(3 + 2 \cdot \floor{((k-1) \cdot (k+1)/2)}/(k-1) )$ FLOPs/DoF 
(for $k+1=n_q$) \cite{Kronbichler2017a}. This operation is known as sum 
factorization and has its origin in the spectral-element 
community~\cite{Deville2003,Melenk99fullydiscrete,Orszag1980}. 
Similarly, the testing with the gradient of the shape functions can be 
performed efficiently with $2d$ sweeps \cite{Kronbichler2017a}.

The same five steps can be performed for all faces $f\in \mathcal{F}$ to compute 
the flux between two neighboring cells once (with the difference that steps 1--2 
and 4--5 are performed for both sides of a face) in a separate loop known as 
``face-centric loop'' (FCL). However, recently the benefits of processing a cell 
and in direct succession processing all its $2d$ faces (i.e., visiting all faces 
twice), as in the case of ``element-centric loops'' (ECL), have become clear for 
modern CPU processor architecture in the literature \cite{Kronbichler2017a},
although this kind of loop implies that fluxes have to be computed twice (for 
each side of an interior face).
The reasons for the advantage of ECL are twofold: On the one hand, entries in the 
solution vector are written exactly only once back to main memory in the case of 
ECL, while in the case of FCL at least once---despite of cache-efficient 
scheduling of cell and face loops---due to cache capacity misses. On the other 
hand, since each entry of the solution vector is accessed only exactly once, no 
synchronization between threads is needed while accessing  the solution vector in 
the case of ECL. This absence of race conditions during writing into the 
destination vector makes ECL particularly suitable for shared-memory 
parallelization. Since the exploitation of shared memory is a key ingredient for 
the reduction of the communication and as a consequence of the memory 
consumption (as already discussed in Subsection~\ref{sec:dealii:comm}), we 
believe that ECL is a suitable approach to enable computations with larger 
problem sizes despite of the possible increase in computations. 
 
One should also note that although fluxes are computed twice in the case of ECL, 
this does not automatically translate into doubling of the computation, since 
values already interpolated to the cell quadrature points can be interpolated to 
a face with a single 1D sweep.

%
%

%

It is common practice to process a batch of $v_{len}$ cells by one instruction, 
known as vectorization over elements. This kind of processing can be accomplished 
by not operating directly on the primitive types \texttt{double}/\texttt{float} 
but on structs built around intrinsic instructions, with each vector lane 
dedicated to a separate cell of mesh. Depending on the given hardware, up to 
$v_{len}$=$2$ (SSE2), =$4$ (AVX), and =$8$ doubles (AVX-512 instruction-set
extension---as most modern Intel-based processors have) can be 
processed by a single instruction. Although the working-set size increases by a 
factor of $v_{len}$ \cite{Kempf2018}, this vectorization strategy showed a better 
performance than alternative vectorization strategies in 2D and 3D (for their 
description see \cite{Kronbichler2017a}) and the auto-vectorization by the 
compiler.

As the conclusion of this section, Table~\ref{tab:dealii:workingset} gives a 
rough estimate of the working sets for the matrix-free evaluation of the 
advection operator at different stages of the algorithm. 
Table~\ref{tab:dealii:mem} shows the estimated minimal memory consumption of the 
complete simulation. The number of ghost degrees of freedom also gives an  
estimate for the amount of data to be communicated.

\begin{table}

\centering 

\caption{Estimated working set of different stages of a matrix-free evaluation of 
the advection operator for ECL and vectorization over elements: (1) includes both 
the source and the destination element vector, (2) includes the buffers needed during 
testing and face evaluation, (3) includes the degrees of freedom of neighboring 
cells, needed during flux computation.}\label{tab:dealii:workingset}

\begin{footnotesize}
\begin{tabular}{llcc}
\toprule
&\textbf{stage} & & \textbf{working set}\\
\midrule
(1)& sum factorization   & $>$   & $ v_{len} \cdot \max\left((k+1)^d,\,n^d_q\right)$ \\
(2)& derived quantities  & $>$   & $  (d+1) \cdot v_{len} \cdot n^d_q$\\
(3)& flux computation    & $\gg$ & $  (2 \cdot d+1) \cdot v_{len} \cdot n^d_q$\\
\bottomrule
\end{tabular}
\end{footnotesize}

\caption{Estimated minimal memory consumption for $d$-dimensional simulations 
with $N$ degrees of freedom on $p$ processes.}\label{tab:dealii:mem}
\begin{footnotesize}
\begin{tabular}{lllclc}
\toprule 
    & \textbf{reason} & & \textbf{times} & & \textbf{amount} \\
\midrule
(1) & ghost DoFs (+buffer) &  & $2$   & $\cdot$ & $2\cdot d \cdot N^{\frac{d-1}{d}} \cdot p^{\frac{1}{d}}$ \\
(2) & vector              & $+$ & $2$ & $\cdot$ & $N$ \\
(3) & mapping             & $+$ &       &         & $d^2 \cdot N$ \\
\bottomrule
\end{tabular}
\end{footnotesize}
\end{table}

\subsection{Potential problems in higher dimensions}\label{sec:dealii:problems}

The descriptions in Sections~\ref{sec:into} and \ref{sec:dealii} regarding 
finite-element methods have been general for $d$-dimensional space $\Omega$. As a 
consequence, all algorithms also work in high dimensions. However, we face two 
major challenges: The first one is the lack of libraries designed for high 
dimensions. For example, the libraries \texttt{deal.II} and \texttt{p4est}  are 
limited to dimensions up to three. If we would ``na\"ively'' extend these 
libraries for high dimensions, the second problem would be that some algorithms
do not scale to higher dimensions. The following 
specific difficulties would arise:
\begin{enumerate}
\item \textbf{Significant memory overhead due to ghost values and mapping}: In 
high dimensions, solution vectors (2--3 are needed, depending on the selected 
time discretization scheme) are  huge ($\mathcal{O}(N^d_{1D})$), with $N_{1D}$ 
the number of degrees of freedom in each direction, which are needed to achieve 
the required resolution. Also the ghost values and the mapping have significant 
memory requirements in high dimensions: The evaluation of the advection cell 
integral on complex geometries needs among other things the Jacobian matrix of 
size $\mathcal{O}(d^2)$ at each quadrature point. If precomputed, this implies an 
at least 36-fold memory consumption as the actual vector in 6D. For high-
dimensional problems, this is not feasible, as only little memory would remain 
for the actual solution vectors and only problems with significantly smaller 
resolutions  could be solved. 
\label{issue:dealli::memory}

\item \textbf{Increased ghost-value exchange due to increased surface-to-volume 
ratio}: 
The communication amount scales---according to 
Equation~\eqref{eq:dealii:comm:tamount}---with 
$2\cdot d \cdot N^{(d-1)/d} \cdot p^{1/d}$. According to 
\cite{Kronbichler2017a}, the MPI ghost-value exchange already leads to a 
noticeable share of time in  purely MPI-parallelized applications (30\% for 
Laplacian) in comparison to the highly efficient matrix-free operator evaluations 
if computations are performed on a single compute node. 
For high dimensions, the situation is even worse: an estimation with $d=6$, 
$N=10^{12}$, $p=1024\cdot 48$ (1024 compute nodes with 48 processes each) gives 
that the size of the ghost values is at least $72\%$ of the size of the actual 
solution vector. The usage of shared memory is inevitable to decrease  the memory 
consumption and the time spent in communication: For the given example, the size 
of ghost values could be halved to $37\%$  if all 48 processes on a compute node 
shared their locally owned values.
\label{issue:dealli::comm}


\item \textbf{Decreased efficiency of the operator evaluation due to exponential increasing size of working sets}: 
The working set of a cell is at least $\mathcal{O}(\max((k+1)^{d},n_q^{d}))$ 
(cf.~Table~\ref{tab:dealii:workingset}) so that for high order and/or 
dimension the data eventually drops out of the cache during each 
sum-factorization sweep of one cell.
\label{issue:dealli::performance}
\end{enumerate}
While this work cannot solve all these difficulties, we will show how it is
possible to mitigate them: We address problem~(\ref{issue:dealli::memory}) by 
restricting ourselves to the tensor product of two grids in 1-3D. This reduces 
the size of the mapping data and makes it, moreover, possible to reuse much of 
the infrastructure available in the baseline library \texttt{deal.II}. We will 
describe in the next section how such a tensor product can be formed. 
Problem~(\ref{issue:dealli::comm}) demonstrates that it is essential to exploit 
shared-memory parallelism in particular in high dimensions. In 
Section~\ref{sec:concept:sm}, we propose a novel shared-memory implementation 
that is based on MPI-3.0. To mitigate problem~(\ref{issue:dealli::performance}),
we try to minimize the number of cache misses due to increased working set sizes
by reorganizing the loops. In this respect, we will demonstrate the benefit of 
vectorizing over fewer elements than the given instruction-set 
extensions allow. We accomplish this by explicitly using narrower instructions, 
such as \texttt{AVX2} or \texttt{SSE2}, instead of \texttt{AVX-512} or by 
working directly with \texttt{double}s and relying on auto-optimization of the compiler. 
We defer the investigation of explicit vectorization within elements to 
future work.

\section{hyper.deal: a tensor product of two meshes}\label{sec:tensorproduct}

In this section, we explain how a tensor product of two meshes can be used to 
solve problems in up to six dimensions.  In our sample application of an 
advection equation in phase space, separating the meshes in configuration space 
as well as in velocity space is natural, which is why we use the indices 
$\vec{x}$ and $\vec{v}$ for the two parts of the dimensions. 

In the following, we assume that our computational domain is given as a tensor 
product of two domains
$\Omega := \Omega_{\vec{x}} \otimes \Omega_{\vec{v}}$.
The boundary of the high-dimensional domain is then described by
$\Gamma := (\Gamma_{\vec{x}} \otimes \Omega_{\vec{v}}) \cup (\Omega_{\vec{x}} \otimes \Gamma_{\vec{v}})$. 
Let us reformulate and simplify the discretized advection equation (see 
Equation.~\eqref{eq:into:adv:ref})
for the phase space, exploiting the fact that the Jacobian matrix $\mathcal{J}$ 
is a block-diagonal matrix in phase space,
\begin{align}
\mathcal{J} = \left(
\begin{array}{cc}
\mathcal{J}_{\vec{x}} & 0 \\
0 & \mathcal{J}_{\vec{v}}
\end{array}
\right),
\end{align}
with the blocks being the respective matrices of the $\vec{x}$-space and the 
$\vec{v}$-space.
As a consequence, the inverse $\mathcal{J}^{-1}$ is the inverse of each of its 
blocks and  
its determinant $|\mathcal{J}|$ is the product of the determinants of each of its 
blocks.
Exploiting the block-diagonal structure of $\mathcal{J}$, some operations in 
Equation~\eqref{eq:into:adv:ref} can be simplified:
\begin{align}
\mathcal{J}^{-1} (\vec{a} f) = 
\left(
\begin{array}{cc}
\mathcal{J}_{\vec{x}}^{-1} & 0 \\
0 & \mathcal{J}_{\vec{v}}^{-1}
\end{array}
\right)
\left(\begin{array}{c}
\vec{a}_{\vec{x}} \\ \vec{a}_{\vec{v}}
\end{array}\right) f
=
\left(\begin{array}{c}
\mathcal{J}_{\vec{x}}^{-1} \vec{a}_{\vec{x}} \\ 
\mathcal{J}_{\vec{v}}^{-1} \vec{a}_{\vec{v}}
\end{array}\right) f.
\end{align}
Furthermore, face integrals over the faces in phase space $\Gamma^{(e)} = 
(\Gamma_{\vec{x}}^{(e)} \otimes \Omega_{\vec{v}}^{(e)})
\;\cup\;
(\Omega_{\vec{x}}^{(e)} \otimes \Gamma_{\vec{v}}^{(e)})$ can be split into 
integration over $\vec{x}$-space faces 
$\Gamma_{\vec{x}}^{(e)} \otimes \Omega_{\vec{v}}^{(e)}$ and integration over 
$\vec{v}$-space faces  
$\Omega_{\vec{x}}^{(e)} \otimes \Gamma_{\vec{v}}^{(e)}$. The following 
relation is true for $\vec{x}$-space faces: Due to 
$\vec{n}^\top=(\vec{n}^\top, 0)$ for $\vec{x}$-space faces, 
$\vec{n} \cdot \vec{a} = \vec{n}_{\vec{x}} \cdot \vec{a}_{\vec{x}}$;
and the same is true for $\vec{v}$-space faces 
$\vec{n} \cdot \vec{a} = \vec{n}_{\vec{v}} \cdot \vec{a}_{\vec{v}}$.
Finally,  Equation~\eqref{eq:into:adv:ref}, which can be solved efficiently 
with \texttt{hyper.deal}, is:
\begin{align} \nonumber
\left(g, \frac{\partial f}{\partial\, t}\right)_{\Omega^{(e)}} &- \bigg(  \nabla_\xi g,  \left(
\begin{array}{c}
\mathcal{J}^{-1}_{\vec{x}} \vec{a}_{\vec{x}} \\
\mathcal{J}^{-1}_{\vec{v}} \vec{a}_{\vec{v}} \\
\end{array}
\right) f \bigg)_{\Omega^{(e)}} \\ &= 
\bigg( g,\, \vec{n}_{\vec{x}} \cdot (\vec{a}_{\vec{x}} f)^* \bigg)_{\Gamma_{\vec{x}}^{(e)} \otimes \Omega_{\vec{v}}^{(e)}}
+
\bigg( g,\, \vec{n}_{\vec{v}} \cdot (\vec{a}_{\vec{v}} f)^* \bigg)_{\Omega_{\vec{x}}^{(e)} \otimes \Gamma_{\vec{v}}^{(e)}}.
\end{align}

\subsection{Triangulation}

\begin{figure}[!]

\centering

\def\svgwidth{0.8\textwidth}{\footnotesize 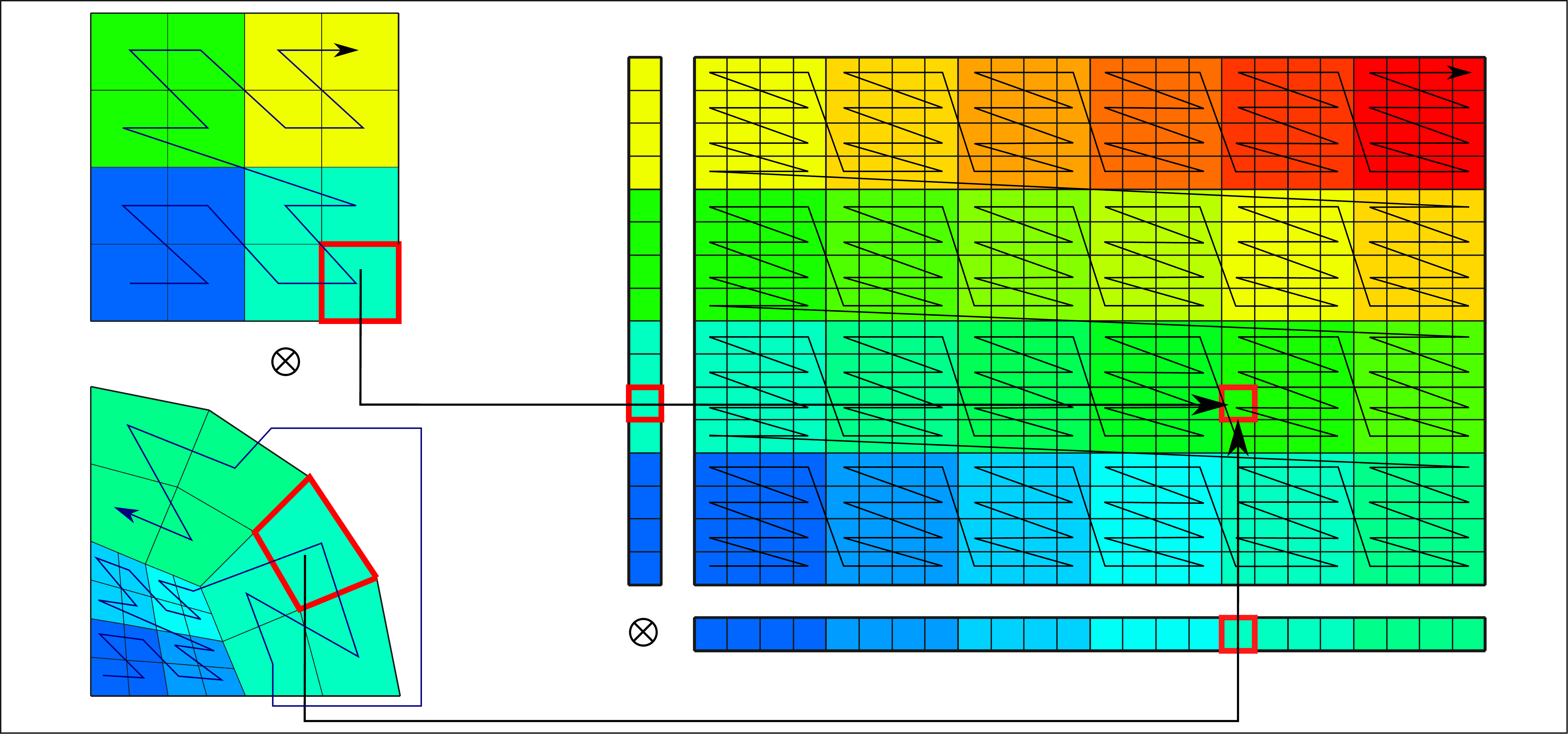}

\caption{On-the-fly mesh generation of a distributed triangulation in
 \texttt{hyper.deal} by taking the tensor product of  two low-dimensional 
triangulations  from the base library \texttt{deal.II}. Cells are ordered 
lexicographically within a process (as are the processes themselves), leading to 
the depicted space-filling curve.}\label{fig:concept:mesh:tp:cells}

\def\svgwidth{0.8\textwidth}{\footnotesize 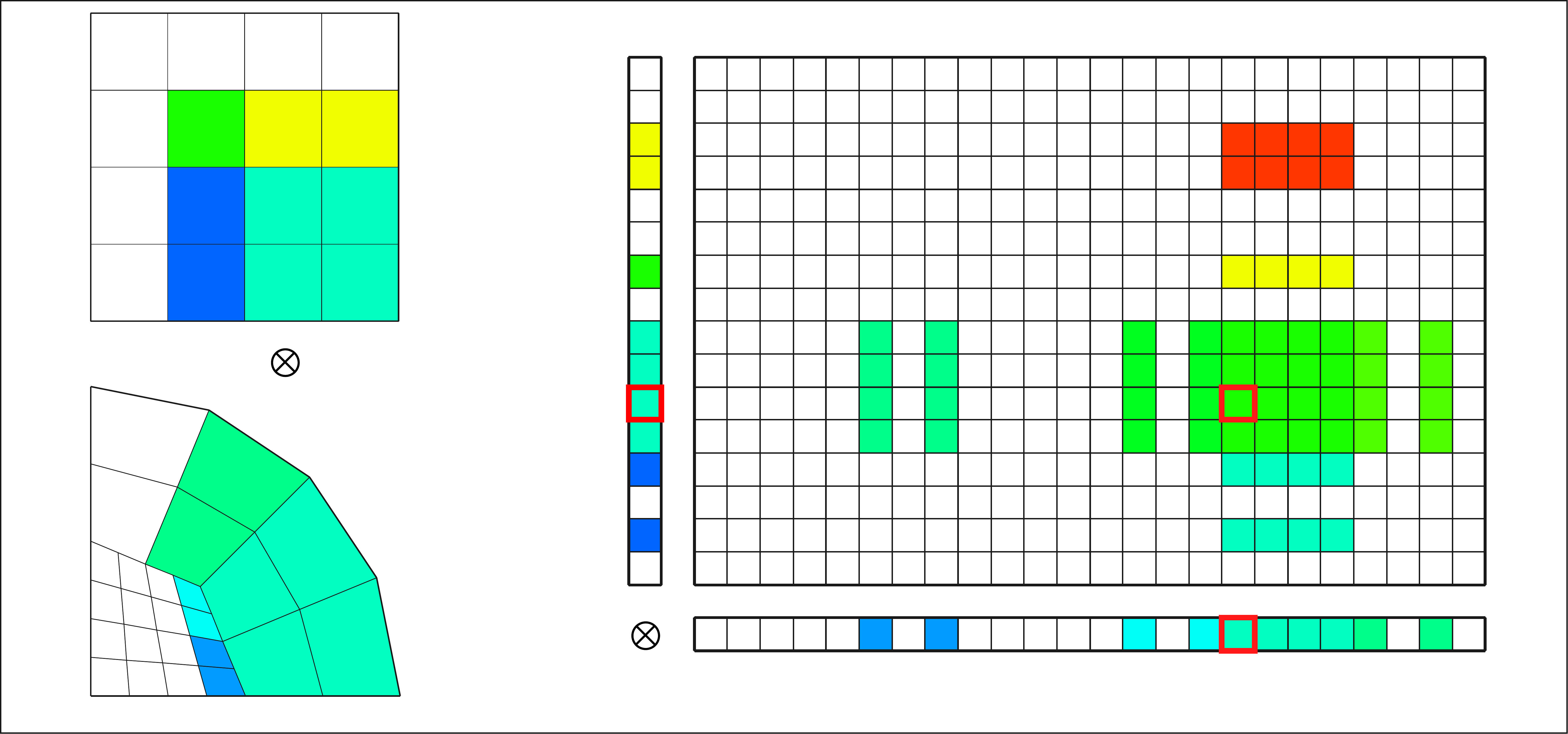}

\caption{Local view of an arbitrary process: local cells and ghost cells}\label{fig:concept:mesh:tp:ghost}

\includegraphics[width=0.8\textwidth]{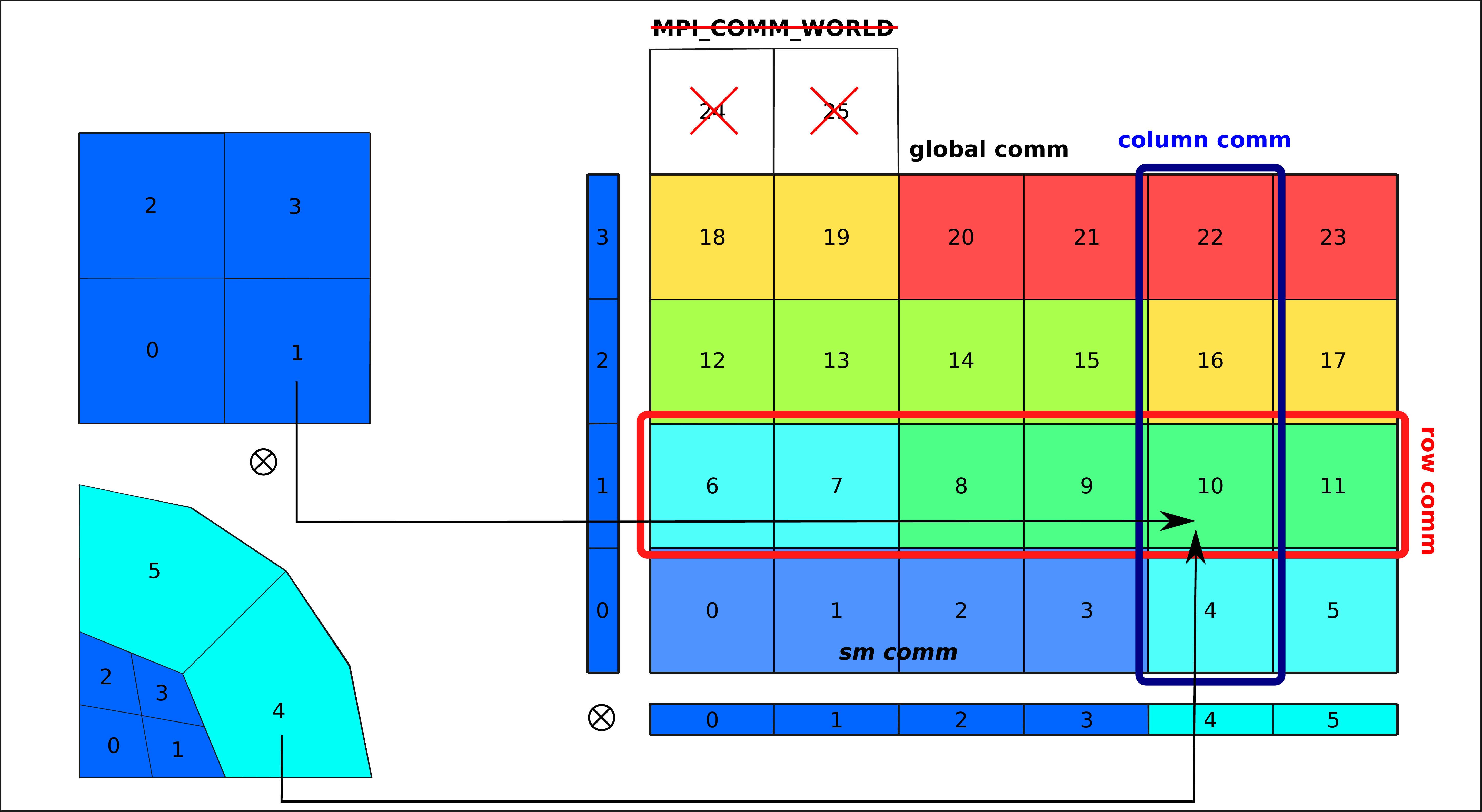}

\caption{Four MPI communicators used in \texttt{hyper.deal} (\texttt{global}, 
\texttt{row}, \texttt{column}, \texttt{sm} comm) for a hypothetical setup of 26 
ranks in \texttt{MPI\_COMM\_WORLD} and of a 6$\times4$ partition of the 4D space. 
}\label{fig:concept:mesh:tp:mpi}

\end{figure}

Naturally, both domains $\Omega_{\vec{x}}$ and $\Omega_{\vec{v}}$ can be meshed 
separately. Note, however, that this restricts the possibilities of mesh 
refinement to the two spaces separately from each other. As a consequence, the 
final triangulation results from the tensor product of the two triangulations 
$\mathcal{T}_{\vec{x}}$ and $\mathcal{T}_{\vec{v}}$ (visualized in 
Figure~\ref{fig:concept:mesh:tp:cells}),
\begin{align}\label{eq:concept:tria:tensor}
\mathcal{T} :=
\mathcal{T}_{\vec{x}} \otimes \mathcal{T}_{\vec{v}}.
\end{align}
In this context, cells $\mathcal{C}$, inner faces $\mathcal{I}$, and 
boundary faces $\mathcal{B}$ are defined as
\begin{align}\label{eq:concept:tria:primitives}
\mathcal{C} := \mathcal{C}_{\vec{x}} \otimes \mathcal{C}_{\vec{v}},\qquad
\mathcal{I} := (\mathcal{I}_{\vec{x}} \otimes \mathcal{C}_{\vec{v}}) \cup (\mathcal{C}_{\vec{x}} \otimes \mathcal{I}_{\vec{v}}),\qquad
\mathcal{B} := (\mathcal{B}_{\vec{x}} \otimes \mathcal{C}_{\vec{v}}) \cup (\mathcal{C}_{\vec{x}} \otimes \mathcal{B}_{\vec{v}}).
\end{align}
The cells in both low-dimensional triangulations are enumerated independently 
along space-filling curves. The enumeration of the cells of $\mathcal{T}$ can be 
chosen arbitrarily. However, in Subsection~\ref{sec:concept:partitioning} we 
propose an order that simplifies the initial setup.

\textit{Note: By simplifying the $d_{\vec{x}}$-dimensional triangulation 
$\mathcal{T}_{\vec{x}}$ and  the $d_{\vec{v}}$-dimensional triangulation 
$\mathcal{T}_{\vec{v}}$ to a 1D space-filling curve in each case and by 
taking the tensor product of these curves, we get---depending on the point of 
view---a 2D Cartesian grid or a matrix. As a result, we use well-known concepts 
from these fields of research in the library \texttt{hyper.deal}. However, it 
should be emphasized that, the analogy is not immediate. While the neighborship 
of a cell on a 2D grid is clear, this is not the case for the high-dimensional 
triangulation that is depicted as a 2D grid: The number of neighbors is 
significantly larger, and neighboring cells might be disjunct in this case (see 
Figure~\ref{fig:concept:mesh:tp:ghost}). What is true for individual cells is 
also true for partitions so that the communication patterns are (structured but) 
significantly more complicated and communication over the boundaries of nodes is 
unavoidable.
}

\subsection{Parallelization: partitioning}\label{sec:concept:partitioning}

Since we are using a domain decomposition from the base library \texttt{deal.II} 
for $\mathcal{T}_{\vec{x}}$ and $\mathcal{T}_{\vec{v}}$, we assume these meshes 
are already split up independently into $p_{\vec{x}}$ and $p_{\vec{v}}$ 
subpartitions with
\begin{align}
\mathcal{T}_{\vec{x}} = \biguplus_{0 \le i < p_{\vec{x}}} \mathcal{T}_{\vec{x}}^i	
\qquad \text{and} \qquad
\mathcal{T}_{\vec{v}} = \biguplus_{0 \le j < p_{\vec{v}}} \mathcal{T}_{\vec{v}}^j.
\end{align}
The subpartition of the phase-space domain belonging to rank $f(i,j)$ is 
constructed by
$\mathcal{T}^{f(i,j)} := \mathcal{T}_{\vec{x}}^i \otimes  \mathcal{T}_{\vec{v}}^j$,
where ranks are enumerated lexicographically according to 
$f(i,j):=j\cdot p_{\vec{x}} + i$ with the $\vec{x}$-rank $i$ being the fastest 
running index. As a consequence, a quasi-checkerboard partitioning is obtained 
(see Figure~\ref{fig:concept:mesh:tp:cells}). Note that it might be advantageous 
in some cases not to use the full number of processes in order to get a more 
symmetric decomposition (cf.~Figure~\ref{fig:concept:mesh:tp:mpi}). To take this 
into account, we are not working on \texttt{MPI\_COMM\_WORLD} but on a 
subcommunicator, which we will call \textit{global comm}(unicator) in the 
following. 

Please note the following relationship:
\begin{align}
\mathcal{T}_{\vec{x}} \otimes \mathcal{T}_{\vec{v}}^j = \biguplus_{f(i,j) / n_{\vec{x}} = j} \mathcal{T}^{f(i,j)}
\qquad\text{and}\qquad
\mathcal{T}_{\vec{x}}^i \otimes \mathcal{T}_{\vec{v}} = \biguplus_{f(i,j) \% n_{\vec{x}} = i} \mathcal{T}^{f(i,j)}
\end{align}
so that parallel reduction to distributed $\Omega_{\vec{x}}$-space and to 
distributed $\Omega_{\vec{v}}$-space becomes a collective communication of 
subsets of processes. Due to the importance of parallel reductions in 
mathematical operations like $\int  \mathrm{d} \Omega_{\vec{x}}$ and 
$\int  \mathrm{d} \Omega_{\vec{v}}$, we make the subsets of processes available 
via the MPI communicators \textit{column comm} and \textit{row comm}. 

We enumerate cells within a subdomain lexicographically so that we get a space-filling curve as depicted in Figure~\ref{fig:concept:mesh:tp:cells}. This enables 
us to determine a globally unique cell ID of locally owned cells and of ghost 
cells by querying the low-dimensional triangulation cells for their IDs and ranks 
without the need for communication.

As a final remark, it should be emphasized that the presented partitioning 
approach delivers good results if the low-dimensional triangulations 
$\mathcal{T}_{\vec{x}}$ and $\mathcal{T}_{\vec{v}}$ have already been partitioned 
well. Depending on the given mesh, a space-filling curve approach as supported by 
\texttt{p4est} or a graph-based approach as supported by \texttt{METIS} and 
\texttt{Zoltan} might be beneficial. 


In Subsection~\ref{sec:dealii:problems}, we have discussed  the importance of the 
usage of shared memory for solving high-dimensional problems. Placing ranks 
according to $\floor*{f(i,j)/p_{node}}$ onto the same compute node (with 
$p_{node}$ being the number of processes per node) leads to striped partitioning 
(see the colors of the blocks in Figure~\ref{fig:concept:mesh:tp:mpi}). This 
results in a suboptimal shape of the union of subpartitions belonging to the same 
compute node, leading to decreased benefit of the usage of shared memory. In 
order to improve the placing of subpartitions onto the compute nodes without 
having to change the function $f$, we are operating on a virtual topology, as 
will be presented in Section~\ref{sec:concept:sm}.

\subsection{Elements, degrees of freedom, and quadrature}\label{sec:concept:fe}

Since we are considering shape functions that are derived by the tensor product 
of 1D-shape functions, i.e., 
$\mathcal{P}_k^{d_{\vec{x}}}=\underbrace{\mathcal{P}_k^{1} \otimes \cdots \otimes \mathcal{P}_k^{1}}_{ \times d_{\vec{x}}}$ 
and  
$\mathcal{P}_k^{d_{\vec{v}}}=\underbrace{\mathcal{P}_k^{1} \otimes \cdots \otimes \mathcal{P}_k^{1}}_{ \times d_{\vec{v}}}$, 
the extension to higher dimensions is trivial
\begin{align}
\mathcal{P}_k^{d_{\vec{x}}+d_{\vec{v}}}
=
\underbrace{\mathcal{P}_k^{1} \otimes \cdots \otimes \mathcal{P}_k^{1}}_{ \times d_{\vec{x}}}
\otimes
\underbrace{\mathcal{P}_k^{1} \otimes \cdots \otimes \mathcal{P}_k^{1}}_{ \times d_{\vec{v}}}
=
\mathcal{P}_k^{d_{\vec{x}}}
\otimes
\mathcal{P}_k^{d_{\vec{v}}}.
\end{align}
This relationship is also true for the quadrature formulas so that a basis change 
can be performed---as usual---with $d_{\vec{x}}+d_{\vec{v}}$ sum-factorization 
sweeps.


%
\begin{table}

\centering

\caption{Comparison of the memory consumption (in doubles) of the mapping data 
(per quadrature point) if the phase-space structure is exploited ($J_{\vec{x}}$ 
and $J_{\vec{v}}$) and if the phase-space structure is not exploited ($J$).}\label{tab:concept:quad:mc}

\begin{footnotesize}
\begin{tabular}{lcccc|ccccc}
\toprule 
& $J_{\vec{x}}$ and $J_{\vec{v}}$ & & $J$ &  & \textbf{example:} $k=3$, $d=6$ \\
\midrule
Jacobian: 
& $\displaystyle{\frac{(k+1)^{d_{\vec{x}}} \cdot d_{\vec{x}}^2 + (k+1)^{d_{\vec{v}}} \cdot d_{\vec{v}}^2}{(k+1)^{d_{\vec{x}}+d_{\vec{v}}}} }$ 
& 
& $\displaystyle{ (d_{\vec{x}}+d_{\vec{v}})^2}$
&&0.28  $\ll$ 36
\\ \bottomrule
\end{tabular}
\end{footnotesize}

\end{table}
%
%
As a consequence, only the mapping data from lower-dimensional spaces (e.g., 
Jacobian matrix and its determinant) have to be precomputed and can be reused, 
which leads to a significantly reduced memory consumption even for complex 
geometries as shown in Table~\ref{tab:concept:quad:mc}.
%
%
%
%

\subsection{Matrix-free operator evaluation}\label{sec:concept:matrixfree}

%
%
%

The extension of operator evaluations with sum factorization to higher dimensions 
is also straightforward. Instead of looping over all cells of a triangulation, it 
requires an iteration over all possible pairs of cells from both low-dimensional 
triangulations 
$(c_{\vec{x}},c_{\vec{v}}) \in \mathcal{C}_{\vec{x}} \times \mathcal{C}_{\vec{v}} $ 
in the case of ECL. In addition to cell pairs, FCL iterates also over cell-face 
and cell-boundary-face pairs as expressed by 
Equation~\eqref{eq:concept:tria:primitives}. This comes in handy, since the 
mapping information of the cells $c_{\vec{x}}$ and $c_{\vec{v}}$ as well as of 
the faces $f_{\vec{x}}$ and $f_{\vec{v}}$ can be queried from the low-dimensional 
library independently and is only combined on the fly. The separate cell IDs 
$c_{\vec{x}}$ and $c_{\vec{v}}$ only have to  be combined when accessing the 
solution vector, which is the only data structure set up for the whole 
high-dimensional space.

\begin{figure}[th]

%
%
%


\SetAlFnt{\small}

\begin{algorithm}[H]
 \texttt{update\_ghost\_values}: Import vector values of $\vec u$ from MPI processes that are adjacent to locally owned cells \\ 
 \tcc{loop over all cell pairs}
  \ForEach{$(e_{\vec{x}}, e_{\vec{v}}) \in \mathcal{C}_{\vec{x}} \times \mathcal{C}_{\vec{v}}$}{
   process\_cell($e_{\vec{x}}$, $e_{\vec{v}}$) \tcc*{e.g., advection cell operator in Algorithm~\ref{algo:concept:cell}}
   }
 \caption{Element-centric loop for arbitrary operators}\label{algo:concept:loop}
\end{algorithm}

\begin{algorithm}[H]
\SetAlgoLined
\tcc{\underline{\smash{step 1: gather values (AoS $\rightarrow$ SoA)}}}
  gather local vector values $u_i^{(e)}$ on the cell from global input vector $\vec u$\label{algo:cell:line:gather} \\
\tcc{\underline{\smash{step 2: apply advection cell contributions}}}
  interpolate local vector values $\vec u^{(e)}$ onto quadrature points, $u_h^{e} (\vec\xi_q) = \sum_i \phi_i u_i^{(e)}$ \label{algo:cell:line:interpolate} \\
  for each quadrature index $q=(q_{\vec{x}}, q_{\vec{v}})$, prepare integrand on each quadrature point by computing $\vec t_q = 
\mathcal{J}_{(e_{\vec{x}})}^{-1} \vec c_{\vec{x}} \left( \hat{x}^{(e_{\vec{x}})}(\vec\xi_{q_{\vec{x}}}),\; \hat{x}^{(e_{\vec{v}})}(\vec\xi_{q_{\vec{v}}}) \right)
u_h^{(e)}(\vec \xi_q ) \underbrace{|\mathcal{J}_{q_{\vec{x}}}| |\mathcal{J}_{q_{\vec{v}}}| w_{q_{\vec{x}}} w_{q_{\vec{v}}}}_{"|\mathcal{J}_{(q)}|w_q"}$ and 
evaluate local integrals by quadrature $b_i = \left( \nabla_{\vec{x}} \phi_i^{co}, \vec{c}_{\vec{x}} u_h^{(e)} \right)_{\Omega_{(e)}} \approx \sum_q \nabla_{\vec{x}} \phi_i^{co} (\vec{\xi}_q) \cdot \vec{t}_q $ \tcc*{buffer $\vec{b}$}
  for each quadrature index $q=(q_{\vec{x}}, q_{\vec{v}})$, prepare integrand on each quadrature point by computing $\vec t_q = 
\mathcal{J}_{(e_{\vec{v}})}^{-1} \vec c_{\vec{v}} \left( \hat{x}^{(e_{\vec{x}})}(\vec\xi_{q_{\vec{x}}}),\; \hat{x}^{(e_{\vec{v}})}(\vec\xi_{q_{\vec{v}}}) \right)
u_h^{(e)}(\vec \xi_q ) \underbrace{|\mathcal{J}_{q_{\vec{x}}}| |\mathcal{J}_{q_{\vec{v}}}| w_{q_{\vec{x}}} w_{q_{\vec{v}}}}_{"|\mathcal{J}_{(q)}|w_q"}$ and 
evaluate local integrals by quadrature $b_i =  b_i + \left( \nabla_{\vec{v}} \phi_i^{co}, \vec{c}_{\vec{x}} u_h^{(e)} \right)_{\Omega_{(e)}} \approx b_i + \sum_q \nabla_{\vec{v}} \phi_i^{co} (\vec{\xi}_q) \cdot \vec{t}_q $ \label{algo:cell:line:quadrature} \\ 
\tcc{\underline{\smash{step 3: apply advection face contributions (loop over all 2$d$ faces of $\Omega_e$)}}} 
 \ForEach{$f \in \mathcal{F}_{(e)}$ }{
  interpolate values from cell array $\vec{u}^{(e)}$ to quadrature points of face \\
  if not on boundary, gather values from neighbor $\Omega_{e^+}$ of current face \label{algo:cell:line:face:loadneigh} \\
  interpolate $u^+$ onto face quadrature points \label{algo:cell:line:facep:interpolate} \\
  compute numerical flux and multiply by quadrature weights 
\tcc*{not shown here}
  evaluate local integrals related to cell $e$ by quadrature and add into cell contribution $b_i$ \label{algo:cell:line:face:quadrature}
 }
\tcc{\underline{\smash{step 4: apply inverse mass matrix}}}
for each quadrature index $q=(q_{\vec{x}}, q_{\vec{v}})$, prepare integrand on each quadrature point by computing $t_q = b_i w_{q_{\vec{x}}}^{-1} w_{q_{\vec{v}}}^{-1}$ and 
evaluate local integrals by quadrature $y_i^{(e)} =  \sum_q \tilde\phi_{iq} \cdot \vec{t}_q $ with $\tilde\phi_{iq} = \mathcal{V}_{iq}^{-1}$ with $\mathcal{V}_{iq}=\phi_i(\vec{\xi}_q)$ \label{algo:cell:line:mass} \\
\tcc{\underline{\smash{step 5: scatter values (SoA $\rightarrow$ AoS)}}} 
 set all contributions of cell, $\vec{y}^{(e)}$, into global result vector $\vec{y}$ \label{algo:cell:line:scatter}
 \caption{DG integration of a cell batch for advection operator evaluation for ECL and for vectorization over elements}\label{algo:concept:cell}
\end{algorithm}


\end{figure}

Algorithm~\ref{algo:concept:loop}--\ref{algo:concept:cell} show the pseudocode of 
a possible matrix-free advection operator evaluation. 
As an example, Algorithm~\ref{algo:concept:loop} contains an element-centric loop 
(ECL) iterating over all cell pairs 
and calling the function that should be evaluated on the cell pairs. Up to here, 
the algorithm is independent of the equation to be solved; the equation comes 
into play (apart from the specific ghost-value update) by the called function, 
which might be the advection operator in Algorithm~\ref{algo:concept:cell}. 

Furthermore, we note that 
lines~\ref{algo:cell:line:interpolate}--\ref{algo:cell:line:quadrature}, 
\ref{algo:cell:line:facep:interpolate}, \ref{algo:cell:line:face:quadrature}, 
\ref{algo:cell:line:mass} in Algorithm~\ref{algo:concept:cell} are evaluated 
with sum factorization. To reduce the working set, we do not compute all 
$(d_{\vec{x}}+d_{\vec{v}})$-derivatives at once, but first we compute 
the contributions from $\vec{x}$-space and then the contributions from 
$\vec{v}$-space. One could reduce the working set even more by loop 
blocking~\cite{Kronbichler2017a}, but we deal with a generic variant here.

\subsection{Implementation of operator evaluations with hyper.deal}

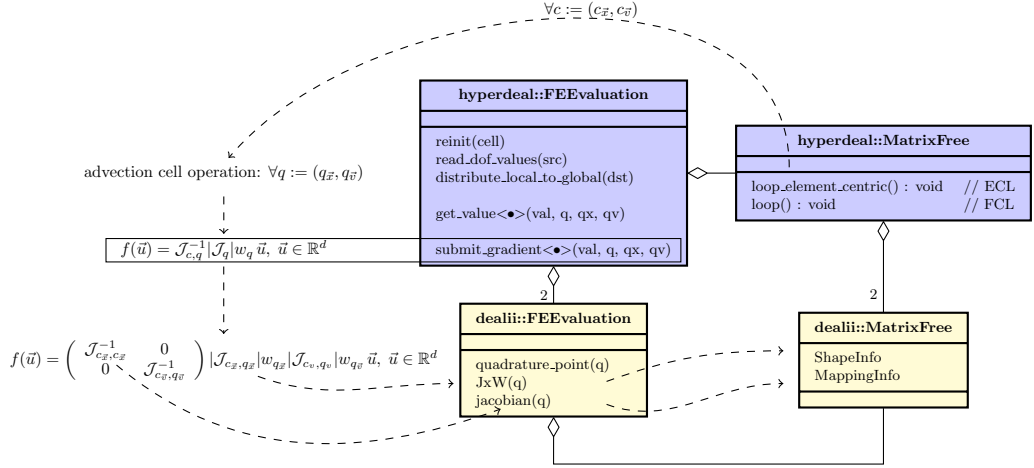
\begin{figure}[!t]

\vspace{-1.2cm}

\begin{center}
\begin{tikzpicture}[thick,scale=0.62, every node/.style={transform shape}]

\node (P) at (-18, -5.0) {advection cell operation: $\forall q:=(q_{\vec{x}},q_{\vec{v}})$};

\node (A) at (-18, -6.65) {$f(\vec{u})=\mathcal{J}^{-1}_{c,q} |\mathcal{J}_q| w_q \, \vec{u},\;\vec{u}\in \mathbb{R}^{d}$};

\node (B) at (-18, -9) {$f(\vec{u})=\left(\begin{array}{cc}\mathcal{J}^{-1}_{c_{\vec{x}},c_{\vec{x}}} & 0 \\ 0 & \mathcal{J}^{-1}_{c_{\vec{v}},q_{\vec{v}}} \end{array} \right) |\mathcal{J}_{c_{\vec{x}},q_{\vec{x}}}| w_{q_{\vec{x}}}   |\mathcal{J}_{c_v,q_v}| w_{q_{\vec{v}}} \, \vec{u},\;\vec{u}\in \mathbb{R}^{d}$};

\umlclass[x=-4,y=-5.0,fill=blue!20]{hyperdeal::MatrixFree}{}{
loop\_element\_centric() : void \quad $\slash\slash$ ECL \\
loop() : void  \qquad\qquad\qquad\qquad $\slash\slash$ FCL \\
}

\umlclass[x=-4,y=-9.0,fill=yellow!20]{dealii::MatrixFree}{ShapeInfo\\ MappingInfo}{}

\umlclass[x=-11,y=-5.0,fill=blue!20]{hyperdeal::FEEvaluation}{}{
reinit(cell)\\ 
read\_dof\_values(src)\\ 
distribute\_local\_to\_global(dst)\\ \\
get\_value$<\!\!\bullet\!\!>$(val, q, qx, qv)\\ \\
submit\_gradient$<\!\!\bullet\!\!>$(val, q, qx, qv) \\
}


\umlclass[x=-11,y=-9.0,fill=yellow!20]{dealii::FEEvaluation}{}{
quadrature\_point(q)\\
JxW(q) \\
jacobian(q)
}


\umlaggreg[mult=2]{hyperdeal::MatrixFree}{dealii::MatrixFree}

\umlaggreg[mult=2]{hyperdeal::FEEvaluation}{dealii::FEEvaluation}

\umlaggreg[]{hyperdeal::FEEvaluation}{hyperdeal::MatrixFree}


\umlaggreg[geometry=|-|,anchor1=-90,anchor2=-90,arm1=-1.0]{dealii::FEEvaluation}{dealii::MatrixFree}

\draw[->, dashed] (A) edge (B);

\node (C) at (-6, -5) {};
\node (D) at (-18.0, -6.3)  {};

\node (Q) at (-18.0, -4.8)  {};

\draw [dashed,->]   (C) to[out=90,in=50] (Q);

\draw [dashed,->]   (-18.0, -5.5) to (-18.0, -6.3);

\node (E) at (-17.4, -9.3) {};
\node (F) at (-13.0, -9.5)  {};

\draw [->, dashed]   (E) to[out=-20,in=-180] (F);

\node (G) at (-20.4, -9.0) {};
\node (H) at (-12.0, -10)  {};

\draw [->, dashed]   (G) to[out=-40,in=-160] (H);

\node (I) at (-10, -9.9) {};
\node (J) at (-6, -9.5)  {};

\draw [->, dashed]   (I) to[out=-20,in=-180] (J);

\node (K) at (-10, -9.5) {};
\node (L) at (-6, -8.8)  {};

\draw [->, dashed]   (K) to[out=+20,in=-180] (L);

\node at (-10.2, -1.5) {$\forall c:=(c_{\vec{x}},c_{\vec{v}})$};

\draw[draw] (-20.5,-6.9) rectangle ++(12.2,0.5);

\end{tikzpicture}
\end{center}
\caption{Class diagram of part of the matrix-free infrastructure of \texttt{hyper.deal}. 
It presents how classes from \texttt{hyper.deal} (namespace \texttt{hyperdeal}---
highlighted blue) and from \texttt{deal.II} (namespace \texttt{dealii}---
highlighted yellow) relate to each other. Only the \texttt{hyper.deal} methods 
are shown  that are relevant for the evaluation of the advection operator and the 
\texttt{deal.II} methods that are used in those.}\label{fig:hyperdeal:matrixfree}

\end{figure}

The library \texttt{hyper.deal} provides classes that contain inter alia 
utility functions needed in Algorithm~\ref{algo:concept:cell} and are 
built around \texttt{deal.II} classes. To enable a smooth start for users 
already familiar with \texttt{deal.II}, we have chosen the same class and 
function names living in the namespace \texttt{hyperdeal}. The relationship 
between some \texttt{hyper.deal} classes and some \texttt{deal.II} classes is 
visualized in the UML diagram in Figure~\ref{fig:hyperdeal:matrixfree}. The class 
\texttt{hyperdeal::MatrixFree} is responsible for looping over cells (and faces) 
as well as for storing precomputed information related to shape functions and 
precomputed quantities at the quadrature points. The classes 
\texttt{hyperdeal::FEEvaluation} and  \texttt{hyperdeal::FEFaceEvaluation} (not shown) 
contain utility functions for the current cells and faces: These utility 
functions include functions to read and write cell-/face-local values from a 
global vector as well as also operations at the quadrature points. {As an example,
Figure~\ref{fig:hyperdeal:matrixfree} shows the implementation of the 
\texttt{hyperdeal::FEEvaluation::submit\_gradient()} method, which uses for the evaluation of $f(\vec{u})=\mathcal{J}^{-1}_{c,q} |\mathcal{J}_q| w_q \, \vec{u},\;\vec{u}\in \mathbb{R}^{d},$ two instances of the \texttt{deal.II} class with the same name---one for $\vec{x}$- and one for $\vec{v}$-space.}

We use ``vectorization over elements''  from \texttt{deal.II} as vectorization 
strategy. To be precise, we vectorize only over elements in $\vec{x}$-space, 
whereas $\vec{v}$-space is not vectorized. Note that in the Vlasov--Maxwell or 
Vlasov--Poisson model, where parts of the code operate on the full phase space and 
other parts on the $\vec{x}$-space only, it is important to vectorize over $\vec{x}$.  
Then, the data structures are already laid out correctly for an efficient matrix-free 
solution of the lower-dimensional problem.

\section{Parallelization by shared-memory MPI}\label{sec:concept:sm}

\begin{figure}[!t]

\centering

\def\svgwidth{0.16\textwidth}{\footnotesize 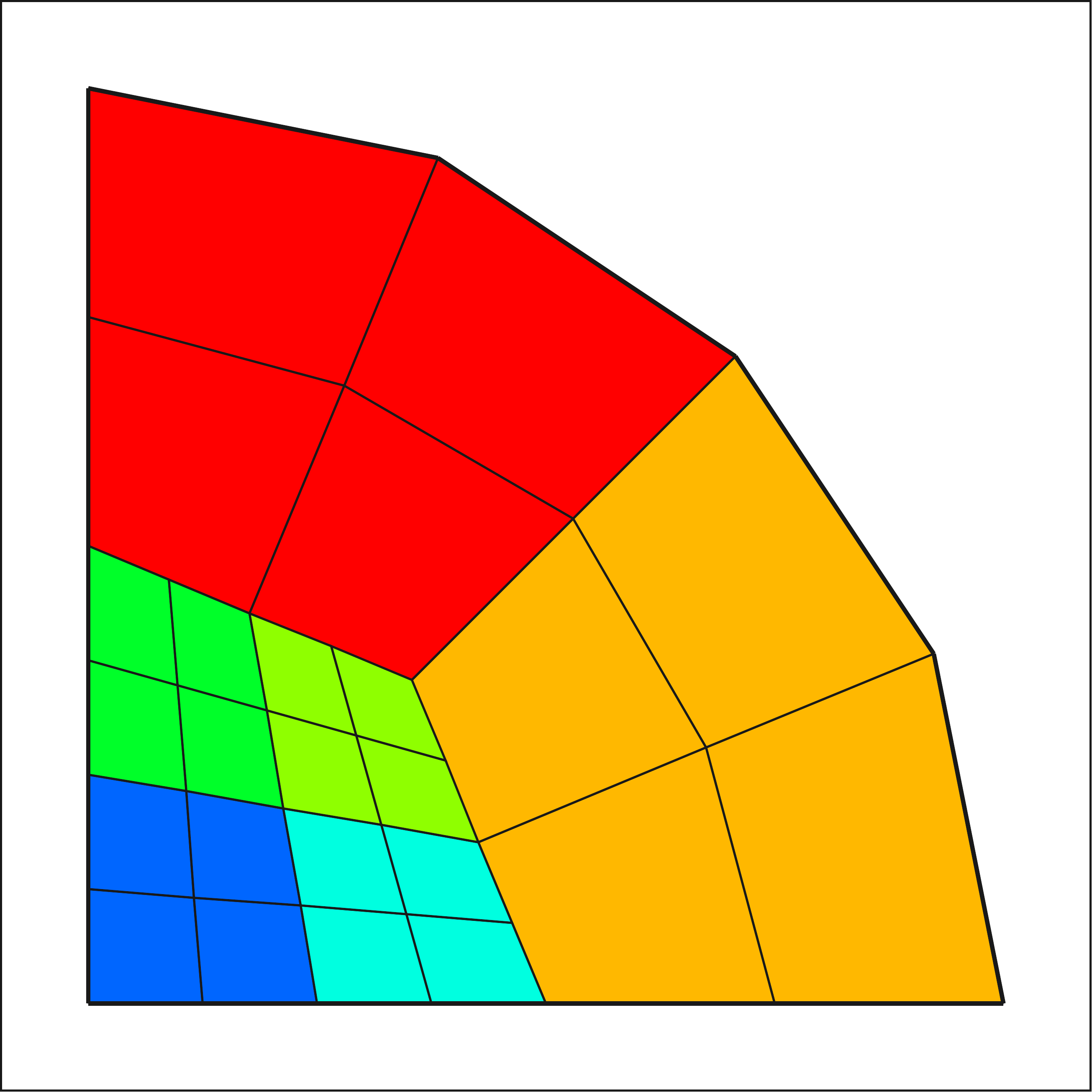}
\def\svgwidth{0.16\textwidth}{\footnotesize 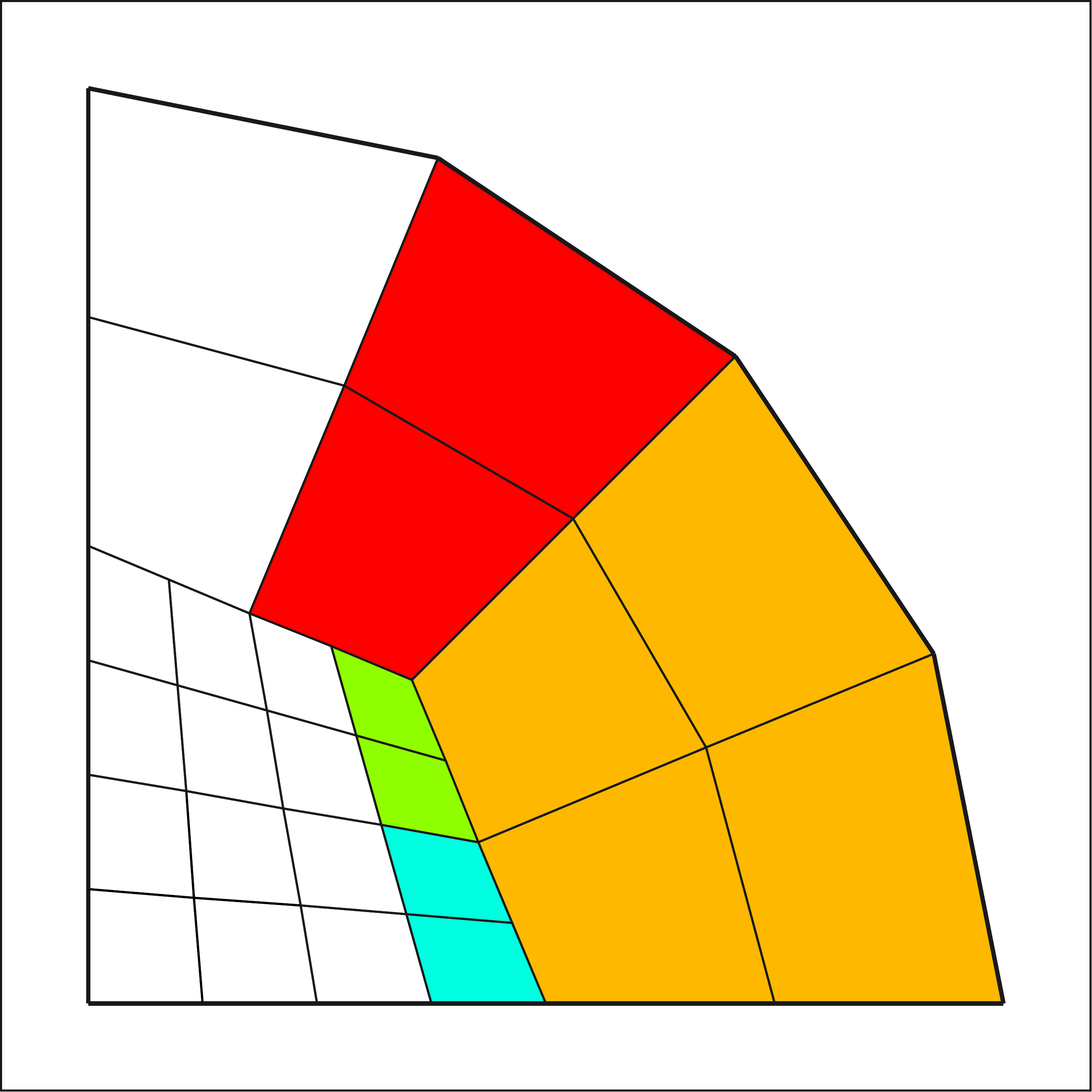}
\def\svgwidth{0.16\textwidth}{\footnotesize 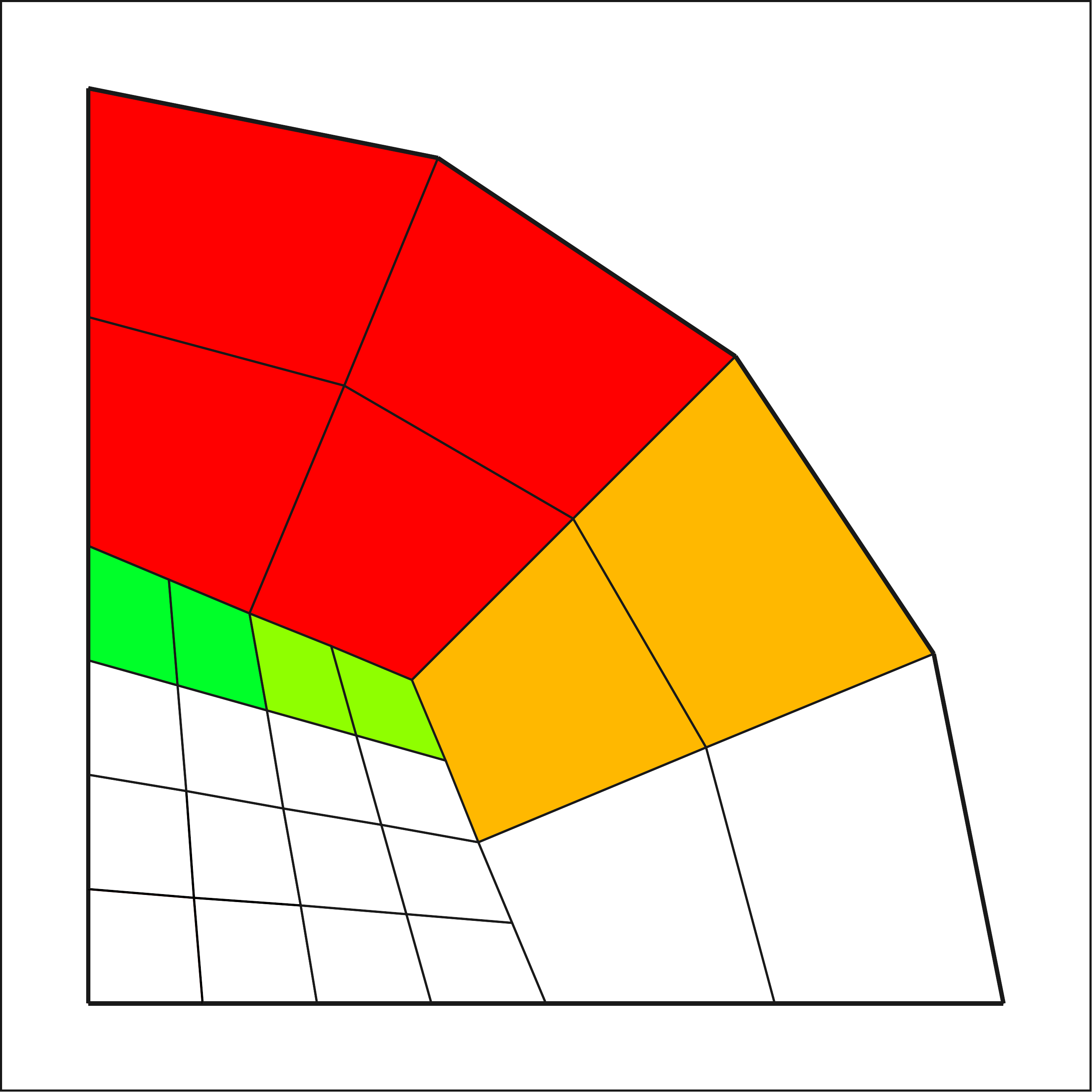}
\def\svgwidth{0.259\textwidth}{\footnotesize 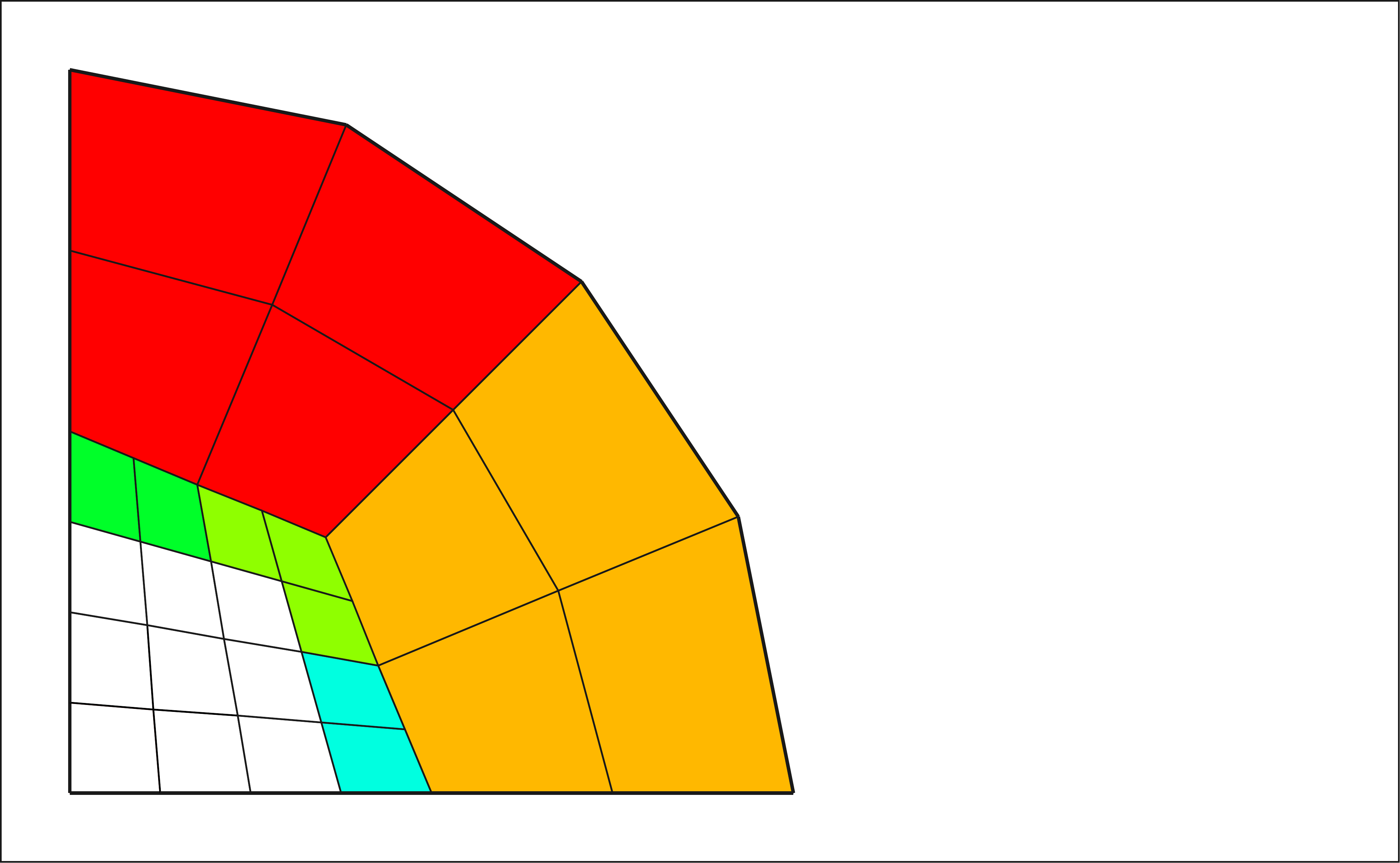}

\caption{Shared-memory domain: a) partitioned distributed triangulation; b-c) 
locally relevant cells of ranks 4 and 5; d) locally relevant cells of 
shared-memory domain (containing ranks 4 and 5) with a shared ghost layer and without 
ghost cells between the processes in the same shared-memory domain.}\label{fig:concept:sharedcells}

%

\centering

%
%
%

\subfloat[A hybrid approach using MPI-3.0 shared-memory features 
(\texttt{buffered mode}). Ghost values are updated via send/recv between nodes or 
explicitly via memory copy within the node. The similarity to 
a standard \texttt{MPI} implementation is clear with the difference that memcpy is 
called directly by the program, making packing/unpacking of data superfluous.
\label{fig:concept:communication:b}]{\includegraphics[width=0.9\textwidth]
{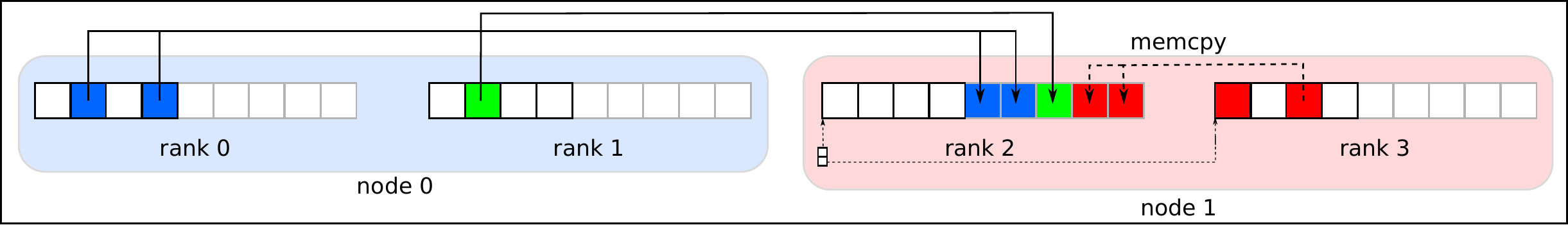}}

\subfloat[A hybrid approach using MPI-3.0 shared-memory features 
(\texttt{non-buffered mode}) similar to \ref{fig:concept:communication:b}, with the difference 
that only ghost values that live in different shared-memory domains are updated. 
Ghost values living in the same shared-memory domain are directly accessed only 
when needed. The similarity to a thread-based implementation is clear with 
the differences that vectors are non-contiguous, requiring an indirect access to 
values owned by other processes, and that each process may manage its own ghost 
values and send/receive its own medium-sized messages needed to fully utilize the 
network controller. \label{fig:concept:communication:d}]
{\includegraphics[width=0.9\textwidth]{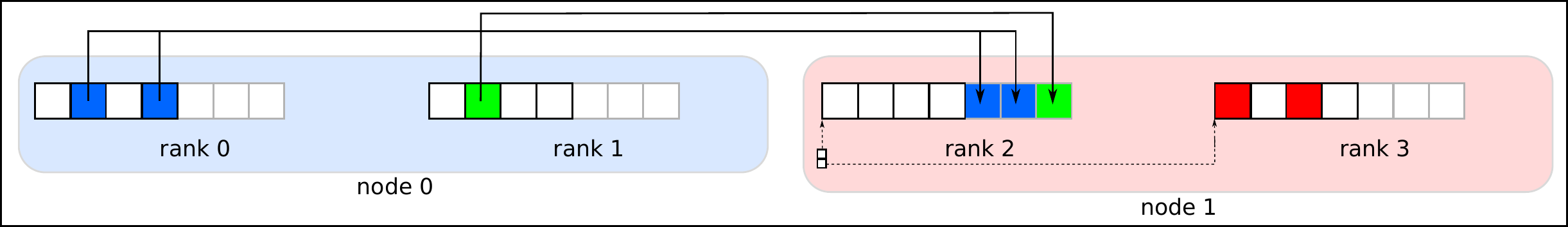}}

\caption{Two hybrid ghost-value-update approaches for a hypothetical setup with 2 nodes, 
each with two cores. Only the communication pattern of rank 2 is considered.}

\end{figure}

The parallelization of the library \texttt{hyper.deal} is purely MPI-based with 
one MPI process per physical core. Each MPI process possesses its own 
subpartition and has a halo of ghost faces (see Figure~\ref{fig:dealii:tria} and 
\ref{fig:concept:mesh:tp:ghost}). 
MPI allows to program for distributed systems, which is crucial for solving high-dimensional 
problems due to their immense memory requirements. A downside of a 
purely MPI-based code is that many data structures (incl. ghost values) are 
created and updated multiple times on the same compute node, although they could 
be shared. In FEM codes, it is widespread that ghost values filled by standard 
\texttt{MPI\_(I)Send}/\texttt{MPI\_(I)Recv} reside in an additional section of 
the solution vector. Depending on 
the MPI implementation, these operations will be replaced by efficient 
alternatives (e.g., based on \texttt{memcpy}---if the calling peers are on the same compute 
node. Nevertheless, the  allocation of additional memory---if main memory is 
scarce---might be unacceptable.

Adding shared-memory libraries like \texttt{TBB} and \texttt{OpenMP} to an 
existing \texttt{MPI} program would allow to use shared memory, however, this comes with an overhead 
for the programmer, since all parallelizable code sections have to be found and 
transformed according to the library used, including the difficulty when some 
third-party numerical library like an iterative solver package only relies on 
\texttt{MPI}. 

Considering a purely MPI-parallelized FEM application, one can identify that the 
major time and memory benefit of using shared memory would come from accessing 
the part of the solution vector owned by the processes on the same compute node 
without the need to make explicit copies and buffering them. This is why we 
propose a new vector class that uses \texttt{MPI-3.0} features to allocate shared 
memory and provides controlled access to it, while retaining the same vector 
interface and adding only a few new methods.


\texttt{MPI} provides the function \texttt{MPI\_WIN\_ALLOCATE\_SHARED} to 
allocate non-contiguous memory that is shared among all processes in the 
subcommunicator  containing all MPI processes on the same compute node. To query 
the beginning of the local array of each process, it provides the function 
\texttt{MPI\_WIN\_SHARED\_QUERY}.

These two functions enable us to allocate memory needed by each process for its 
locally owned cells and for the ghost faces that are not owned by any process 
on the same compute node as well as to get hold of the beginning of the arrays of 
the other processes. Appendix~\ref{sec:appendix:sm} provides further 
implementation details of the allocation/deallocation process of the shared 
memory in the vector class. With basic preprocessing steps, the address of each 
cell residing on the same compute node can be determined so that the processes 
have access to all degrees of freedom owned by that compute node (see 
Figure~\ref{fig:concept:sharedcells}). 
A natural way to access the solution vector is by specifying vector entry indices and the 
cell ID for degrees of freedom owned by a cell or by specifying a pair of a cell 
ID and a face number ($< 2d$) for degrees of freedom owned by faces.


In \texttt{hyper.deal}, the class
\texttt{dealii::LinearAlgebra::SharedMPI::Vector} manages the shared memory and 
also provides access to the values of the degrees of freedom of the local and the 
ghost cells: It returns either pointers to buffers or the shared memory, 
depending on the cell type (see Figure~\ref{fig:concept:communication:d}). In 
this way, the user of the vector class gets the illusion of a pure MPI program, 
since the new vector has to be added only at a single place and only a few 
functions querying  values from the vector (e.g., \texttt{read\_dof\_values} and 
\texttt{distribute\_local\_to\_global} in Figure~
\ref{fig:hyperdeal:matrixfree})---oblivious to the user--- have to be 
specialized.





\begin{figure}[!t]

%
%

\centering

\includegraphics[width=1.0\textwidth]{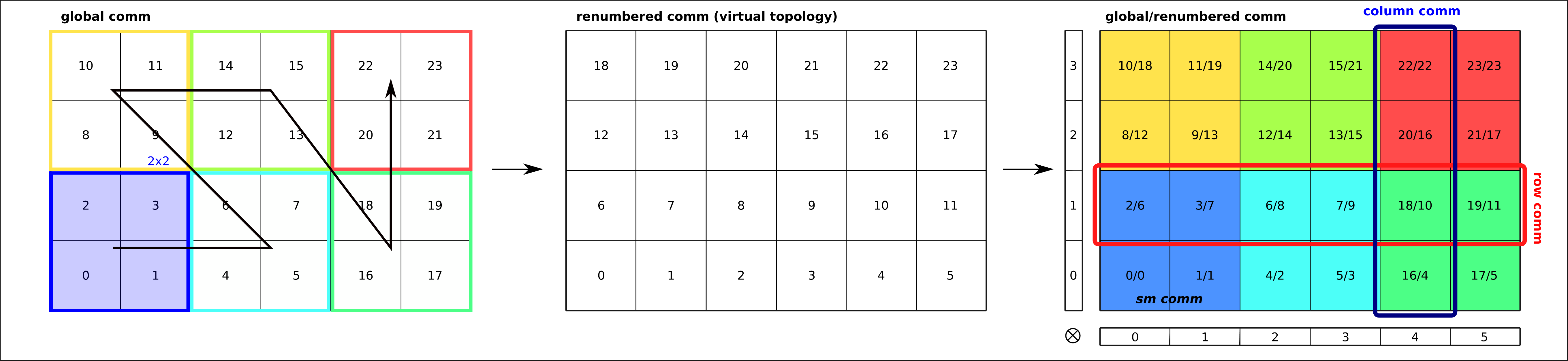}

\caption{ Renumbering of ranks in the global communicator (via 
\texttt{MPI\_Comm\_split}): For a hypothetical setup 
of 26 ranks in \texttt{MPI\_COMM\_WORLD} and of a 6$\times4$ partition, 
processes are grouped in $2\times2$ blocks and these blocks are ordered along
a z-curve.
Based on this new global communicator, the partitioning as described in 
Subsection~\ref{sec:concept:partitioning} is applied. A checkerboard partitioning 
 with a better data locality for the shared-memory domains is obtained.}\label{fig:hyperdeal:virtual}

\end{figure}

%

We provide two operation modes: 
\begin{itemize}
\item In the \texttt{buffered mode} (see 
Figure~\ref{fig:concept:communication:b}), memory is allocated also for ghost 
values owned by the same compute node; these ghost values are updated directly 
via \texttt{memcpy}, without an intermediate step via \texttt{MPI}. This mode is 
necessary if ghost values are modified, as it takes place in face-centric loops. 
It promises some performance benefit, since data packing/unpacking can be 
skipped.
\item The \texttt{non-buffered mode} (see 
Figure~\ref{fig:concept:communication:d}) does not allocate any redundant memory 
for ghost values owned by the same compute node. This mode works perfectly with 
ECL, since it is by design free  of race conditions. Due to the harmony of ECL 
and the \texttt{non-buffered mode} of the shared-memory vector, we rely on this 
vector mode in the rest of this paper. 
\end{itemize}

In order to make this type of hybrid parallelization approach successful, the 
union of the subpartitions of all processes on a compute node should build a 
well-formed subpartition on its own with an improved surface-to-volume ratio. We 
achieve this via blocking within a Cartesian virtual topology (see also 
Figure~\ref{fig:hyperdeal:virtual}), e.g., by 48 process blocks with 8 processes 
in $\vec{x}$-space and with 6 processes in $\vec{v}$-space. Compute nodes are ordered along 
a $z$-curve.

As a final remark, we emphasize that a vector built around \texttt{MPI-3.0} shared-memory 
features is not limited to high dimensions, but can be used also in lower dimensions; 
in particular, if not only a single ``ghost layer'' has to be exchanged (as in the case 
of an advection operator) but multiple ``ghost layers'' (as in the case of a Laplace 
operator discretized with the interior penalty method~\cite{kronbichler2019hermite}). 
\section{Application: high-dimensional scalar transport}\label{sec:performance}

In the following section, we show results of the solution of a high-dimensional 
scalar transport problem. These results confirm the suitability of the underlying 
concepts and the implementation of the library \texttt{hyper.deal} for high 
orders and high dimensions. 
Both node-level performance and parallel performance results are shown, including 
the findings of strong and weak scaling analyses with up to 147,456 processes on 
3,072 compute nodes. 

%
%
%
%
%

\subsection{Experimental setup and performance metrics}\label{sec:mem:performance}

The setup of the simulations is as follows. We consider the computational domains 
$\Omega_{\vec{x}}$=$[0,1]^{d_{\vec{x}}}$ and 
$\Omega_{\vec{v}}$=$[0,1]^{d_{\vec{v}}}$ with the following decomposition of the 
dimensions $d=d_{\vec{x}}+d_{\vec{v}}$: $2=1+ 1$, $3=2+ 1$, $4=2+ 2$, $5=3+ 2$, 
$6=3+ 3$.
The computational domains are initially meshed separately  with 
subdivided $d_{\vec{x}}$/$d_{\vec{v}}$-dimensional hyperrectangles--~with
$(2^{l_1}, \dots, 2^{l_{d_{\vec{x}}}}) \in \mathbb{N}^{d_{\vec{x}}}$ and 
$(2^{l_{d_{\vec{x}}+1}}, \dots, 2^{l_{d_{\vec{x}}+d_{\vec{v}}}}) \in \mathbb{N}^{d_{\vec{v}}}$ 
hexahedral elements in each direction and with a difference in the
mesh size of at most two, i.e., meshed for 4D from the mesh sequence 
$(l_1, l_2, l_3, l_4)$:  $(1,1,1,1)$, $(2,1,1,1)$, $(2,2,1,1)$, $(2,2,2,1)$, 
$(2,2,2,2)$, $(3,2,2,2)$. The number of elements is selected for each 
``dimension $d$ / polynomial degree $k$'' configuration in such a way that the 
solution vectors do not fit into the cache. To obtain unique Jacobian matrices at 
each quadrature points ($\mathcal{J}=\mathcal{J}(\vec{x})$) and to prevent the 
usage of algorithms explicitly designed for affine meshes, we deform the 
Cartesian mesh slightly. The velocity $\vec{a}$ in 
Equation~\eqref{equ:intro:advection} is set constant and uniform over the 
whole domain ($\vec{a}\neq \vec{a}(t,\vec{x})$).

The measurement data have been gathered either with user-defined timers or with 
the help of the script \texttt{likwid-mpirun} from the \texttt{LIKWID suite} and 
with suitable in-code \texttt{LIKWID API} annotations 
\cite{Treibig2010,Rohl2015}. 
The following metrics are used to quantify the quality of the implementations:
\begin{itemize}
\item \textbf{throughput}: processed degrees of freedom per time unit
\begin{align}
\text{throughput} = \frac{\text{processed DoFs}}{\text{time}} \stackrel{\text{Eqn.~\eqref{eq:dealii:totalunkonwns}}}{=} \frac{|\mathcal{C}| \cdot (k+1)^d }{\text{time}}
\end{align}
(In Subsections~\ref{sec:performance:celllocal}--\ref{sec:performance:alternatives}, 
we consider the throughput for the application of the advection operator, while a 
single Runge--Kutta stage, i.e., the evaluation of the advection operator plus 
vector updates, is considered in Subsection~\ref{sec:performance:scaling}.);
\item \textbf{performance}: maximum number of floating-point operations per 
second;
\item \textbf{data volume}: the amount of data transferred within the memory 
hierarchy (Here we consider the transfer between the L1-, L2-, and L3-caches as 
well as the main memory.);
\item \textbf{bandwidth}: data volume transferred between the levels in the 
memory hierarchy  per time.
\end{itemize}

Our main objective is to decrease the time-to-solution and to increase the 
throughput (at constant error, not considered in this paper). The measured 
quantities ``performance'', ``data volume'', and ``bandwidth''  are useful, since 
they show how well the given hardware is utilized and how much additional work or 
memory transfer is performed compared to the theoretical requirements of the 
mathematical algorithm.

\begin{table}
\centering

\caption{Degrees of freedom per cell: $(k+1)^d$. The $k$-$d$ configurations with 
working-set size $v_{len}  \cdot (k+1)^d < L_1$ are highlighted in italics and 
configurations with working-set size $L_1 \le v_{len} \cdot (k+1)^d < L_2$ in 
bold. Hardware characteristics ($v_{len}$, $L_1$, $L_2$) are taken from Table~\ref{tab:systems}.}\label{tab:performance:ws}
\begin{footnotesize}
\begin{tabular}{c|ccccc}
\toprule
$k$/$d$ & 2 & 3 & 4 & 5 & 6 \\ \midrule
  2 & \textit{ 9}  &   \textit{ 27} &  \textit{  81} &  \textit{243} &   \textbf{  729}\\
  3 & \textit{16} &    \textit{ 64} &  \textit{ 256} &  \textbf{1024} &  \textbf{ 4096}\\
  4 & \textit{25} &    \textit{125} &  \textbf{ 625} &  \textbf{3125} &  \textbf{15625}\\
  5 & \textit{36} &    \textit{216} &  \textbf{1296} &  \textbf{7776} &          46656\\ \bottomrule
\end{tabular}
\end{footnotesize}
\end{table}

The focus of this study is on the performance of \texttt{hyper.deal} for high-order 
and high-dimensional problems that have a large working set $(k+1)^d$. The 
evaluation of this expression for $2\le k \le 5$ and $2 \le d \le 6$ is presented 
in Table~\ref{tab:performance:ws}. The $k$-$d$ configurations with working-set 
size of $v_{len}  \cdot (k+1)^d < L_1$, fitting into L1 cache (highlighted in 
italics), are expected to show good performance; the $k$-$d$ configurations with 
working-set size of $L_1 \le v_{len} \cdot (k+1)^d < L_2$ (highlighted in 
bold) are expected to be performance-critical, since each sum-factorization sweep 
might drop out of the cache. These latter configurations are, however, the most 
relevant with regard to high-order and high-dimensional problems. 

\begin{table}
  \caption{Specification of the hardware system used for evaluation with 
  turbo mode enabled. Memory
    bandwidth is according to the STREAM triad benchmark (optimized variant 
    without
read for ownership transfer involving two reads and one write), and GFLOP/s are 
based
on the theoretical maximum at the AVX-512 frequency. The \texttt{dgemm} 
performance is measured for $m=n=k=12{,}000$ with Intel MKL 18.0.2. 
We measured a frequency of 2.3 GHz with AVX-512 dense code for
the current experiments. The empirical machine balance is computed as the ratio 
of measured
\texttt{dgemm} performance and STREAM bandwidth from RAM memory.}
    \label{tab:systems}
  {
	\footnotesize
	\strut\hfill
    \begin{tabular}{lc}
      \hline
      & Intel Skylake   Xeon Platinum 8174 \\
      \hline
      cores       & $2\times 24$ \\
      frequency base (max AVX-512 frequency) & 2.3 GHz (2.3 GHz) \\
      SIMD width    & 512 bit \\
      arithmetic peak (\texttt{dgemm} performance)  & 4147 GFLOP/s (2920 GFLOP/s)\\
      memory interface & DDR4-2666, 12 channels \\
      STREAM memory bandwidth & 205 GB/s\\
      empirical machine balance & 14.3 FLOP/Byte  \\
      L1-/L2-/L3-/MEM size & 32kB/1MB/66MB (shared)/96GB(shared) \\
      \hline
      compiler + compiler flags & \texttt{g++}, version 9.1.0, \texttt{-O3 -funroll-loops -march=skylake-avx512}\\
      \hline
    \end{tabular}
    \hfill\strut
  }
\end{table}

All performance measurements have been conducted on the SuperMUC-NG 
supercomputer. Its compute nodes have 2 sockets (each with 24 cores of Intel Xeon 
Skylake), a measured bandwidth of 205GB/s to main memory, and the AVX-512 ISA 
extension so that 8 doubles can be processed per instruction.  A detailed 
specification of the hardware is given in Table~\ref{tab:systems}.
An island consists of 792 compute nodes. The maximum network bandwidth per node 
within an island is  100GBit/s=12.5GB/s\footnote{\url{https://doku.lrz.de/
display/PUBLIC/SuperMUC-NG}} due to the fat-tree network topology. Islands are 
connected via a pruned tree network architecture (pruning factor 1:4). 

The library \texttt{hyper.deal} has been configured in the following way: All 
processes of a node are grouped, and they build blocks of the size of 
48=8$\times$6. All processes in these blocks share their values via the 
shared-memory vector. We use the highest ISA extension AVX-512 so that 8 cells 
are processed at once. Jacobian matrix and its determinant are precomputed for 
$\vec{x}$- and $\vec{v}$-space and combined on the fly. The quadrature is 
based on the Gauss-Lagrange formula with $n_q=k+1$.  
In the following, we refer to this configuration as ``default configuration''.


\subsection{Cell-local computation}\label{sec:performance:celllocal}

This subsection takes an in-depth look at the cell-local computation in the 
element-centric evaluation of the advection operator (see 
Algorithm~\ref{algo:concept:cell}). Cell-local computations do not access 
non-cached values of neighboring cells, making it easier to reason about the 
number of sweeps and the working-set size (see the first two working-set sizes in 
Table~\ref{tab:dealii:workingset}), which increase with the dimension.

\subsubsection{Cell integrals}\label{sec:performance:celllocal:a}

We first consider all steps in Algorithm~\ref{algo:concept:cell} related to the 
cell integrals 
(lines~\ref{algo:cell:line:gather}-\ref{algo:cell:line:quadrature}, 
\ref{algo:cell:line:mass}, \ref{algo:cell:line:scatter}), skipping the loops over 
faces and ignoring the flux computation.

During the cell integrals, values are read from the global vector, a basis change 
to the Gauss--Legendre quadrature points is performed (with $\approx d$ data 
sweeps for reading and $\approx d$ for writing), the values obtained are 
multiplied with the velocities at the quadrature points ($\approx 2d$) and tested 
by the gradient of the collocation functions ($\approx 3d$), the inverse mass 
matrix is applied ($\approx 2d$), and finally the results are written back to the 
global vector. A total of $\approx 9d$ data sweeps are necessary if reading and 
writing are counted separately. The working set of sum factorization is 
$v_{len}\cdot (k+1)^{d_1+d_2}$, and the working set of the intermediate values is 
$v_{len}\cdot\max(d_1, d_2) \cdot (k+1)^{d_1+d_2}$. A comparison with hardware 
statistics shows that the working set of sum factorization exceeds the size of 
the L1 cache for configurations $k=3$~/~$d=5$ and  $k=5$~/~$d=4$ so that every 
data sweep has to fetch the data from the L2 cache. 

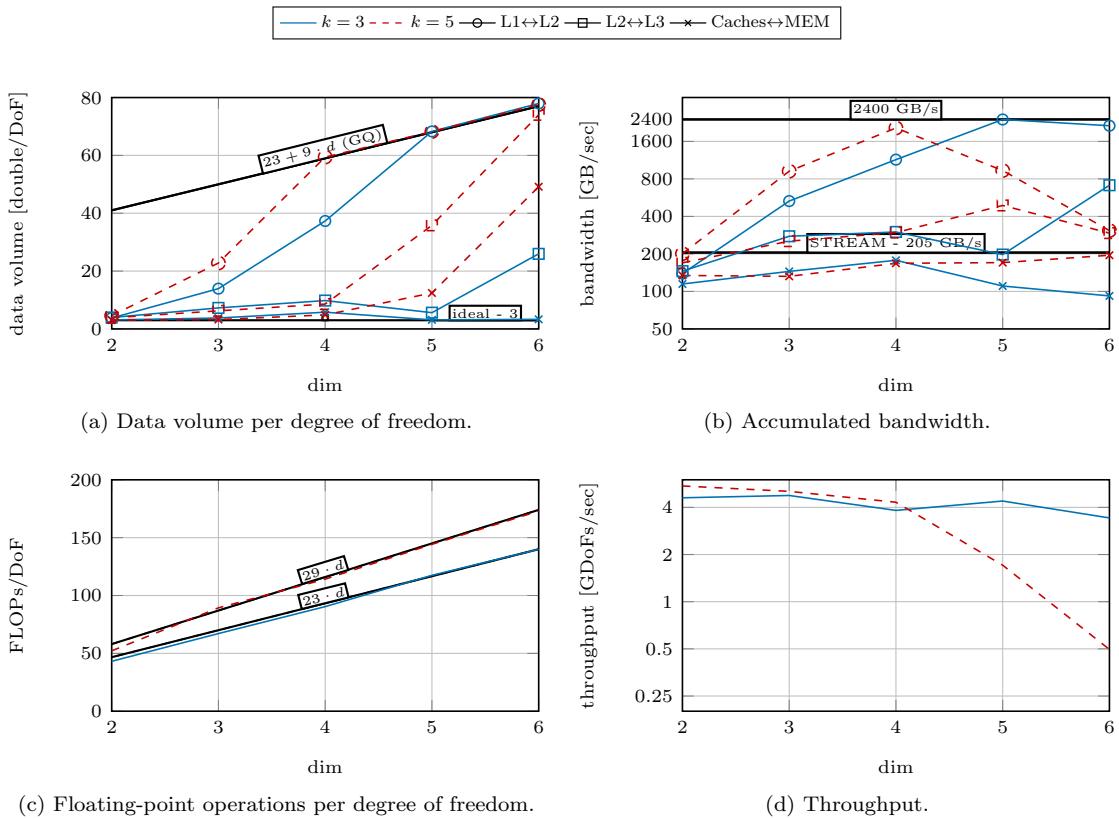
\begin{figure}



\begin{center}
\begin{tikzpicture}[scale=0.8]
    \begin{axis}[%
    hide axis,
    legend style={font=\footnotesize},
    xmin=10,
    xmax=50,
    ymin=0,
    ymax=0.4,
    semithick,
    legend style={draw=white!15!black,legend cell align=left},legend columns=-1
    ]

    \addlegendimage{gnuplot@darkblue,every mark/.append style={fill=gnuplot@red!50!black}}
    \addlegendentry{$k=3$};
    \addlegendimage{gnuplot@red,every mark/.append style={fill=gnuplot@red!50!black}, dashed}
    \addlegendentry{$k=5$};

    \addlegendimage{black,mark=o}    
    \addlegendentry{L1$\leftrightarrow$L2};
    \addlegendimage{black,mark=square}
    \addlegendentry{L2$\leftrightarrow$L3};
    \addlegendimage{black,mark=x}
    \addlegendentry{Caches$\leftrightarrow$MEM};
    \end{axis}
\end{tikzpicture}
\end{center}

\vspace{-4.2cm}

\subfloat[{Data volume per degree of freedom.}\label{fig:performance:cell:only:data}]
{
  \begin{tikzpicture}
    \begin{axis}[
      width=0.48\textwidth,
      height=0.31\textwidth,
      title style={font=\footnotesize},
      xlabel={dim},
      ylabel={data volume [double/DoF]},
      legend pos={south west},
      legend cell align={left},
      cycle list name=colorGPL,
      tick label style={font=\scriptsize},
      label style={font=\scriptsize},
      legend style={font=\scriptsize},
      grid,
      semithick,
      ymin=0,
      ymax=80,
      xmin=2,xmax=6,
      ]

	\draw [draw=black,thick] (axis cs: 2, 41) -- node[above,rotate=18,inner sep=2pt,outer sep=0.5pt]{} (axis cs: 6, 77);
  \draw[draw=black,thick] (axis cs:2, 41) -- node[above,rotate=14,inner sep=1pt,outer sep=0pt, fill=white,draw]{{\tiny \color{black} $23+ 9\cdot d$ (GQ)}}(axis cs:6, 77);


	\draw [draw=black,thick] (axis cs: 2, 3) -- node[above,rotate=0,inner sep=1pt,outer sep=0pt, fill=white,draw]{} (axis cs: 6, 3);
	\draw [draw=black,thick] (axis cs: 5, 3) -- node[above,rotate=0,inner sep=1pt,outer sep=0pt, fill=white,draw]{{\tiny \color{black} ideal - 3}} (axis cs: 6, 3);

      \addplot +[gnuplot@darkblue,every mark/.append style={fill=gnuplot@darkblue!80!black},mark=o, solid] table [x=dim, y=da1] {results/cell_cache/post/all1_8_3_0.csv};
      \addplot +[gnuplot@darkblue,every mark/.append style={fill=gnuplot@darkblue!80!black},mark=square, solid] table [x=dim, y=da2] {results/cell_cache/post/all1_8_3_0.csv};
      \addplot +[gnuplot@darkblue,every mark/.append style={fill=gnuplot@darkblue!80!black},mark=x, solid] table [x=dim, y=da3] {results/cell_cache/post/all1_8_3_0.csv};

      \addplot +[gnuplot@red,every mark/.append style={fill=gnuplot@darkblue!80!black},mark=o, dashed] table [x=dim, y=da1] {results/cell_cache/post/all1_8_5_0.csv};
      \addplot +[gnuplot@red,every mark/.append style={fill=gnuplot@darkblue!80!black},mark=square, dashed] table [x=dim, y=da2] {results/cell_cache/post/all1_8_5_0.csv};
      \addplot +[gnuplot@red,every mark/.append style={fill=gnuplot@darkblue!80!black},mark=x, dashed] table [x=dim, y=da3] {results/cell_cache/post/all1_8_5_0.csv};
%
%
%
%
%
%
%


    \end{axis}
  \end{tikzpicture}
}
\subfloat[{Accumulated bandwidth.}\label{fig:performance:cell:only:bw}]
{
  \begin{tikzpicture}
    \begin{semilogyaxis}[
      width=0.48\textwidth,
      height=0.31\textwidth,
      title style={font=\footnotesize},
      xlabel={dim},
      ylabel={bandwidth [GB/sec]},
      legend pos={south west},
      legend cell align={left},
      cycle list name=colorGPL,
      tick label style={font=\scriptsize},
      label style={font=\scriptsize},
      legend style={font=\scriptsize},
      grid,
      semithick,
      ymin=50,ymax=3600,
      ytick={50,100,200,400,800,1600,2400},
      yticklabels={50,100,200,400,800,1600,2400},
      xmin=2,xmax=6,
      ]
      
	\draw [draw=black,thick] (axis cs: 2, 2400) -- node[above,rotate=0,inner sep=1pt,outer sep=0pt, fill=white,draw]{} (axis cs: 6, 2400);
	\draw [draw=black,thick] (axis cs: 2, 2400) -- node[above,rotate=0,inner sep=1pt,outer sep=0pt, fill=white,draw]{{\tiny \color{black} 2400 GB/s}} (axis cs: 6, 2400);

	\draw [draw=black,thick] (axis cs: 2, 205) -- node[above,rotate=0,inner sep=1pt,outer sep=0pt, fill=white,draw]{} (axis cs: 6, 205);
	\draw [draw=black,thick] (axis cs: 2, 205) -- node[above,rotate=0,inner sep=1pt,outer sep=0pt, fill=white,draw]{{\tiny \color{black} STREAM - 205 GB/s}} (axis cs: 6, 205);	


      \addplot +[gnuplot@darkblue,every mark/.append style={fill=gnuplot@darkblue!80!black},mark=o,solid] table [x=dim, y=bw1] {results/cell_cache/post/all1_8_3_0.csv};
      \addplot +[gnuplot@darkblue,every mark/.append style={fill=gnuplot@darkblue!80!black},mark=square,solid] table [x=dim, y=bw2] {results/cell_cache/post/all1_8_3_0.csv};
      \addplot +[gnuplot@darkblue,every mark/.append style={fill=gnuplot@darkblue!80!black},mark=x,solid] table [x=dim, y=bw3] {results/cell_cache/post/all1_8_3_0.csv};

      \addplot +[gnuplot@red,every mark/.append style={fill=gnuplot@darkblue!80!black},mark=o,dashed] table [x=dim, y=bw1] {results/cell_cache/post/all1_8_5_0.csv};
      \addplot +[gnuplot@red,every mark/.append style={fill=gnuplot@darkblue!80!black},mark=square,dashed] table [x=dim, y=bw2] {results/cell_cache/post/all1_8_5_0.csv};
      \addplot +[gnuplot@red,every mark/.append style={fill=gnuplot@darkblue!80!black},mark=x,dashed] table [x=dim, y=bw3] {results/cell_cache/post/all1_8_5_0.csv};


    \end{semilogyaxis}
  \end{tikzpicture}
}  
 
\subfloat[{Floating-point operations per degree of freedom.}\label{fig:performance:cell:only:flops}]
{ 
  \begin{tikzpicture}
    \begin{axis}[
      width=0.48\textwidth,
      height=0.31\textwidth,
      title style={font=\footnotesize},
      xlabel={dim},
      ylabel={FLOPs/DoF},
      legend pos={south west},
      legend cell align={left},
      cycle list name=colorGPL,
      tick label style={font=\scriptsize},
      label style={font=\scriptsize},
      legend style={font=\scriptsize},
      grid,
      semithick,
      ymin=0,
      ymax=200,
      xmin=2,xmax=6,
      ]

	
	\draw [-, dashed] (axis cs: 0, 0) -- node[above,rotate=11,inner sep=2pt,outer sep=0.5pt]{} (axis cs: 6, 174);
  \draw[draw=black,thick] (axis cs:2, 58) -- node[above,rotate=17,inner sep=1pt,outer sep=0pt, fill=white,draw]{{\tiny \color{black} $29 \cdot d$}}(axis cs:6, 174);

	\node[anchor=east] at (axis cs: 0, 0) (nodeCE) {};
	\node[anchor=west] at (axis cs: 6, 140) (nodeCF) {};
	\draw [-, dashed] (axis cs: 0, 0) -- node[above,rotate=0,inner sep=2pt,outer sep=0.5pt]{} (axis cs: 6, 140);  
  \draw[draw=black,thick] (axis cs: 2, 46.667) -- node[above,rotate=14,inner sep=1pt,outer sep=0pt, fill=white,draw]{{\tiny \color{black}  $23 \cdot d$}}(axis cs: 6, 140);

      \addplot +[gnuplot@darkblue,every mark/.append style={fill=gnuplot@darkblue!80!black},mark=none, solid] table [x=dim, y=tflops] {results/cell_cache/post/all1_8_3_0.csv};
      
      
      \addplot +[gnuplot@red,every mark/.append style={fill=gnuplot@darkblue!80!black},mark=none,dashed] table [x=dim, y=tflops] {results/cell_cache/post/all1_8_5_0.csv};


%
%

    \end{axis}
  \end{tikzpicture}  
}
\subfloat[{Throughput.}\label{fig:performance:cell:only:throughput}]
{
  \begin{tikzpicture}
    \begin{semilogyaxis}[
      width=0.48\textwidth,
      height=0.31\textwidth,
      title style={font=\footnotesize},
      xlabel={dim},
      ylabel={throughput [GDoFs/sec]},
      legend pos={south west},
      legend cell align={left},
      cycle list name=colorGPL,
      tick label style={font=\scriptsize},
      label style={font=\scriptsize},
      legend style={font=\scriptsize},
      grid,
      semithick,
      ytick={0.25,0.5,1,2,4,8},
      ymin=0.2,ymax=6,
      yticklabels={0.25,0.5,1,2,4,8},
      xmin=2,xmax=6,
      ]

      \addplot +[gnuplot@darkblue,every mark/.append style={fill=gnuplot@darkblue!80!black},mark=none] table [x=dim, y=throughput] {results/cell_cache/post/all1_8_3_0.csv};

      \addplot +[gnuplot@red,every mark/.append style={fill=gnuplot@darkblue!80!black},mark=none,dashed] table [x=dim, y=throughput] {results/cell_cache/post/all1_8_5_0.csv};

%

    \end{semilogyaxis}
  \end{tikzpicture}
}


\caption{Node-level analysis of the cell integrals of the advection operator}
\label{fig:performance:cell:only}

\end{figure}

The theoretical considerations made above are supported by the measurement 
results in Figure~\ref{fig:performance:cell:only}, which shows the data traffic 
between the memory hierarchy levels (data volume per DoF and bandwidth), the 
floating-point operations per DoF, and the throughput for $k=3$ and $k=5$ for 
$2\le d \le 6$.

In Figure~\ref{fig:performance:cell:only:data}, one can observe that with 
increasing dimension the data traffic between the memory hierarchy levels 
increases as the data volume and the corresponding bandwidth increase.  Beginning 
from the configurations mentioned above ($k=3$~/~$d=5$ and  $k=5$~/~$d=4$), the 
data has to be fetched from and written back to L2 cache again during every 
sweep, resulting in a data volume traffic between L1 and L2 caches that linearly 
increases with the number of sweeps. The constant offset of 23 double/DoF is 
mainly related to the access to the global solution vector. However, data also 
has to be loaded from L2 cache for smaller working-set sizes than the ones of the 
$k$-$d$ configurations mentioned above; the main reason for this is that the 
intermediate values do not fit into the cache any more.

For even higher dimensions, the L3 cache and main memory have to be accessed 
during the sweeps. While this operation  is negligible for $k=3$ (see 
Figure~\ref{fig:performance:cell:only:throughput}), it is performance-limiting in 
the case of $k=5$: For $k=5$~/~$d=5$, the bandwidth to L1 cache is limited by the 
access to L2 cache (see Figure~\ref{fig:performance:cell:only:bw}); for 
$k=5$~/~$d=6$, it is even limited by the main memory. In the latter case, the 
caches are hardly utilized any more and the data has to be fetched from/written 
back to main memory during every sweep, leading to a bandwidth close to the 
values measured for the STREAM benchmark, however, also resulting in a 
significant performance drop.

\begin{figure}



\begin{center}
\begin{tikzpicture}[scale=0.8]
    \begin{axis}[%
    hide axis,
    legend style={font=\footnotesize},
    xmin=10,
    xmax=50,
    ymin=0,
    ymax=0.4,
    semithick,
    legend style={draw=white!15!black,legend cell align=left},legend columns=-1
    ]

    \addlegendimage{gnuplot@darkblue,every mark/.append style={fill=gnuplot@red!50!black}}
    \addlegendentry{$k=3$};
    \addlegendimage{gnuplot@red,every mark/.append style={fill=gnuplot@red!50!black},dashed}
    \addlegendentry{$k=5$};

    \addlegendimage{black,mark=o}    
    \addlegendentry{L1$\leftrightarrow$L2};
    \addlegendimage{black,mark=square}
    \addlegendentry{L2$\leftrightarrow$L3};
    \addlegendimage{black,mark=x}
    \addlegendentry{Caches$\leftrightarrow$MEM};
    \end{axis}
\end{tikzpicture}
\end{center}

\vspace{-4.0cm}

\subfloat[{Data volume per degree of freedom.}\label{fig:performance:cell:all:data}]
{
  \begin{tikzpicture}
    \begin{semilogyaxis}[
      width=0.48\textwidth,
      height=0.31\textwidth,
      title style={font=\footnotesize},
      xlabel={dim},
      ylabel={data volume [double/DoF]},
      legend pos={south west},
      legend cell align={left},
      cycle list name=colorGPL,
      tick label style={font=\scriptsize},
      label style={font=\scriptsize},
      legend style={font=\scriptsize},
      grid,
      semithick,
      ymin=2,ymax=250,
      ytick={3,5,10,20,40,80,160},
      yticklabels={3,5,10,20,40,80,160},
      xmin=2,xmax=6,
      ]
      
	\draw [draw=black,thick] (axis cs: 2, 3) -- node[above,rotate=0,inner sep=1pt,outer sep=0pt, fill=white,draw]{} (axis cs: 6, 3);
	\draw [draw=black,thick] (axis cs: 2, 3) -- node[above,rotate=0,inner sep=1pt,outer sep=0pt, fill=white,draw]{{\tiny \color{black} ideal - 3}} (axis cs: 6, 3);

     
      \addplot +[gnuplot@darkblue,every mark/.append style={fill=gnuplot@darkblue!80!black},mark=o, solid] table [x=dim, y=da1] {results/cell_cache/post/all3_8_3_0.csv};
      \addplot +[gnuplot@darkblue,every mark/.append style={fill=gnuplot@darkblue!80!black},mark=square, solid] table [x=dim, y=da2] {results/cell_cache/post/all3_8_3_0.csv};
      \addplot +[gnuplot@darkblue,every mark/.append style={fill=gnuplot@darkblue!80!black},mark=x, solid] table [x=dim, y=da3] {results/cell_cache/post/all3_8_3_0.csv};

      \addplot +[gnuplot@red,every mark/.append style={fill=gnuplot@darkblue!80!black},mark=o, dashed] table [x=dim, y=da1] {results/cell_cache/post/all3_8_5_0.csv};
      \addplot +[gnuplot@red,every mark/.append style={fill=gnuplot@darkblue!80!black},mark=square, dashed] table [x=dim, y=da2] {results/cell_cache/post/all3_8_5_0.csv};
      \addplot +[gnuplot@red,every mark/.append style={fill=gnuplot@darkblue!80!black},mark=x, dashed] table [x=dim, y=da3] {results/cell_cache/post/all3_8_5_0.csv};

    \end{semilogyaxis}
  \end{tikzpicture}
}
\subfloat[{Accumulated bandwidth.}\label{fig:performance:cell:all:bw}]
{
  \begin{tikzpicture}
    \begin{semilogyaxis}[
      width=0.48\textwidth,
      height=0.31\textwidth,
      title style={font=\footnotesize},
      xlabel={dim},
      ylabel={bandwidth [GB/sec]},
      legend pos={south west},
      legend cell align={left},
      cycle list name=colorGPL,
      tick label style={font=\scriptsize},
      label style={font=\scriptsize},
      legend style={font=\scriptsize},
      grid,
      semithick,
      ymin=50,ymax=3600,
      ytick={50,100,200,400,800,1600,2400},
      yticklabels={50,100,200,400,800,1600,2400},
      xmin=2,xmax=6,
      ]
      
	\draw [draw=black,thick] (axis cs: 2, 2400) -- node[above,rotate=0,inner sep=1pt,outer sep=0pt, fill=white,draw]{} (axis cs: 6, 2400);
	\draw [draw=black,thick] (axis cs: 2, 2400) -- node[above,rotate=0,inner sep=1pt,outer sep=0pt, fill=white,draw]{{\tiny \color{black} 2400 GB/s}} (axis cs: 6, 2400);

	\draw [draw=black,thick] (axis cs: 2, 205) -- node[above,rotate=0,inner sep=1pt,outer sep=0pt, fill=white,draw]{} (axis cs: 6, 205);
	\draw [draw=black,thick] (axis cs: 2, 205) -- node[above,rotate=0,inner sep=1pt,outer sep=0pt, fill=white,draw]{{\tiny \color{black} STREAM - 205 GB/s}} (axis cs: 6, 205);	


      \addplot +[gnuplot@darkblue,every mark/.append style={fill=gnuplot@darkblue!80!black},mark=o, solid] table [x=dim, y=bw1] {results/cell_cache/post/all3_8_3_0.csv};
      \addplot +[gnuplot@darkblue,every mark/.append style={fill=gnuplot@darkblue!80!black},mark=square, solid] table [x=dim, y=bw2] {results/cell_cache/post/all3_8_3_0.csv};
      \addplot +[gnuplot@darkblue,every mark/.append style={fill=gnuplot@darkblue!80!black},mark=x, solid] table [x=dim, y=bw3] {results/cell_cache/post/all3_8_3_0.csv};

      \addplot +[gnuplot@red,every mark/.append style={fill=gnuplot@darkblue!80!black},mark=o, dashed] table [x=dim, y=bw1] {results/cell_cache/post/all3_8_5_0.csv};
      \addplot +[gnuplot@red,every mark/.append style={fill=gnuplot@darkblue!80!black},mark=square, dashed] table [x=dim, y=bw2] {results/cell_cache/post/all3_8_5_0.csv};
      \addplot +[gnuplot@red,every mark/.append style={fill=gnuplot@darkblue!80!black},mark=x, dashed] table [x=dim, y=bw3] {results/cell_cache/post/all3_8_5_0.csv};


    \end{semilogyaxis}
  \end{tikzpicture}
}
  
\subfloat[{Floating-point operations per degree of freedom.}\label{fig:performance:cell:all:flops}]
{  
  \begin{tikzpicture}
    \begin{axis}[
      width=0.48\textwidth,
      height=0.31\textwidth,
      title style={font=\footnotesize},
      xlabel={dim},
      ylabel={FLOPs/DoF},
      legend pos={south west},
      legend cell align={left},
      cycle list name=colorGPL,
      tick label style={font=\scriptsize},
      label style={font=\scriptsize},
      legend style={font=\scriptsize},
      grid,
      semithick,
      ymin=0,
      xmin=2,xmax=6,
      ]

	


      \addplot +[gnuplot@darkblue,every mark/.append style={fill=gnuplot@darkblue!80!black},mark=none, solid] table [x=dim, y=tflops] {results/cell_cache/post/all3_8_3_0.csv};
            
      \addplot +[gnuplot@red,every mark/.append style={fill=gnuplot@darkblue!80!black},mark=none,dashed] table [x=dim, y=tflops] {results/cell_cache/post/all3_8_5_0.csv};

    \end{axis}
  \end{tikzpicture}    
}
\subfloat[{Throughput.}\label{fig:performance:cell:all:throughput}]
{
  \begin{tikzpicture}
    \begin{semilogyaxis}[
      width=0.48\textwidth,
      height=0.31\textwidth,
      title style={font=\footnotesize},
      xlabel={dim},
      ylabel={throughput [GDoFs/sec]},
      legend pos={south west},
      legend cell align={left},
      cycle list name=colorGPL,
      tick label style={font=\scriptsize},
      label style={font=\scriptsize},
      legend style={font=\scriptsize},
      grid,
      semithick,
      ytick={0.25,0.5,1,2,4,8},
      ymin=0.2,ymax=6,
      yticklabels={0.25,0.5,1,2,4,8},
      xmin=2,xmax=6,
      ]

      \addplot +[gnuplot@darkblue,every mark/.append style={fill=gnuplot@darkblue!80!black},mark=none] table [x=dim, y=throughput] {results/cell_cache/post/all3_8_3_0.csv};

      \addplot +[gnuplot@red,every mark/.append style={fill=gnuplot@darkblue!80!black},mark=none,dashed] table [x=dim, y=throughput] {results/cell_cache/post/all3_8_5_0.csv};


    \end{semilogyaxis}
  \end{tikzpicture}
}

\caption{Full operator evaluation without loading values from neighboring cells}
\label{fig:performance:cell:all}

\end{figure}
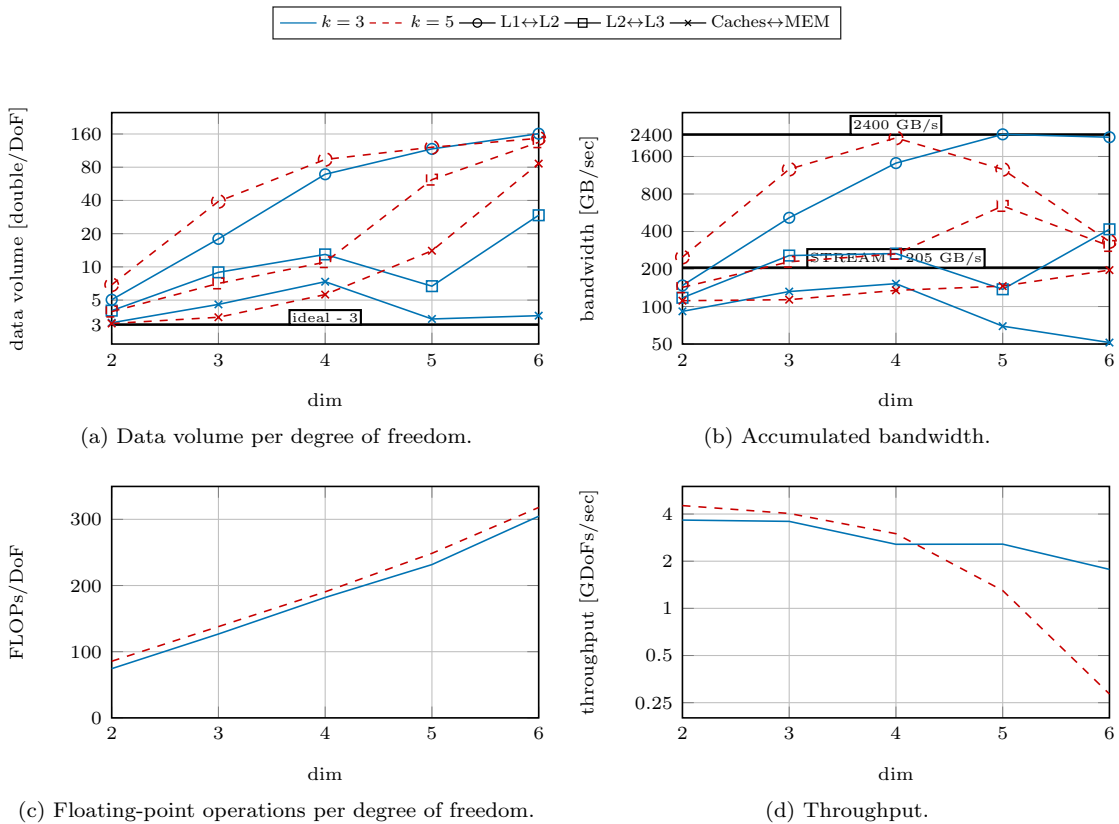

Figure~\ref{fig:performance:cell:only:flops} also shows the number of 
floating-point 
operations performed per degree of freedom. It increases linearly with the 
dimension $d$---with higher polynomial degrees requiring more work. It is clear 
that the arithmetic intensity will also increase linearly as long as the data 
stays in the cache (see also Subsection~\ref{sec:performance:advection}).

\subsubsection{Local cell and face integrals:}\label{sec:performance:celllocal:b}

In this subsection, we consider all computation steps in 
Algorithm~\ref{algo:concept:cell}, but ignore the data access to neighboring 
cells (line~\ref{algo:cell:line:face:loadneigh}). This means that face values  
from neighboring cells are not gathered and face buffers for exterior values are 
left unchanged. In this way, we are able to demonstrate the effects of increased 
working sets (of both face buffers) and of the increased number of sweeps. 
Additional $\ge 2 d$ sweeps have to be performed for interpolating values from 
the cell quadrature points to the quadrature points of the $2d$ faces (as well as 
for interpolating the values of the neighboring cells onto the quadrature points 
and for the flux computation).

Figure~\ref{fig:performance:cell:all} shows the data traffic between the memory 
hierarchy levels (data volume per DoF and bandwidth), the floating-point 
operations per DoF, and the throughput for $k=3$ and $k=5$ for $2\le d \le 6$. In 
comparison to the results of the experiments that only consider the cell 
integrals in Figure~\ref{fig:performance:cell:only}, the following observations 
can be made: As expected, the data volume transferred between the cache levels 
(Figure~\ref{fig:performance:cell:only:data} and 
\ref{fig:performance:cell:all:data}) and the number of floating-point operations 
approximately double (Figure~\ref{fig:performance:cell:only:flops} and 
\ref{fig:performance:cell:all:flops}). However, the configurations at which the 
traffic to the next cache level increases have not changed, indicating that the 
increase in working set is not limiting the performance here. 

The doubling of the data volume to be transferred for $k=5$ and high dimensions 
naturally leads to half the throughput (see 
Figure~\ref{fig:performance:cell:all:throughput}). In the case of $k=3$, we can 
also observe a drop of performance in high dimensions. This has another cause:
The memory transfer between L1 and L2 caches is up to 2,400GB/s and between L2 
and L3 caches about 400GB/s. The latter value is the half of that observed for the 
cell-integral-only run, indicating  that the data for the computation is mainly 
delivered form the L2 cache. This fact leads to an under-utilization of the L3 
cache and of the main-memory bandwidth, resulting in the drop of the overall 
performance by 50\% for high dimensions.




\subsection{Full advection operator}\label{sec:performance:advection}

\input{chapters/performance_advection_figure}


This subsection considers the application of the full advection operator as shown in Algorithm~\ref{algo:concept:cell}, including the access to neighboring cells during the computation of the numerical flux. Figure~\ref{fig:performance:ecl} presents the results of parameter studies of the dimension $2 \le d \le 6$ for different polynomial degrees $2\le k \le 5$.

Please note that the number of degrees of freedom per cell,
$(k+1)^{d}$,
is utilized as $x$-axis. This is done because the working set of this size is a suitable indicator of the overall performance of the operator evaluation (see Subsection~\ref{sec:performance:celllocal}). Also results taken from parameter studies of the polynomial degree $k$ are comparable to results obtained from parameter studies of the dimension $d$. 

The following observations can be  made in Figure~\ref{fig:performance:ecl}: For working sets that fit into the L1 cache, a higher polynomial degree leads to a higher throughput. For working sets exceeding the size of the L1 cache and only fitting into the L2 cache, the throughput drops. In this region, the curves overlap and we conclude that the throughput is indeed a function of the working-set size $\approx (k+1)^d$ and independent of the polynomial degree $k$ and the dimension $d$ individually.

Comparing these findings with the results presented in Subsection~\ref{sec:performance:celllocal:a} and \ref{sec:performance:celllocal:b}, an averaged performance drop of 54\% and 23\%, respectively, can be observed (see Table~\ref{tab:performance:advection:cmp:cell} and ~\ref{tab:performance:advection:cmp:noneigh}). Looking at the results for high dimensions, it becomes clear that processing the faces is more expensive than loading the actual values from the neighbors.

\pgfplotstableread{
a	b	c	d	e
2	187.4685554	0.6086133349	0.6105043262	2.1188980753
3	249.2821186	1.0469299708	1.0529824229	2.945271555
4	287.754456	1.2809114069	1.3459211919	4.2951218482
5	289.2086372	0.3948558362	1.6748650864	5.2567195592
6	280.2479469	0.2110942521	1.1126678389	7.2831981819
7	204.4116985	0.2161822663	0.2236526827	11.5873106494
8	99.1044422	0.2111527302	0.2137795902	1.1065111845
}\rooflineLaplaceThreeD

\pgfplotstableread{
a	b	c	d	e
2	199.7553059	1.9601791234	1.9711579881	80.8805440065
3	347.9614939	3.0698223003	3.3036382153	160.6026950268
4	311.7701722	0.4077780576	4.4670760088	167.0151690317
5	261.2815116	0.213689806	1.6504250368	177.7979930337
6	210.4110418	0.1763955965	0.316865132	36.0134130943
7	50.1353539	0.1747888893	0.1793850174	0.4610884748
8	28.5199677	0.1731614937	0.1773463697	0.256800029
}\rooflineLaplaceThreeE

%

  \begin{figure}[t]

  \begin{minipage}[c]{0.49\textwidth}

    \centering

      \captionof{table}{Relative performance due to flux computation (ratio Fig~\ref{fig:performance:ecl}-\ref{fig:performance:cell:only}).}\label{tab:performance:advection:cmp:cell}
\begin{footnotesize}
\begin{tabular}{c|ccccc}
\toprule
k/d & 2 & 3 & 4 & 5 & 6 \\ \midrule
3 & 58\%&	58\%&	51\%&	45\%&	41\%\\
4 & 66\%&	60\%&	53\%&	41\%&	54\%\\
5 & 68\%&	66\%&	54\%&	54\%&	52\%\\
 \bottomrule
\end{tabular}
\end{footnotesize}

\vspace{0.7cm}    
    
      \captionof{table}{Relative performance due to access to the values of neighboring cells (ratio Fig~\ref{fig:performance:ecl}-\ref{fig:performance:cell:all}).}\label{tab:performance:advection:cmp:noneigh}
\begin{footnotesize}
\begin{tabular}{c|ccccc}
\toprule
k/d & 2 & 3 & 4 & 5 & 6 \\ \midrule
3 & 73\%&	77\%&	76\%&	77\%&	79\%\\
4 & 79\%&	75\%&	77\%&	67\%&	80\%\\
5 & 82\%&	83\%&	78\%&	71\%&	91\%\\
 \bottomrule
\end{tabular}
\end{footnotesize}
    \end{minipage}
    \hfill  
  \begin{minipage}[c]{0.49\textwidth}
    \centering
\begin{tikzpicture}

\begin{loglogaxis}[
  at={(0.5\linewidth,0)},
  grid=both,
  width=0.95\textwidth,
  height=0.95\textwidth,
  xlabel=FLOP/byte ratio,
  ylabel=GFLOP/s,
  axis lines=left,
  ytick={2,4,8,16,32,64,128,256,512,1024,2048, 4096},
  yticklabels={2,4,8,16,32,64,128,256,512,1024,2048, 4096},
  xmin=0.125, xmax=48,ymin=16,ymax=5500,
  xtick={0.25, 0.5,1,2,4,8,16,32},
  xticklabels={$\frac 14$,$\frac 12$,1,2,4,8,16,32},
  tick label style={font=\scriptsize},
  legend pos = south west,
  semithick,
  legend columns=3, 
  legend style={font=\tiny}
  ]
  
    \addlegendimage{gnuplot@darkblue,every mark/.append style={fill=gnuplot@red!50!black}}
    \addlegendentry{$k=3$};
    \addlegendimage{gnuplot@red,every mark/.append style={fill=gnuplot@red!50!black},dashed}
    \addlegendentry{$k=5$};

    \addlegendimage{black,mark=*}    
    \addlegendentry{L1$\leftrightarrow$L2};
    \addlegendimage{black,mark=x}
    \addlegendentry{L2$\leftrightarrow$L3};
    \addlegendimage{black,mark=square}
    \addlegendentry{L3$\leftrightarrow$MEM};
  
    \begin{scope}[on background layer]
    \addplot[thick, color=black!10, fill=black!45, fill opacity=0.4,forget plot,overlay,draw=none] coordinates{ (0.125,16) (0.125,25) (20.75,4150) (48,4150) (48,16) (0.125,16)} \closedcycle;
  \end{scope}
  
  \draw[draw=black,thick] (axis cs:0.125,25) -- node[above,rotate=45,inner sep=2pt,outer sep=0.5pt]{\tiny Pure load memory bw 200 GB/s}(axis cs:20.75,4150);


  \draw[draw=black,thick] (axis cs:0.125,300) -- node[above,rotate=45,inner sep=2pt,outer sep=0.5pt]{\tiny 2400 GB/s (L1$\leftrightarrow$L2)}(axis cs:1.73,4150);    
  
  \draw[draw=black,thick] (axis cs:1.73, 4150) -- (axis cs:48, 4150) node[anchor=south east,inner sep=1pt,outer sep=0.5pt]{\tiny Peak DP};

\addplot +[gnuplot@darkblue,every mark/.append style={fill=gnuplot@darkblue},mark=*] table [x=ai3, y=flops] {results/cell/post/ecl_8_3_0.csv};

\addplot +[gnuplot@darkblue,every mark/.append style={fill=gnuplot@red!50!black},mark=x] table [x=ai2, y=flops] {results/cell/post/ecl_8_3_0.csv};

\addplot +[gnuplot@darkblue,every mark/.append style={fill=gnuplot@red!50!black},mark=square] table [x=ai1, y=flops] {results/cell/post/ecl_8_3_0.csv};

\addplot +[gnuplot@red,every mark/.append style={fill=gnuplot@red},mark=*,dashed] table [x=ai3, y=flops] {results/cell/post/ecl_8_5_0.csv};

\addplot +[gnuplot@red,every mark/.append style={fill=gnuplot@red!50!black},mark=x,dashed] table [x=ai2, y=flops] {results/cell/post/ecl_8_5_0.csv};

\addplot +[gnuplot@red,every mark/.append style={fill=gnuplot@red!50!black},mark=square,dashed] table [x=ai1, y=flops] {results/cell/post/ecl_8_5_0.csv};
  
\end{loglogaxis}

\end{tikzpicture}
    \captionof{figure}{Roofline model for $k=3/5$.}\label{fig:performance:roofline}
  \end{minipage}
  \end{figure}

Figure~\ref{fig:performance:roofline} shows a Roofline model~\cite{Williams2009} for $k=3$ and $k=5$.  In this model, the measured performance is plotted over the measured arithmetic intensity
\begin{align}
\text{(measured arithmetic intensity)}_i = \frac{\text{measured performance}}{\text{(measured bandwidth)}_i},
\end{align}
with $i \in \{ \text{L1 $\leftrightarrow$ L2}, \text{L2 $\leftrightarrow$ L3}, \text{Caches $\leftrightarrow$ MEM} \}$. 
We can compute the arithmetic intensity of each level of memory hierarchy as we measure the necessary bandwidth with \texttt{LIKWID}. The diagram confirms the observation made above: A high arithmetic intensity and consequently high performance can only be reached if the caches (L1 and L2) are utilized well. As the working sets get too large, L1 and L2 caches are under-utilized, the arithmetic intensity on the other levels drops and new hard (bandwidth) ceilings limit the maximal possible performance.

\subsection{Alternative implementations}\label{sec:performance:alternatives}

In the following subsection, we compare the performance of the default setup of the library \texttt{hyper.deal} (tensor product of mappings, ECL, Gauss quadrature, vectorization over elements with highest ISA extension for vectorization--see also Subsection~\ref{sec:mem:performance}) with the performance of alternative algorithms and/or configurations.

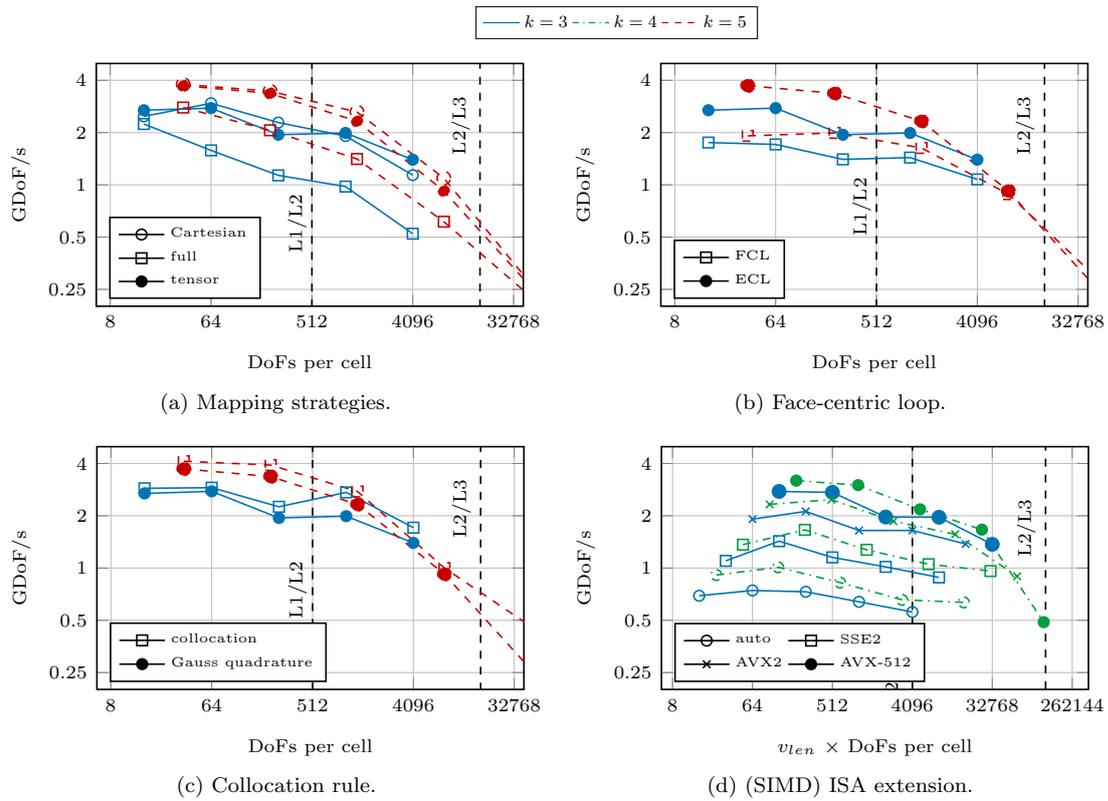
\begin{figure}

\centering

\begin{tikzpicture}[scale=0.8]
    \begin{axis}[%
    hide axis,
    legend style={font=\footnotesize},
    xmin=10,
    xmax=1000,
    ymin=0,
    ymax=0.4,
    semithick,
    legend style={draw=white!15!black,legend cell align=left},legend columns=-1
    ]

    \addlegendimage{gnuplot@darkblue,every mark/.append style={fill=gnuplot@red!50!black}}
    \addlegendentry{$k=3$};
    \addlegendimage{gnuplot@green,every mark/.append style={fill=gnuplot@red!50!green},dash dot}
    \addlegendentry{$k=4$};
    \addlegendimage{gnuplot@red,every mark/.append style={fill=gnuplot@red!50!black},dashed}
    \addlegendentry{$k=5$};

    \end{axis}
\end{tikzpicture}

\vspace{-4.1cm}



%
%
%


\subfloat[{Mapping strategies.}\label{fig:performance:alternatives:mapping}]
{
  \begin{tikzpicture}
    \begin{loglogaxis}[
      width=0.48\textwidth,
      height=0.32\textwidth,
      title style={font=\footnotesize},every axis title/.style={above left,at={(1,1)},draw=black,fill=white},
      xlabel={DoFs per cell}, ylabel near ticks,
      ylabel={GDoF/s},
      legend pos={south west},
      legend cell align={left},
      cycle list name=colorGPL,
      tick label style={font=\scriptsize},
      label style={font=\scriptsize},
      legend style={font=\scriptsize},
      grid,
      semithick,
      ytick={0.25,0.5,1,2,4,8},
      ymin=0.2,ymax=5,
      yticklabels={0.25,0.5,1,2,4,8},
      xmin=6,xmax=40000,
      xtick={8, 64,512, 4096, 32768},
      xticklabels={8, 64,512, 4096, 32768},
  legend columns=1, legend style={font=\tiny}
      ]

    \addlegendimage{mark=o, solid}
    \addlegendentry{Cartesian};
    \addlegendimage{mark=square}
    \addlegendentry{full};
    \addlegendimage{mark=*}
    \addlegendentry{tensor};

%
%
    
	\node[anchor=south] at (axis cs: 512, 0.1) (nodeFE) {};
	\node[anchor=north] at (axis cs: 512, 1000) (nodeFF) {};
	\draw [-, dashed] (nodeFE) -- (nodeFF) ;
    \node[anchor=south, rotate=90] at (axis cs:  560, 0.6) (nodeDE) {{\scriptsize L1/L2}};
    
	\node[anchor=south] at (axis cs: 16384, 0.1) (nodeIE) {};
	\node[anchor=north] at (axis cs: 16384, 10000) (nodeIF) {};
	\draw [-, dashed] (nodeIE) -- (nodeIF) ;
    \node[anchor=south, rotate=90] at (axis cs:  16384, 2.2) (nodeIE) {{\scriptsize L2/L3}};


      \addplot +[gnuplot@darkblue,every mark/.append style={fill=gnuplot@darkblue},mark=*] table [x=dofs, y=throughput] {results/cell/post/ecl_8_3_0.csv};
      

      \addplot +[gnuplot@red,every mark/.append style={fill=gnuplot@red},mark=*,dashed] table [x=dofs, y=throughput] {results/cell/post/ecl_8_5_0.csv};


      \addplot +[gnuplot@darkblue,every mark/.append style={fill=gnuplot@red},mark=o, solid] table [x=dofs, y=throughput] {results/cart/post/ecl_8_3_0.csv};
      

      \addplot +[gnuplot@red,every mark/.append style={fill=gnuplot@orange},mark=o, solid,dashed] table [x=dofs, y=throughput] {results/cart/post/ecl_8_5_0.csv};


      \addplot +[gnuplot@darkblue,every mark/.append style={fill=gnuplot@red!50!black},mark=square, solid] table [x=dofs, y=throughput] {results/full/post/ecl_8_3_0.csv};
      

      \addplot +[gnuplot@red,every mark/.append style={fill=gnuplot@red!50!black},mark=square,dashed] table [x=dofs, y=throughput] {results/full/post/ecl_8_5_0.csv};

    \end{loglogaxis}
  \end{tikzpicture}
}
\subfloat[{Face-centric loop.}\label{fig:performance:alternatives:fcl}]
{
  \begin{tikzpicture}
    \begin{loglogaxis}[
      width=0.48\textwidth,
      height=0.32\textwidth,
      title style={font=\footnotesize},
      xlabel={DoFs per cell}, ylabel near ticks,
      ylabel={GDoF/s},
      legend pos={south west},
      legend cell align={left},
      cycle list name=colorGPL,
      tick label style={font=\scriptsize},
      label style={font=\scriptsize},
      legend style={font=\scriptsize},
      grid,
      semithick,
      ytick={0.25,0.5,1,2,4,8},
      ymin=0.2,ymax=5,
      yticklabels={0.25,0.5,1,2,4,8},
      xmin=6,xmax=40000,
      xtick={8, 64,512, 4096, 32768},
      xticklabels={8, 64,512, 4096, 32768},
  legend columns=1, legend style={font=\tiny}
      ]

%
%
    
	\node[anchor=south] at (axis cs: 512, 0.1) (nodeFE) {};
	\node[anchor=north] at (axis cs: 512, 1000) (nodeFF) {};
	\draw [-, dashed] (nodeFE) -- (nodeFF) ;
    \node[anchor=south, rotate=90] at (axis cs:  560, 0.75) (nodeDE) {{\scriptsize L1/L2}};
    
	\node[anchor=south] at (axis cs: 16384, 0.1) (nodeIE) {};
	\node[anchor=north] at (axis cs: 16384, 10000) (nodeIF) {};
	\draw [-, dashed] (nodeIE) -- (nodeIF) ;
    \node[anchor=south, rotate=90] at (axis cs:  16384, 2.2) (nodeIE) {{\scriptsize L2/L3}};
      
    \addlegendimage{mark=square, black}
    \addlegendentry{FCL};
    \addlegendimage{mark=*, black}
    \addlegendentry{ECL};


      \addplot +[gnuplot@darkblue,every mark/.append style={fill=gnuplot@red!50!black},mark=square] table [x=dofs, y=throughput] {results/fcl/post/fcl_8_3_0.csv};
      

      \addplot +[gnuplot@red,every mark/.append style={fill=gnuplot@red!50!black},mark=square,dashed] table [x=dofs, y=throughput] {results/fcl/post/fcl_8_5_0.csv};


      \addplot +[gnuplot@darkblue,every mark/.append style={fill=gnuplot@darkblue},mark=*] table [x=dofs, y=throughput] {results/cell/post/ecl_8_3_0.csv};
      

      \addplot +[gnuplot@red,every mark/.append style={fill=gnuplot@red},mark=*,dashed] table [x=dofs, y=throughput] {results/cell/post/ecl_8_5_0.csv};


    \end{loglogaxis}
  \end{tikzpicture}
}

\subfloat[{Collocation rule.}\label{fig:performance:alternatives:collocation}]
{
  \begin{tikzpicture}
    \begin{loglogaxis}[
      width=0.48\textwidth,
      height=0.32\textwidth,
      title style={font=\footnotesize},
      xlabel={DoFs per cell},
      ylabel={GDoF/s}, ylabel near ticks,
      legend pos={south west},
      legend cell align={left},
      cycle list name=colorGPL,
      tick label style={font=\scriptsize},
      label style={font=\scriptsize},
      legend style={font=\scriptsize},
      grid,
      semithick,
      ytick={0.25,0.5,1,2,4,8},
      ymin=0.2,ymax=5,
      yticklabels={0.25,0.5,1,2,4,8},
      xmin=6,xmax=40000,
      xtick={8, 64,512, 4096, 32768},
      xticklabels={8, 64,512, 4096, 32768},
  legend columns=1, legend style={font=\tiny}
      ]

%
%
    
	\node[anchor=south] at (axis cs: 512, 0.1) (nodeFE) {};
	\node[anchor=north] at (axis cs: 512, 1000) (nodeFF) {};
	\draw [-, dashed] (nodeFE) -- (nodeFF) ;
    \node[anchor=south, rotate=90] at (axis cs:  560, 0.75) (nodeDE) {{\scriptsize L1/L2}};
    
	\node[anchor=south] at (axis cs: 16384, 0.1) (nodeIE) {};
	\node[anchor=north] at (axis cs: 16384, 10000) (nodeIF) {};
	\draw [-, dashed] (nodeIE) -- (nodeIF) ;
    \node[anchor=south, rotate=90] at (axis cs:  16384, 2.2) (nodeIE) {{\scriptsize L2/L3}};

    \addlegendimage{mark=square, black}
    \addlegendentry{collocation};
    \addlegendimage{mark=*, black}
    \addlegendentry{Gauss quadrature};
      

      \addplot +[gnuplot@darkblue,every mark/.append style={fill=gnuplot@red!50!black},mark=square] table [x=dofs, y=throughput] {results/collocation/post/ecl_8_3_0.csv};
      

      \addplot +[gnuplot@red,every mark/.append style={fill=gnuplot@red!50!black},mark=square,dashed] table [x=dofs, y=throughput] {results/collocation/post/ecl_8_5_0.csv};


      \addplot +[gnuplot@darkblue,every mark/.append style={fill=gnuplot@darkblue},mark=*] table [x=dofs, y=throughput] {results/cell/post/ecl_8_3_0.csv};
      

      \addplot +[gnuplot@red,every mark/.append style={fill=gnuplot@red},mark=*,dashed] table [x=dofs, y=throughput] {results/cell/post/ecl_8_5_0.csv};

    \end{loglogaxis}
  \end{tikzpicture}
} 
\subfloat[{(SIMD) ISA extension.}\label{fig:performance:simd}]
{
  \begin{tikzpicture}
    \begin{loglogaxis}[
      width=0.48\textwidth,
      height=0.32\textwidth,
      title style={font=\footnotesize},
      xlabel={$v_{len}$ $\times$ DoFs per cell},
      ylabel={GDoF/s},ylabel near ticks,
      legend pos={south west},
      legend cell align={left},
      cycle list name=colorGPL,
      tick label style={font=\scriptsize},
      label style={font=\scriptsize},
      legend style={font=\scriptsize},
      grid,
      semithick,
      ytick={0.25,0.5,1,2,4,8},
      ymin=0.2,ymax=5.0,
      yticklabels={0.25,0.5,1,2,4,8},
      xmin=6,xmax=400000,
      xtick={8, 64,512, 4096, 32768, 262144},
      xticklabels={8, 64,512, 4096, 32768,262144},
  legend columns=2, legend style={font=\tiny}
      ]
      
    \addlegendimage{black, solid, mark=o}
    \addlegendentry{auto};
    \addlegendimage{black, solid, mark=square}
    \addlegendentry{SSE2};
    \addlegendimage{black, solid, mark=x}
    \addlegendentry{AVX2};
    \addlegendimage{black, solid, mark=*}
    \addlegendentry{AVX-512};

%
%
    
	\node[anchor=south] at (axis cs: 4096, 0.01) (nodeFE) {};
	\node[anchor=north] at (axis cs: 4096, 1000) (nodeFF) {};
	\draw [-, dashed] (nodeFE) -- (nodeFF) ;
    \node[anchor=south, rotate=90] at (axis cs:  4096, 0.15) (nodeDE) {{\scriptsize L1/L2}};
    
	\node[anchor=south] at (axis cs: 131072, 0.01) (nodeIE) {};
	\node[anchor=north] at (axis cs: 131072, 10000) (nodeIF) {};
	\draw [-, dashed] (nodeIE) -- (nodeIF) ;
    \node[anchor=south, rotate=90] at (axis cs:  131072, 1.7) (nodeIE) {{\scriptsize L2/L3}};

      \addplot +[gnuplot@darkblue,every mark/.append style={fill=gnuplot@darkblue!80!black},mark=o] table [x=dofs, y=throughput,x expr=\thisrow{dofs}*1] {results/simd/post/ecl_1_3_0.csv};

      \addplot +[gnuplot@darkblue,every mark/.append style={fill=gnuplot@red!50!black},mark=square] table [x=dofs, y=throughput,x expr=\thisrow{dofs}*2] {results/simd/post/ecl_2_3_0.csv};
      
      \addplot +[gnuplot@darkblue,every mark/.append style={fill=gnuplot@red!50!black},mark=x] table [x=dofs, y=throughput,x expr=\thisrow{dofs}*4] {results/simd/post/ecl_4_3_0.csv};      

      \addplot +[gnuplot@darkblue,every mark/.append style={fill=gnuplot@darkblue},mark=*] table [x=dofs, y=throughput,x expr=\thisrow{dofs}*8] {results/simd/post/ecl_8_3_0.csv};

      \addplot +[gnuplot@green,every mark/.append style={fill=gnuplot@green},mark=o, dash dot] table [x=dofs, y=throughput,x expr=\thisrow{dofs}*1] {results/simd2/post/ecl_1_4_0.csv};

      \addplot +[gnuplot@green,every mark/.append style={fill=gnuplot@green},mark=square, dash dot] table [x=dofs, y=throughput,x expr=\thisrow{dofs}*2] {results/simd2/post/ecl_2_4_0.csv};
      
      \addplot +[gnuplot@green,every mark/.append style={fill=gnuplot@green},mark=x, dash dot] table [x=dofs, y=throughput,x expr=\thisrow{dofs}*4] {results/simd2/post/ecl_4_4_0.csv};      

      \addplot +[gnuplot@green,every mark/.append style={fill=gnuplot@green},mark=*, dash dot] table [x=dofs, y=throughput,x expr=\thisrow{dofs}*8] {results/simd2/post/ecl_8_4_0.csv};

    \end{loglogaxis}
  \end{tikzpicture}
  
}
  
\caption{a-c)Comparison of the performance of different algorithms and the default configuration for $k=3$ and $k=5$. d) Comparison of the performance of different (SIMD) ISA extensions and the auto-vectorization for $k=3$ and $k=4$. }\label{fig:performance:alternatives}

\end{figure}

The library \texttt{hyper.deal} has been developed to be able to compute efficiently on complex geometries both in geometric and velocity space. 
This is achieved by combining the Jacobians of the mapping of the individual spaces on the fly. The upper limit of the performance of this approach is given by the consideration of the tensor product of two Cartesian grids, which leads to the same constant and diagonal Jacobian matrix at all quadrature points. As a lower limit, one can consider the case that each quadrature point has a unique Jacobian of size $d \times d$. Figure~\ref{fig:performance:alternatives:mapping} shows that the behavior of the default tensor-product setup is similar to that of a pure Cartesian grid simulation with only a small averaged performance penalty of approx. 4\%. This observation matches our expectations expressed in Subsection~\ref{sec:concept:fe} and means that in high dimensions the evaluation of curved meshes in the tensor-product factors is essentially for free, compared to storing the full Jacobian matrices. 

In the default configuration of \texttt{hyper.deal}, we have been using element-centric loops (ECL). They show significantly better performance than face-centric loops (FCL) as shown in Figure~\ref{fig:performance:alternatives:fcl}. We have not implemented any advanced blocking schemes---as it is available in \texttt{deal.II}---for neither ECL nor FCL. The fact that ECL nevertheless shows a better performance demonstrates the natural cache-friendly property of ECL. Furthermore, ECL is well-suited for shared-memory computation and a single communication step is sufficient. The benefit of ECL decreases for high dimensions due to the increased number of sweeps, related to the repeated evaluation of the flux terms. Nevertheless, we propose to use ECL for high-dimension problems because of its suitability for shared-memory computations that reduce the allocated memory.

We favor the Gauss--Legendre quadrature method over the collocation methods due to its higher numerical accuracy. This benefit comes at the price of a basis change from the Gauss--Lobatto points to the Gauss quadrature points and vice versa. Figure~\ref{fig:performance:alternatives:collocation} shows a performance drop of on average 15\% due to these basis changes as long as the data that should be interpolated remains in cache.

In the library \texttt{hyper.deal}, we currently only support ``vectorization over elements''. As a default, the highest instruction-set extension is selected, i.e., the maximum number of cells is processed at once by a core. Since the number of lanes to be used is templated, the user can reduce the number of elements that are processed at once, as it is demonstrated in Figure~\ref{fig:performance:simd}. It can be observed that, in general, the usage of higher instruction-set extensions leads to a better throughput. However, once the working set of a cell batch exceeds the size of the cache, a performance drop can be observed. The performance drop leads to the fact that in 6D with cubic elements the throughputs of AVX-512 and of AVX2 are comparable and in 6D with quartic elements SSE2 shows the best performance.


In this subsection, we have demonstrated that the chosen default configuration of the library \texttt{hyper.deal} has a competitive throughput compared to less memory-expensive  and computationally demanding algorithms, which are numerically inferior.


%
%

\subsection{Strong and weak scaling}\label{sec:performance:scaling}

\begin{table}[t]

\centering

\caption{Partitioning the $\vec{x}$- and $\vec{v}$-triangulations on up to 3,072 
nodes with 48 cores}\label{tab:performance:sw:configurations}

\begin{footnotesize}
\begin{tabular}{cccccccccccccccccccccc}
\toprule
nodes & & 1 & 2 & 4 & 8 & 16 & 32 & 64 & 128 & 256 & 512 & 1,024 & 2,048 & 3,072 \\
\midrule
$p_{\vec{x}}$ & & 8 & 12 & 16 & 24 & 32 & 48 & 64 & 96 & 128 & 192 & 256 & 384 & 384 \\
$p_{\vec{v}}$ & & 6 & 8 & 12 & 16 & 24 & 32 & 48 & 64 & 96 & 128 & 192 & 256 & 384 \\
\bottomrule
\end{tabular}
\end{footnotesize}

\caption{Strong and weak scaling configurations: left) $d=4$ / $k=5$;  right) $d=6$ / $k=3$}\label{tab:performance:sw:partitions}

\begin{footnotesize}
\begin{tabular}{lc}
\toprule 
DoFs & configuration \\
\midrule
5.4GDoFs & $384^2 \cdot 192^2$\\
21.7GDoFs & $384^4$\\
268MDoFs/node & $192^2 \cdot 96^2$\\
1.1GDoFs/node & $192^4$ \\
\bottomrule
\end{tabular}
\begin{tabular}{lc}
\toprule 
DoFs & configuration \\
\midrule
2.1GDoFs & $32^5 \cdot 64^1$\\
17.2GDoFs & $32^2 \cdot 64^4$\\
268MDoFs/node & $16^2 \cdot 32^4$\\
1.1GDoFs/node & $32^6$ \\
\bottomrule
\end{tabular}
\end{footnotesize}

\end{table}


In this subsection, we examine the parallel efficiency of the library 
\texttt{hyper.deal}. For this study, we consider the advection operator embedded 
into a low-storage Runge--Kutta scheme of order 4 with 5 stages, which uses two 
auxiliary vectors besides the solution vector \cite{Kennedy00}. From these three 
vectors, only one (auxiliary) vector is ghosted.

\begin{figure}

\centering

  \strut\hfill
\subfloat[{Configuration ``$d=4$~/~$k=5$''.}\label{fig:performance:scaling:time:a}]
{  
  \begin{tikzpicture}
    \begin{loglogaxis}[
      width=0.48\textwidth,
      height=0.38\textwidth,
      title style={font=\tiny},every axis title/.style={above left,at={(1,1)},draw=black,fill=white},
      xlabel={Nodes ($\times$ 48 CPUs)},
      ylabel={Time per Runge--Kutta stage [s]},
      legend pos={south west},
      legend cell align={left},
      cycle list name=colorGPL,
      tick label style={font=\scriptsize},
      label style={font=\scriptsize},
      legend style={font=\scriptsize},
      grid,
      semithick,
      ymin=0.0008,ymax=6,
      xmin=0.8,xmax=4096,
      xtick={1, 2,4, 8, 16,32,64,128,256,512, 1024, 2048, 3072},
      xticklabels={1, ,4, , 16,,64,,256,, 1024, ,3072},
      legend style={at={(1.9,0.8)},
	anchor=north,legend columns=4},
      ]
      
	\node[anchor=south] at (axis cs: 768, 1e-5) (nodeAE) {};
	\node[anchor=north] at (axis cs: 768, 1e+2) (nodeAF) {};
	\draw [-, dashed] (nodeAE) -- (nodeAF) ;
	
	\node[anchor=south] at (axis cs: 1536, 1e-5) (nodeBE) {};
	\node[anchor=north] at (axis cs: 1536, 1e+2) (nodeBF) {};
	\draw [-, dashed] (nodeBE) -- (nodeBF) ;	


  \draw[draw=black,thick] (axis cs:0.1, 0.19) -- node[above,rotate=-0,inner sep=2pt,outer sep=0.5pt]{}(axis cs:4096, 0.19);       
  
  \draw[draw=black,thick] (axis cs:128, 0.19) -- node[above,rotate=-0,inner sep=1pt,outer sep=-6pt, fill=white, draw]{{\tiny  \color{black} \textbf{340MDoFs/node}}}(axis cs:4096, 0.19);

  \draw[draw=black,thick] (axis cs:0.1, 0.79) -- node[above,rotate=-0,inner sep=2pt,outer sep=0.5pt]{}(axis cs:4096, 0.79);       
  
  \draw[draw=black,thick] (axis cs:128, 0.79) -- node[above,rotate=-0,inner sep=1pt,outer sep=-6pt, fill=white, draw]{{\tiny  \color{black} \textbf{1.4GDoFs/node}}}(axis cs:4096, 0.79);

  \draw[draw=black,thick] (axis cs:0.5, 6.2301) -- node[above,rotate=-35,inner sep=1pt,outer sep=-6pt, fill=white]{}(axis cs:4096, 7.6052e-04);  
  \draw[draw=black,thick] (axis cs:16, 0.19469) -- node[above,rotate=-35,inner sep=1pt,outer sep=-6pt, fill=white,draw]{{\tiny  \color{black} \textbf{5.4GDoFs}}}(axis cs:4096, 7.6052e-04);

  \draw[draw=black,thick] (axis cs:0.5, 25.331) -- node[above,rotate=-35,inner sep=1pt,outer sep=-6pt, fill=white]{}(axis cs:4096, 0.0030922);  
  \draw[draw=black,thick] (axis cs:16, 0.79160) -- node[above,rotate=-35,inner sep=1pt,outer sep=-6pt, fill=white,draw]{{\tiny  \color{black} \textbf{21.7GDoFs}}}(axis cs:4096, 0.0030922);


	\addplot table [x index=0,y index=1] {results/scaling-advection-4-5/scaling.14.csv};
	
	\addplot table [x index=0,y index=1] {results/scaling-advection-4-5/scaling.16.csv};	
	
	\addplot table [x index=0,y index=1] {results/scaling-advection-4-5/weak.5.csv};
	\addplot table [x index=0,y index=1] {results/scaling-advection-4-5/weak.9.csv};


      


    \end{loglogaxis}
  \end{tikzpicture}
}
\subfloat[{Configuration ``$d=6$~/~$k=3$''.}\label{fig:performance:scaling:time:b}]
{
  \begin{tikzpicture}
    \begin{loglogaxis}[
      width=0.48\textwidth,
      height=0.38\textwidth,
      title style={font=\tiny},every axis title/.style={above left,at={(1,1)},draw=black,fill=white},
      xlabel={Nodes ($\times$ 48 CPUs)},
      legend pos={south west},
      legend cell align={left},
      cycle list name=colorGPL,
      tick label style={font=\scriptsize},
      label style={font=\scriptsize},
      legend style={font=\scriptsize},
      grid,
      semithick,
      ymin=0.0008,ymax=6,
      xmin=0.8,xmax=4096,
      xtick={1, 2,4, 8, 16,32,64,128,256,512, 1024, 2048, 3072},
      xticklabels={1, ,4, , 16,,64,,256,, 1024, ,3072},
      ]

	\node[anchor=south] at (axis cs: 768, 1e-5) (nodeAE) {};
	\node[anchor=north] at (axis cs: 768, 1e+2) (nodeAF) {};
	\draw [-, dashed] (nodeAE) -- (nodeAF) ;
	
	\node[anchor=south] at (axis cs: 1536, 1e-5) (nodeBE) {};
	\node[anchor=north] at (axis cs: 1536, 1e+2) (nodeBF) {};
	\draw [-, dashed] (nodeBE) -- (nodeBF) ;	


  \draw[draw=black,thick] (axis cs:0.1, 0.95516) -- node[above,rotate=-0,inner sep=2pt,outer sep=-12pt]{}(axis cs:4096, 0.95516);
      
  \draw[draw=black,thick] (axis cs:128, 0.95516) -- node[above,rotate=-0,inner sep=1pt,outer sep=-6pt, fill=white, draw]{{\tiny \color{black} \textbf{1.1GDoFs/node}}}(axis cs:4096, 0.95516);       
  
  \draw[draw=black,thick] (axis cs:0.1, 0.25549) -- node[above,rotate=-0,inner sep=2pt,outer sep=-12pt]{}(axis cs:4096, 0.25549);
      
  \draw[draw=black,thick] (axis cs:128, 0.25549) -- node[above,rotate=-0,inner sep=1pt,outer sep=-6pt, fill=white, draw]{{\tiny \color{black} \textbf{268MDoFs/node}}}(axis cs:4096, 0.25549);       
  
  \draw[draw=black,thick] (axis cs:0.5, 40.332) -- node[above,rotate=-35,inner sep=1pt,outer sep=-6pt, fill=white]{}(axis cs:4096, 0.0049233);    
  \draw[draw=black,thick] (axis cs:16, 1.2604) -- node[above,rotate=-35,inner sep=1pt,outer sep=-6pt, fill=white,draw]{{\tiny \color{black} \textbf{17.2GDoFs}}}(axis cs:4096, 0.0049233);  

  \draw[draw=black,thick] (axis cs:0.5, 8.7832) -- node[above,rotate=-35,inner sep=1pt,outer sep=-6pt, fill=white]{}(axis cs:4096, 0.0010722);    
  \draw[draw=black,thick] (axis cs:16, 0.27448) -- node[above,rotate=-35,inner sep=1pt,outer sep=-6pt, fill=white,draw]{{\tiny \color{black} \textbf{2.1GDoFs}}}(axis cs:4096, 0.0010722);

\addplot table [x index=0,y index=1] {results/scaling-advection-curved/scaling.12.csv};     
\addplot table [x index=0,y index=1] {results/scaling-advection-curved/scaling.14.csv};

\addplot table [x index=0,y index=1] {results/scaling-advection-curved/weak.14.csv};  

\addplot table [x index=0,y index=1] {results/scaling-advection-curved/weak.18.csv};

\addplot table [x index=0,y index=1] {results/scaling-advection/scaling.22.csv};


      \node[] at (axis cs: 500, 4.0) {{\scriptsize \color{black} \textbf{4.4TDoFs (Cartesian)}}};

\addplot +[black,every mark/.append style={fill=black}, dash dot, mark=x] table [x index=0, y index=8] {results/scaling-advection-range2/final.result2};    

\legend{,,,,,alternative partitioning (see Fig.~\ref{fig:performance:scaling:communication})}

    \end{loglogaxis}
  \end{tikzpicture}
}
  \hfill\strut
  
\vspace{-0.3cm}

\caption{Strong and weak scaling of one Runge--Kutta step with the advection operator as right-hand side with up to 147,456 processes on 3,072 compute nodes.}
\label{fig:performance:scaling:time}
\end{figure}
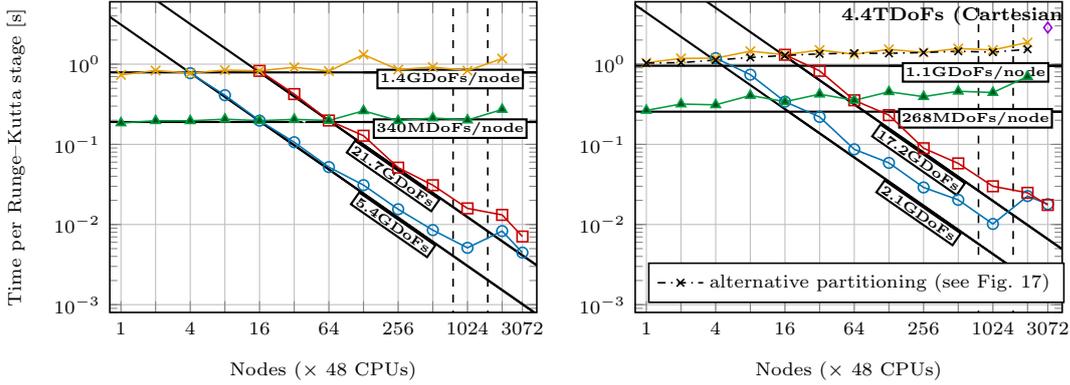

Figure~\ref{fig:performance:scaling:time} shows strong and weak scaling results 
of runs on SuperMUC-NG with up to 3072 nodes with a total of 147,456 cores. We 
consider two configurations: ``$d=4$~/~$k=5$'', an easy configuration, and  
``$d=6$~/~$k=3$'', a demanding configuration. As examples, we present for each 
configuration two strong and two weak scaling curves (see 
Table~\ref{tab:performance:sw:configurations}). 
Table~\ref{tab:performance:sw:partitions} shows the considered process 
decomposition $p=p_{\vec{x}} \cdot p_{\vec{v}}$, which has not been optimized for 
the given mesh configurations.

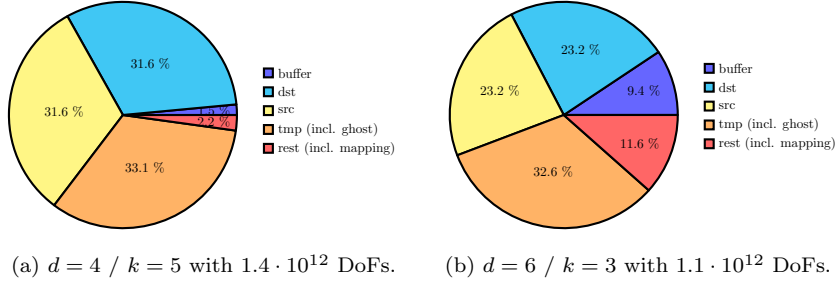
\begin{figure}

\vspace{-0.7cm}

\centering

\subfloat[{{$d=4$~/~$k=5$} with $1.4\cdot 10^{12}$ DoFs.}\label{fig:performace:adv:mem:a}]
{
\begin{tikzpicture}[scale=0.5, every node/.style={scale=0.5}]
\pie [ text = legend ]{1.5 / buffer , 31.6 / dst , 31.6 / src , 33.1 / tmp (incl. ghost) , 2.2 / rest (incl. mapping) }
\end{tikzpicture}
}
~
\subfloat[{{$d=6$~/~$k=3$}  with $1.1\cdot 10^{12}$ DoFs.}\label{fig:performace:adv:mem:b}]
{
\begin{tikzpicture}[scale=0.5, every node/.style={scale=0.5}]
\pie [ text = legend ]{9.4 / buffer , 23.2 / dst , 23.2 / src , 32.6 / tmp (incl. ghost) , 11.6 / rest (incl. mapping) }
\end{tikzpicture}
}

\caption{Memory consumption for 48$\times$1024 cores.}\label{fig:performace:adv:mem}

\end{figure}

For the ``$d=4$~/~$k=5$'' configuration, we observe excellent weak-scaling 
behavior with parallel efficiencies of 92\% and 89\% for up to 2,048 nodes on the 
large and small setup, respectively. We get more than 70/80\% efficiency for 
strong scaling up to the increase in the number of nodes by a factor of 128/256.
For the ``$d=6$~/~$k=3$'' configuration, we see parallel efficiencies of 38/56\% 
for weak scaling.  
These values are lower than the ones in the 4D case, however, they are still very 
good in the light of the immense communication amount in the 6D case: As shown in 
Fig~\ref{fig:performace:adv:mem}, the ghost data amount to 40\% of the solution 
vector in 6D and only 5\% in  4D.

The largest simulation with curved mesh, which contains 
$128^5 \cdot 64^1 = 2.2\cdot10^{12}$ degrees of freedom, reaches a total 
throughput of 1.2PDoFs/s(=$1.2\cdot10^{12}$DoFs/s)  on 2048 compute nodes and 
1.7PDoFs/s on 3072 compute nodes. This means that each Runge-Kutta stage is 
processed in 1.8s/1.3s, and a complete time step consisting of 5 Runge-Kutta 
stages takes 9.2s/6.4s.

Figure~\ref{fig:performace:adv:mem:b} shows the approximated memory consumption 
for a large-scale simulation (1024 nodes, $d=6$~/~$k=3$, $1.1\cdot 10^{12}$ 
DoFs). A total of 34.4PB main memory from available 98PB is used. The largest 
amount of memory is occupied by the three solution vectors (each. 23.2\%). The 
two buffers for MPI communication occupy each 9.4\%. One of the buffers is 
attributed to the ghost-value section of the vector called $tmp$. The remaining 
data structures, which include inter alia the mapping data, occupy only a small 
share (11.6\%) of the main memory, illustrating the benefit of the tensor-product  
approach employed by the library \texttt{hyper.deal} in constructing a 
memory-efficient algorithm for arbitrary complex geometries for high dimensions. 
As reference, the memory consumption for a 4D simulation is shown in 
Figure~\ref{fig:performace:adv:mem:a}. Obviously, the additional memory 
consumption due to the precomputed mapping slightly reduces the largest possible 
problem size fitting to a certain number of nodes. For example, the largest 4.4 
trillion DoFs simulation of Figure~\ref{fig:performance:scaling:time} only fits 
into memory on 3072 compute nodes for a Cartesian mesh.




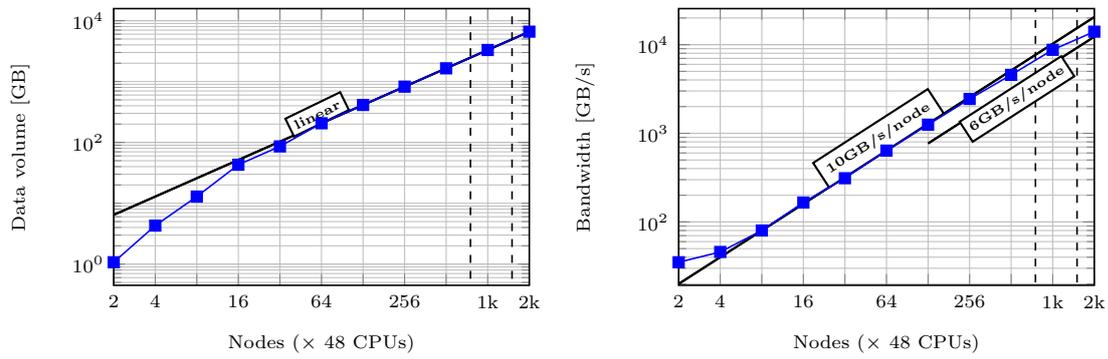
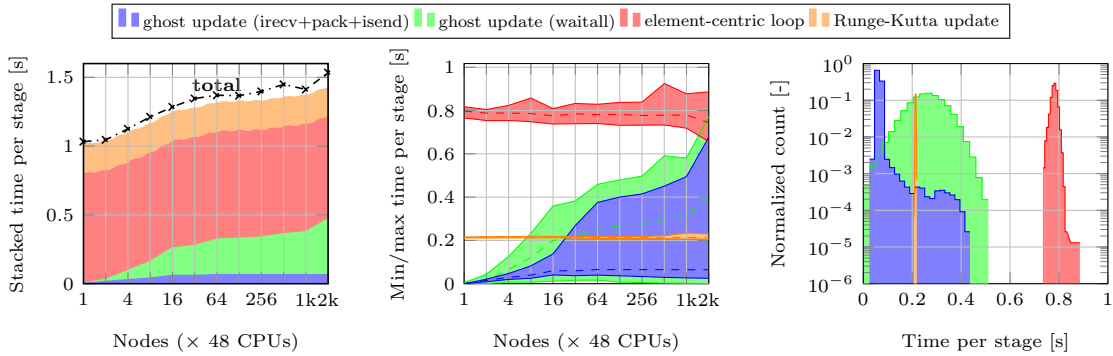
\begin{figure}

\centering

\subfloat[{Total communication data volume.}\label{fig:performance:scaling:communication:data}]
{
  \begin{tikzpicture}
    \begin{loglogaxis}[
      width=0.47\textwidth,
      height=0.35\textwidth,
      title style={font=\tiny},every axis title/.style={above left,at={(1,1)},draw=black,fill=white},
      xlabel={Nodes ($\times$ 48 CPUs)},
      ylabel={Data volume [GB]},
      legend pos={south west},
      legend cell align={left},
      cycle list name=colorGPL,
      tick label style={font=\scriptsize},
      label style={font=\scriptsize},
      legend style={font=\scriptsize},
      grid,
      semithick,
      ymin=0,
      minor ytick={1e-1,2e-1,3e-1,4e-1,5e-1,6e-1,7e-1,8e-1,9e-1,
                   1e-0,2e-0,3e-0,4e-0,5e-0,6e-0,7e-0,8e-0,9e-0,1e+1,2e+1,3e+1,4e+1,5e+1,6e+1,7e+1,8e+1,9e+1,
                   1e+2,2e+2,3e+2,4e+2,5e+2,6e+2,7e+2,8e+2,9e+2,1e+3,2e+3,3e+3,4e+3,5e+3,6e+3,7e+3,8e+3,9e+3,
                   1e+4,2e+4,3e+4,4e+4,5e+4,6e+4,7e+4,8e+4,9e+4
                   },
      xmin=2,xmax=2048,
      xtick={1, 2,4, 8, 16,32,64,128,256,512, 1024, 2048},
      xticklabels={1, 2,4, , 16,,64,,256,, 1k, 2k},
      legend style={at={(1.9,0.8)},
	anchor=north,legend columns=4}, yminorgrids 
      ] 
      
  \draw[draw=black,thick] (axis cs:2, 6.4424) -- node[above,rotate=-0,inner sep=2pt,outer sep=0.5pt]{}(axis cs:2048, 6.597039999999999964e+03);

  \draw[draw=black,thick] (axis cs:2, 6.4424) -- node[above,rotate=26,inner sep=2pt,outer sep=+0pt, fill=white, draw]{{\tiny  \color{black} \textbf{linear}}}(axis cs:2048, 6.597039999999999964e+03);

      
\addplot +[name path=I,blue,every mark/.append style={fill=blue}, solid, mark=square*] table [x index=0,y index=17] {results/scaling-advection-range2/final.result};

          
	\node[anchor=south] at (axis cs: 768, 1e-5) (nodeAE) {};
	\node[anchor=north] at (axis cs: 768, 1e+5) (nodeAF) {};
	\draw [-, dashed] (nodeAE) -- (nodeAF) ;
	
	\node[anchor=south] at (axis cs: 1536, 1e-5) (nodeBE) {};
	\node[anchor=north] at (axis cs: 1536, 1e+5) (nodeBF) {};
	\draw [-, dashed] (nodeBE) -- (nodeBF) ;	
           
    \end{loglogaxis}
  \end{tikzpicture}  
} 
\subfloat[{Accumulated network bandwidth.}\label{fig:performance:scaling:communication:bw}]
{
  \begin{tikzpicture}
    \begin{loglogaxis}[
      width=0.47\textwidth,
      height=0.35\textwidth,
      title style={font=\tiny},every axis title/.style={above left,at={(1,1)},draw=black,fill=white},
      xlabel={Nodes ($\times$ 48 CPUs)},
      ylabel={Bandwidth [GB/s]},
      legend pos={south west},
      legend cell align={left},
      cycle list name=colorGPL,
      tick label style={font=\scriptsize},
      label style={font=\scriptsize},
      legend style={font=\scriptsize},
      grid,
      semithick,
      ymin=0,
      xmin=2,xmax=2048,
      xtick={1, 2,4, 8, 16,32,64,128,256,512, 1024, 2048},
      xticklabels={1, 2,4, , 16,,64,,256,, 1k, 2k},
      legend style={at={(1.9,0.8)},
	anchor=north,legend columns=4}, yminorgrids 
      ] 
      
	\node[anchor=south] at (axis cs: 768, 1e-5) (nodeAE) {};
	\node[anchor=north] at (axis cs: 768, 1e+5) (nodeAF) {};
	\draw [-, dashed] (nodeAE) -- (nodeAF) ;
	
	\node[anchor=south] at (axis cs: 1536, 1e-5) (nodeBE) {};
	\node[anchor=north] at (axis cs: 1536, 1e+5) (nodeBF) {};
	\draw [-, dashed] (nodeBE) -- (nodeBF) ;	
      
  \draw[draw=black,thick] (axis cs:2, 20) -- node[above,rotate=-0,inner sep=3pt,outer sep=0.5pt]{}(axis cs:2048, 20480);

  \draw[draw=black,thick] (axis cs:2, 20) -- node[above,rotate=33,inner sep=3pt,outer sep=+0pt, fill=white, draw]{{\tiny  \color{black} \textbf{10GB/s/node}}}(axis cs:2048, 20480);

  \draw[draw=black,thick] (axis cs:128, 768) -- node[above,rotate=-0,inner sep=2pt,outer sep=0.5pt]{}(axis cs:2048, 12288);             
      
\addplot +[name path=I,blue,every mark/.append style={fill=blue}, solid, mark=square*] table [x index=0,y index=18] {results/scaling-advection-range2/final.result};

  \draw[draw=black,thick] (axis cs:128, 768) -- node[above,rotate=33,inner sep=2pt,outer sep=-8.5pt, fill=white, draw]{{\tiny  \color{black} \textbf{6GB/s/node}}}(axis cs:2048, 12288);

    \end{loglogaxis}
  \end{tikzpicture}
}

\vspace{0.2cm}

\begin{tikzpicture}[scale=0.8]
    \begin{axis}[%
    ybar,
    height=45pt,
    hide axis,
    legend style={font=\footnotesize},
    xmin=10,
    xmax=1000,
    ymin=0,
    ymax=0.4,
    semithick,
    legend style={draw=white!15!black,legend cell align=left},legend columns=-1
    ]

    \addlegendimage{blue!50, fill}
    \addlegendentry{ghost update (irecv+pack+isend)};
    \addlegendimage{green!50, fill}
    \addlegendentry{ghost update (waitall)};
    \addlegendimage{red!50, fill}
    \addlegendentry{element-centric loop};
    \addlegendimage{orange!50, fill}
    \addlegendentry{Runge-Kutta update};
    
    \end{axis}
\end{tikzpicture}

\vspace{-0.4cm}

\subfloat[{Averaged runtime distribution.}\label{fig:performance:scaling:communication:total}]
{
  \begin{tikzpicture}
    \begin{semilogxaxis}[
      width=0.32\textwidth,
      height=0.30\textwidth,
      title style={font=\tiny},every axis title/.style={above left,at={(1,1)},draw=black,fill=white},
      xlabel={Nodes ($\times$ 48 CPUs)},ylabel near ticks,
      ylabel={Stacked time per stage [s]},
      legend pos={south west},
      legend cell align={left},
      cycle list name=colorGPL,
      tick label style={font=\scriptsize},
      label style={font=\scriptsize},
      legend style={font=\scriptsize},
      grid,
      semithick,
      ymin=0,ymax=1.6,
      xmin=1,xmax=2048,
      xtick={1, 2,4, 8, 16,32,64,128,256,512, 1024, 2048},
      xticklabels={1, ,4, , 16,,64,,256,, 1k, 2k},
      legend style={at={(1.9,0.8)},
	anchor=north,legend columns=4}, yminorgrids ,
	stack plots=y,
		area style,
      ]

	\node[anchor=south] at (axis cs: 768, 0.0) (nodeAE) {};
	\node[anchor=north] at (axis cs: 768, 1.6) (nodeAF) {};
	\draw [-, dashed] (nodeAE) -- (nodeAF) ;
	
	\node[anchor=south] at (axis cs: 1536, 0.0) (nodeBE) {};
	\node[anchor=north] at (axis cs: 1536, 1.6) (nodeBF) {};
	\draw [-, dashed] (nodeBE) -- (nodeBF) ;

\addplot +[blue!50,every mark/.append style={fill=blue}, mark=none] table [x index=0,y index=2] {results/scaling-advection-range2/final.result}\closedcycle;


\addplot +[green!50,every mark/.append style={fill=green}, mark=none] table [x index=0,y index=5] {results/scaling-advection-range2/final.result} \closedcycle;


%
%

\addplot +[red!50,every mark/.append style={fill=blue}, dashed, mark=none] table [x index=0,y index=8] {results/scaling-advection-range2/final.result}\closedcycle;


\addplot +[orange!50,every mark/.append style={fill=orange}, dashed, mark=none] table [x index=0,y index=11] {results/scaling-advection-range2/final.result}\closedcycle;

%



\addplot +[black,fill=none,every mark/.append style={fill=black}, dash dot, mark=x] table [x index=0, y index=7] {results/scaling-advection-range2/final.result2};

    
      \node[] at (axis cs:64,1.45) {\scriptsize \textbf{total}}; 
           
    \end{semilogxaxis}
  \end{tikzpicture}
  
}~
\subfloat[{Averaged minimum and maximum runtime.}\label{fig:performance:scaling:communication:share}]
{
  \begin{tikzpicture}
    \begin{semilogxaxis}[
      width=0.32\textwidth,
      height=0.30\textwidth,
      title style={font=\tiny},every axis title/.style={above left,at={(1,1)},draw=black,fill=white},
      xlabel={Nodes ($\times$ 48 CPUs)},ylabel near ticks,
      ylabel={Min/max time per stage [s]},
      legend pos={south west},
      legend cell align={left},
      cycle list name=colorGPL,
      tick label style={font=\scriptsize},
      label style={font=\scriptsize},
      legend style={font=\scriptsize},
      grid,
      semithick,
      ymin=0,
      xmin=1,xmax=2048,
      xtick={1, 2,4, 8, 16,32,64,128,256,512, 1024, 2048},
      xticklabels={1, ,4, , 16,,64,,256,, 1k, 2k},
      legend style={at={(1.9,0.8)},
	anchor=north,legend columns=4}, yminorgrids 
      ] 
      

\addplot +[name path=G,green,every mark/.append style={fill=green}, solid, mark=none] table [x index=0,y index=6] {results/scaling-advection-range2/final.result};
\addplot +[name path=H,green,every mark/.append style={fill=green}, solid, mark=none] table [x index=0,y index=7] {results/scaling-advection-range2/final.result};
\addplot +[green,every mark/.append style={fill=green}, dashed, mark=none] table [x index=0,y index=5] {results/scaling-advection-range2/final.result};

\addplot[green!50] fill between[of=G and H,on layer=axis background];

\addplot +[name path=A,blue,every mark/.append style={fill=blue}, solid, mark=none] table [x index=0,y index=3] {results/scaling-advection-range2/final.result};
\addplot +[name path=B,blue,every mark/.append style={fill=blue}, solid, mark=none] table [x index=0,y index=4] {results/scaling-advection-range2/final.result};
\addplot +[blue,every mark/.append style={fill=blue}, dashed, mark=none] table [x index=0,y index=2] {results/scaling-advection-range2/final.result};

\addplot[fill=blue!50,draw=none,forget plot,on layer=axis background] fill between[of=A and B,on layer=axis background];

%
%

\addplot +[name path=C,red,every mark/.append style={fill=blue}, solid, mark=none] table [x index=0,y index=9] {results/scaling-advection-range2/final.result};
\addplot +[name path=D,red,every mark/.append style={fill=blue}, solid, mark=none] table [x index=0,y index=10] {results/scaling-advection-range2/final.result};
\addplot +[red,every mark/.append style={fill=blue}, dashed, mark=none] table [x index=0,y index=8] {results/scaling-advection-range2/final.result};

\addplot[red!50] fill between[of=C and D,on layer=axis background];

\addplot +[name path=E,orange,every mark/.append style={fill=orange}, solid, mark=none] table [x index=0,y index=12] {results/scaling-advection-range2/final.result};
\addplot +[name path=F,orange,every mark/.append style={fill=orange}, solid, mark=none] table [x index=0,y index=13] {results/scaling-advection-range2/final.result};
\addplot +[orange,every mark/.append style={fill=orange}, dashed, mark=none] table [x index=0,y index=11] {results/scaling-advection-range2/final.result};

\addplot[orange!50] fill between[of=E and F,on layer=axis background];
%



    \end{semilogxaxis}
  \end{tikzpicture}
  
}~
\subfloat[{Histograms of step runtime (64 nodes).}\label{fig:performance:scaling:communication:histogram}]
{
  \begin{tikzpicture}
    \begin{semilogyaxis}[
      width=0.32\textwidth,
      height=0.30\textwidth,
      title style={font=\tiny},every axis title/.style={above left,at={(1,1)},draw=black,fill=white},
      xlabel={Time per stage [s]},ylabel near ticks,
      ylabel={Normalized count [-]},
      legend pos={south west},
      legend cell align={left},
      cycle list name=colorGPL,
      tick label style={font=\scriptsize},
      label style={font=\scriptsize},
      legend style={font=\scriptsize},
      grid,
      semithick,
      ymin=1e-06,ymax=1,
      ytick={1e-6,1e-5,1e-4,1e-3,1e-2,1e-1,1e-0},minor tick num=0,
      xmin=0,xmax=1,
      legend style={at={(1.9,0.8)},
	anchor=north,legend columns=4}, 
      ]

\addplot +[name path=A,green,every mark/.append style={fill=green}, solid, mark=none] table [x index=3,y index=4] {results/scaling-advection-range2-hist/hist.csv};
\addplot +[name path=B,green,every mark/.append style={fill=green}, solid, mark=none] table [x index=3,y index=5] {results/scaling-advection-range2-hist/hist.csv};

\addplot[green!50] fill between[of=A and B,on layer=axis background];

\addplot +[name path=G,blue,every mark/.append style={fill=blue}, solid, mark=none] table [x index=0,y index=1] {results/scaling-advection-range2-hist/hist.csv};
\addplot +[name path=H,blue,every mark/.append style={fill=blue}, solid, mark=none] table [x index=0,y index=2] {results/scaling-advection-range2-hist/hist.csv};

\addplot[blue!50] fill between[of=G and H,on layer=axis background];

\addplot +[name path=C,red,every mark/.append style={fill=blue}, solid, mark=none] table [x index=9,y index=10] {results/scaling-advection-range2-hist/hist.csv};
\addplot +[name path=D,red,every mark/.append style={fill=blue}, solid, mark=none] table [x index=9,y index=11] {results/scaling-advection-range2-hist/hist.csv};

\addplot[red!50] fill between[of=C and D,on layer=axis background];

\addplot +[name path=E,orange,every mark/.append style={fill=orange}, solid, mark=none] table [x index=12,y index=13] {results/scaling-advection-range2-hist/hist.csv};
\addplot +[name path=F,orange,every mark/.append style={fill=orange}, solid, mark=none] table [x index=12,y index=14] {results/scaling-advection-range2-hist/hist.csv};

\addplot[orange!50] fill between[of=E and F,on layer=axis background];

    \end{semilogyaxis}
  \end{tikzpicture}
}


  \caption{Weak scaling of a single Runge--Kutta stage of the solution of the advection equation. }\label{fig:performance:scaling:communication}
  

\end{figure}

Finally, we discuss the drop in parallel efficiency of the weak-scaling runs of 
the large-scale simulations. For this, we have slightly modified the setup: We 
start from a configuration of $8^6$ cells with $k=3$ on one node (32 degrees of 
freedom in each direction). When doubling the number of processes, we double the 
number of cells in one direction, starting from direction 1 to direction 6 (and 
starting over at direction 1). Each time, we double the number of cells along a 
$\vec{x}$-direction, we also double the number of processes $p_{\vec{x}}$, 
keeping $p_{\vec{v}}$ constant and vice versa. In this way, the number of DoFs 
per process along $\vec{x}$ and $\vec{v}$
as well as the number of ghost degrees of freedom per process remain constant 
once all cells at the boundary have periodic neighbors residing on other nodes 
(number of nodes $\ge 2^{d_{\vec{x}}+d_{\vec{v}}}$). As a consequence, the 
computational work load and the communication amount of each node are constant. 
The total time per Runge--Kutta stage is shown in 
Figure~\ref{fig:performance:scaling:time} with dashed lines. 

The total communication amount of the considered setup increases linearly with 
the number of processes as shown in 
Figure~\ref{fig:performance:scaling:communication:data}. Measurements in 
Figure~\ref{fig:performance:scaling:communication:bw} show that the network 
can handle this increase: The data can be sent with a constant network bandwidth 
of 10GB/s per node, which is close to the theoretical 100Gbit/s, as long as the job 
stays on an island due to the fat-tree network topology and with a smaller 
bandwidth once the job stretches over multiple islands due to the pruned tree 
network architecture. In the latter case, we observed a bandwidth of 6GB/s, which 
is related to the fact that only a small ratio of the messages crosses island 
boundaries.

Figure~\ref{fig:performance:scaling:communication:total} and 
\ref{fig:performance:scaling:communication:share} show the time spent in 
different sections (in the following referred to as steps) of the advection 
operator. We consider the following four steps:
\begin{enumerate}
\item Start a ghost-value update by calling \texttt{MPI\_Irecv} as well as pack 
and send (vie \texttt{MPI\_Isend}) messages to each neighboring process residing 
on remote compute nodes.
\item Finish the ghost-value update by waiting (with \texttt{MPI\_Waitall}) until 
all messages have been sent and received.
\item Evaluate the right-hand-side term of the advection equation by performing 
an element-centric loop on locally owned cells.
\item Perform the remaining Runge--Kutta update steps.
\end{enumerate}
Besides averaged times, the minimum and maximum times encountered on any process 
are shown for each step. The times have been averaged over all Runge--Kutta 
stages.

Not surprisingly, the imbalance in the element-centric loop and in the 
Runge--Kutta update is small, which can be explained with the access to the 
shared recourses (mainly RAM). Furthermore, the ratio of communication in the 
overall run time increases with increasing number of nodes---in particular, the 
time increases for low numbers of nodes, while new periodic neighbors are still 
added and for high node numbers when communication to other islands is required. 

What is surprising is that the minimum and the maximum time spent in updating the 
ghost values differ significantly: The difference increases even with increasing 
number of nodes. The detailed analysis of the histograms of the times on 
different processors presented in 
Figure~\ref{fig:performance:scaling:communication:histogram} shows that 
communications are processed according to a Gaussian distribution, resulting in a 
few processes finishing much earlier and in a few ones much later. Overall, 
however, this imbalance caused by the MPI communication is not performance-hindering 
as was demonstrated by the fact that a significant portion of the 
network bandwidth is used. However, one should keep this imbalance in mind and 
not attribute it accidentally to other sections of the code.


\section{Application: Vlasov--Poisson}\label{sec:vp}

We now study the Vlasov--Poisson equations as an example of an application 
scenario for our library. We consider the Vlasov equation for electrons in a 
neutralizing background in the absence of magnetic fields,
\begin{subequations}
\begin{align}
\frac{\partial\,f}{\partial\, t} +
\left(
\begin{array}{c}
\vec{v} \\
-\vec{E}(t,\vec{x})
\end{array}
\right)
\cdot
\nabla f = 0,
\end{align}
where the electric field can be obtained from the Poisson problem
\begin{align}\label{equ:vp:close}
\rho(t,\vec{x})=1-\int f(t,\vec{x},\vec{v})\,\mathrm{d}v,
\qquad
-\nabla_{\vec{x}}^2\phi(t,x)=\rho(t,\vec{x}),
\qquad
\vec{E}(t,\vec{x})=-\nabla_{\vec{x}}\phi(t,\vec{x}).
\end{align}
\end{subequations}
For the solution of this problem, we use the low-storage Runge--Kutta method of 
order 4 with 5 stages. Each stage contains the following five steps for 
evaluating the right-hand side:
\begin{enumerate}
	\item Compute the degrees of freedom of the charge density via integration over the velocity space.
	\item Compute the right-hand side for the Poisson equation.
	\item Solve the Poisson equation for $\phi$.
	\item Compute $\vec{E}$ from $\phi$.
	\item Apply the advection operator. \label{algo:vp:advection}
\end{enumerate}
Step~(\ref{algo:vp:advection}) is a $d_{\vec{x}}+d_{\vec{v}}$-dimensional 
problem, and
Steps~(2)--(4) are $d_{\vec{x}}$-dimensional problems. Step~(1) reduces 
information from the phase space to the configuration space.

\subsection{Implementation details}

The advection step (Step~(\ref{algo:vp:advection})) relies on the operator 
analyzed in Section~\ref{sec:performance}. The constant velocity field function $\vec{a}$ is replaced by the function
$
\vec{a}(t,\vec{x},\vec{v})^\top
=
\left(
\vec{v}^\top ,\;
-\vec{E}(t,\vec{x})^\top
\right)
$.
The following two points should be noted: Firstly, $\vec{v}$ evaluated at a 
quadrature point corresponds exactly  to a quadrature point in the $\vec{v}$-
domain and can hence be queried from a low-dimensional FEM 
library. Similarly, $\vec{E}$ is independent of the velocity $\vec{v}$ and can be 
precomputed once at each Runge--Kutta stage for all quadrature points in the 
$\vec{x}$-space. Exploiting these relations, we never compute the 
$d_{\vec{x}}+d_{\vec{v}}$-dimensional velocity field, but compose the 
$d_{\vec{x}}$- and $d_{\vec{v}}$-information on the fly, just as we did in the 
case of the mapping. Since the data to be loaded per quadrature point is 
negligible (see also the reasoning regarding the Jacobian matrices in 
Subsection~\ref{sec:concept:fe}), the throughput of the advection operator is 
weakly effected by the need to evaluate the velocity field at every quadrature 
point.

For the solution of the Poisson problem
\begin{align}
(\nabla_{\vec{x}}  \psi, \nabla_{\vec{x}} \phi )_{\vec{x}} = (\psi,\,\rho)_{\vec{x}},
\end{align}
we utilize a matrix-free geometric multigrid 
solver~\cite{Kronbichler2016,Fehn2019HybridMM} from \texttt{deal.II}, which 
uses a Chebyshev smoother~\cite{Adams03}. The Poisson problem is solved by 
each process group with a constant velocity grid (see \texttt{row\_comm} in 
Subsection~\ref{sec:concept:partitioning}). The result of this is that the 
solution $\phi$ is available on each process without the need of an additional 
broadcast step.

The integration of $f$ over the velocity space is implemented via an 
\texttt{MPI\_Allreduce} operation over all processes with the same $\vec{x}$ 
grid partition (i.e., \texttt{colomn\_comm}) so that the resulting $\rho$ 
is available on all processes.

We have verified our implementation with a simulation of the Landau 
damping problem as in~\cite{Kormann2019}.

\begin{figure}

\centering

\qquad\qquad\begin{tikzpicture}[scale=0.8]
    \begin{axis}[%
    ybar,
    height=45pt,
    hide axis,
    legend style={font=\footnotesize},
    xmin=10,
    xmax=1000,
    ymin=0,
    ymax=0.4,
    semithick,
    legend style={draw=white!15!black,legend cell align=left},legend columns=-1
    ]

    \addlegendimage{blue!50, fill}
    \addlegendentry{Step 1};
    \addlegendimage{red!50, fill}
    \addlegendentry{Step 2};
    \addlegendimage{orange!50, fill}
    \addlegendentry{Step 3};
    \addlegendimage{green!50, fill}
    \addlegendentry{Step 4};
    \addlegendimage{black!50, fill}
    \addlegendentry{Step 5};

    \end{axis}
\end{tikzpicture}\;\;\;\;

  \begin{tikzpicture}
    \begin{loglogaxis}[
      width=0.48\textwidth,
      height=0.38\textwidth,
      title style={font=\tiny},every axis title/.style={above left,at={(1,1)},draw=black,fill=white},
      xlabel={Nodes ($\times$ 48 CPUs)},
      ylabel={Ave/min/max time per stage [s]},
      legend pos={south west},
      legend cell align={left},
      cycle list name=colorGPL,
      tick label style={font=\scriptsize},
      label style={font=\scriptsize},
      legend style={font=\scriptsize},
      grid,
      semithick,
      ymin=1e-5,ymax=11,
      minor ytick={1e-5,2e-5,3e-5,4e-5,5e-5,6e-5,7e-5,8e-5,9e-5,
                   1e-4,2e-4,3e-4,4e-4,5e-4,6e-4,7e-4,8e-4,9e-4,1e-3,2e-3,3e-3,4e-3,5e-3,6e-3,7e-3,8e-3,9e-3,
                   1e-2,2e-2,3e-2,4e-2,5e-2,6e-2,7e-2,8e-2,9e-2,1e-1,2e-1,3e-1,4e-1,5e-1,6e-1,7e-1,8e-1,9e-1,
                   1e-0,2e-0,3e-0,4e-0,5e-0,6e-0,7e-0,8e-0,9e-0
                   },
      xmin=1,xmax=2048,
      xtick={1, 2,4, 8, 16,32,64,128,256,512, 1024, 2048},
      xticklabels={1, ,4, , 16,,64,,256,, 1k, 2k},
      legend style={at={(1.9,0.8)},
	anchor=north,legend columns=4}, yminorgrids
      ]

      \node[] at (axis cs:64,3) {\scriptsize \textbf{total}};

\addplot +[black,every mark/.append style={fill=black}, solid, mark=*] table [x index=0, y index=2] {results/scaling-vp-adaptive-range/final.result};

\addplot +[name path=A,blue,every mark/.append style={fill=blue}, solid, mark=none] table [x index=0,y index=4] {results/scaling-vp-adaptive-range/final.result};
\addplot +[name path=B,blue,every mark/.append style={fill=blue}, solid, mark=none] table [x index=0,y index=5] {results/scaling-vp-adaptive-range/final.result};
\addplot +[blue,every mark/.append style={fill=blue}, dashed, mark=none] table [x index=0,y index=3] {results/scaling-vp-adaptive-range/final.result};

\addplot[fill=blue!50,draw=none,forget plot,on layer=axis background] fill between[of=A and B,on layer=axis background];

\addplot +[name path=G,green,every mark/.append style={fill=green}, solid, mark=none] table [x index=0,y index=13] {results/scaling-vp-adaptive-range/final.result};
\addplot +[name path=H,green,every mark/.append style={fill=green}, solid, mark=none] table [x index=0,y index=14] {results/scaling-vp-adaptive-range/final.result};
\addplot +[green,every mark/.append style={fill=green}, dashed, mark=none] table [x index=0,y index=12] {results/scaling-vp-adaptive-range/final.result};

\addplot[green!50] fill between[of=G and H,on layer=axis background];

\addplot +[name path=C,red,every mark/.append style={fill=blue}, solid, mark=none] table [x index=0,y index=7] {results/scaling-vp-adaptive-range/final.result};
\addplot +[name path=D,red,every mark/.append style={fill=blue}, solid, mark=none] table [x index=0,y index=8] {results/scaling-vp-adaptive-range/final.result};
\addplot +[red,every mark/.append style={fill=blue}, dashed, mark=none] table [x index=0,y index=6] {results/scaling-vp-adaptive-range/final.result};

\addplot[red!50] fill between[of=C and D,on layer=axis background];

\addplot +[name path=E,orange,every mark/.append style={fill=orange}, solid, mark=none] table [x index=0,y index=10] {results/scaling-vp-adaptive-range/final.result};
\addplot +[name path=F,orange,every mark/.append style={fill=orange}, solid, mark=none] table [x index=0,y index=11] {results/scaling-vp-adaptive-range/final.result};
\addplot +[orange,every mark/.append style={fill=orange}, dashed, mark=none] table [x index=0,y index=9] {results/scaling-vp-adaptive-range/final.result};

\addplot[orange!50] fill between[of=E and F,on layer=axis background];

\addplot +[name path=I,black,every mark/.append style={fill=black}, solid, mark=none] table [x index=0,y index=16] {results/scaling-vp-adaptive-range/final.result};
\addplot +[name path=J,black,every mark/.append style={fill=black}, solid, mark=none] table [x index=0,y index=17] {results/scaling-vp-adaptive-range/final.result};
\addplot +[black,every mark/.append style={fill=black}, dashed, mark=none] table [x index=0,y index=15] {results/scaling-vp-adaptive-range/final.result};

\addplot[black!50] fill between[of=I and J,on layer=axis background];

      \addplot [black, mark=none]coordinates {(0.1, 0.95516) (4096, 0.95516)};

    \end{loglogaxis}
  \end{tikzpicture}

  \caption{Weak scaling of a single Runge--Kutta stage of the solution of the 
  Vlasov--Poisson equations. The average (dashed), averaged minimum and averaged 
  maximum runtime of each step are indicated.}\label{fig:vp:scalingprofile}

\end{figure}
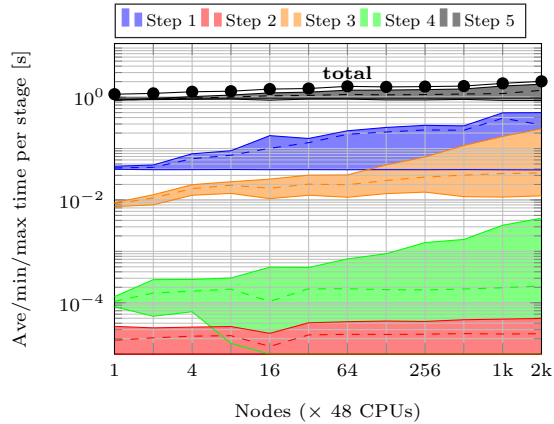

\subsection{Weak scaling}

We perform a weak-scaling experiment for the 6D Vlasov equation, starting from a 
configuration of $8^6$ cells with $k=3$ on one node. When doubling the number of 
processes, we double the number of cells in one direction, starting from 
direction 1 to direction 6 (and starting over at direction 1). Each time, we 
double the number of cells along a $\vec{x}$-direction, we also double the number 
of processes $p_{\vec{x}}$ keeping $p_{\vec{v}}$ constant and vice versa. In this 
way, the number of DoFs per process along $\vec{x}$ and $\vec{v}$ remains 
constant.

Figure~\ref{fig:vp:scalingprofile} shows the scaling of Steps 1-5 of one Runge--Kutta 
stage. We can see that the total computing time is dominated by the 
6D-advection step, which we have analyzed earlier.

Step 1, which reduces $f$ to $\rho$, becomes increasingly important as the 
problem size and parallelism increase. In this step, an all-reduction is 
performed over process groups with constant $p_{\vec{x}}$ coordinate in the 
process grid (called \texttt{comm\_column} in 
Subsection~\ref{sec:concept:partitioning}). The amount of data sent by each 
process corresponds to the number of DoFs in ${\vec{x}}$-direction of this 
process and is thus the same in every experiment.
The total amount of data sent/received is thus proportional to the total number 
of processes, while the size of the subcommunicators---and hence the number of 
reduction steps---only depends on $p_{\vec{v}}$ ($\log(p_{\vec{v}})$).
The scaling experiment shows that the time needed by step 1 generally increases 
with the number of nodes and that this step has the worst scaling behavior. This 
can be explained by the fact that this step contains a global communication 
within the subcommunicators, while the other steps only contain nearest-neighbor 
communication. We also note that the process grid is designed to optimize the 
communication scheme of the advection step. In step 1, this results in the fact 
that each subcommunicator might involve all the nodes.

The steps 2 to 4 are three-dimensional problems that are mostly negligible, only 
the Poisson solver (step 3) has some impact on the total computing time. Let us 
note that the 3D parts are solved $p_{\vec{v}}$ times on all subcommunicators 
(called \texttt{comm\_row} in Subsection~\ref{sec:concept:partitioning}) with 
constant $p_{\vec{v}}$ coordinate. Between the first and the forth data point as 
well as between the seventh and the tenth data point, the size of the Poisson 
problem increases, which results in a slight increase of the computing time. 
Between data point 4 and 7 as well as 10 and 11, the problem size stays constant 
and only the number of process groups performing these operations increases, 
which is why the CPU time stays approximately constant between these steps.

%
%
%


\section{Summary and outlook}\label{sec:outlook}

We have presented the finite-element library \texttt{hyper.deal}. It efficiently 
solves high-dimensional partial differential equations on complex geometries with 
high-order discontinuous Galerkin methods. It constructs a high-dimensional 
triangulation via the tensor product of distributed triangulations with dimensions up to 
three from the low-dimensional FEM library \texttt{deal.II} and solves the given 
partial differential equation with sum-factorization-based matrix-free operator 
evaluation. To reduce the memory consumption and the communication overhead, we 
use a new vector type, which is built around the shared-memory features from 
MPI-3.0.
%

We have compared the node-level performance with alternative algorithms, which 
are specialized for Cartesian and affine meshes or use collocation integration 
schemes. These studies revealed that loading less mapping data and reducing the 
number of sweeps is beneficial to improve the performance. We plan to look into 
the benefits of computing the low-dimensional mapping information on the fly and 
the benefits of increasing the cache locality during sum-factorization sweeps by 
a suitable hierarchical cache-oblivious blocking strategy.

Furthermore, we have studied the reduction of the working set of 
``vectorization over elements'' by processing fewer cells in parallel as SIMD 
lanes would allow. Since we observed the benefit of this approach for 6D and 
polynomial orders higher than three, we intend to investigate ``vectorization 
within an element'' as an alternative vectorization approach in the future.

In this work, we have addressed neither overlapping communication and computation 
nor graph-based partitioning of the mesh. Both points will supposedly mitigate 
the scaling limits due to the huge amount of communication that we have 
encountered especially in 6D.

\appendix
\section{APPENDIX}\label{sec:appendix:sm}
\setcounter{section}{1}

The following code snippets give implementation details on the new vector 
class, which
is built around MPI-3.0 features (see Section~\ref{sec:concept:sm}).
A new MPI communicator \texttt{comm\_sm}, which consists of processes from the 
communicator \texttt{comm} that have access to the same shared memory, can be 
created via:

\vspace{-0.3cm}

\begin{lstlisting}[language=C++, basicstyle=\footnotesize]
MPI_Comm_split_type(comm, MPI_COMM_TYPE_SHARED, rank, MPI_INFO_NULL, &comm_sm);
\end{lstlisting}
We recommend to create this communicator only once globally during setup and pass 
it to the vector.
The following code snippet shows a simplified vector class. It is a container for 
the (local and ghosted) data and for the partitioner class as well as it provides 
linear algebra operations (not shown here).

\vspace{-0.3cm}

\begin{lstlisting}[language=C++]
class Vector
{
  Vector(std::shared_ptr<Partitioner> partitioner)
    : data_this((T *)malloc(0))
  {
    this->partitioner = partitioner;
    this->partitioner->initialize_dof_vector(data_this, data_others, win);
  }

  void
  clear() { MPI_Win_free(&win); /*free window*/ }

  void
  update_ghost_values_start() const {
    partitioner->update_ghost_values_start(this->data_this, this->data_others);
  }

  void
  update_ghost_values_finish() const {
    partitioner->update_ghost_values_finish(this->data_this, this->data_others);
  }

  std::shared_ptr<Partitioner> partitioner;

  MPI_Win                  win;
  mutable T *              data_this;
  mutable std::vector<T *> data_others;
};
\end{lstlisting}
The partitioner class stores the information on which cells the current process 
owns and which ghost faces it needs. Based on this information, it is able to 
determine which cells/faces are shared and to precompute communication patterns 
with other compute nodes. As a consequence, the partitioner class is able to 
allocate the memory of the vector class.

\vspace{-0.3cm}

\begin{lstlisting}[language=C++]
class Partitioner
{
  void
  reinit(const vector<unsigned int> &                    local_cells,
         const vector<pair<unsigned int, unsigned int>> &ghost_faces) {
    // based on the IDs of local cells and ghost faces set up the access to the
    // shared memory and the communication patterns to other compute nodes
    // as well as allocate buffer
  }

  void
  initialize_dof_vector(T *& data, vector<T *> &data_others, MPI_Win & win) {
    // configure shared memory 
    MPI_Info info;
    MPI_Info_create(&info);
    MPI_Info_set(info, "alloc_shared_noncontig", "true");

    // allocate shared memory
    MPI_Win_allocate_shared((_local_size + _ghost_size) * sizeof(T), sizeof(T),
                            info, comm_sm, data, &win);

    // get pointers to the 
    data_others.resize(size_sm);
    for (int i = 0; i < size_sm; i++)
      {
        int      disp_unit;
        MPI_Aint ssize;
        MPI_Win_shared_query(win, i, &ssize, &disp_unit, &data_others[i]);
      }

    data_this = data_others[rank_sm];
  }

  void
  update_ghost_values_start(T *&data_this, vector<T *> &data_others) const {
    MPI_Barrier(comm_sm); // sync processes in same shared memory communicator
    // start communication with other nodes and fill buffer (not shown)
  }

  void
  update_ghost_values_finish(T *&data_this, vector<T *> &data_others) const {
    // finish communication with other nodes(not shown)
    MPI_Barrier(comm_sm); // sync processes in same shared memory communicator
  }

  const MPI_Comm &comm; // global communicator
  const MPI_Comm &comm_sm; // shared memory communicator
  int rank_sm, size_sm;
};
\end{lstlisting}
The partitioner class only has to be created once and can be shared by multiple 
vectors. In our simulations, it was shared by three vectors, with only one vector 
being ghosted.



\ifwithacks

\begin{acks}
This work was supported by the German Research Foundation (DFG) under
    the project ``High-order discontinuous Galerkin for the exa-scale''
    (ExaDG) within the priority program ``Software for Exascale Computing''
    (SPPEXA), grant agreement no.~KO5206/1-1 and KR4661/2-1.
The authors gratefully acknowledge the Gauss Centre for Supercomputing e.V.~(\texttt{www.gauss-centre.eu}) for funding this project
by providing computing time on the GCS Supercomputer SuperMUC-NG at Leibniz Supercomputing Centre (LRZ, \texttt{www.lrz.de})
through project id pr83te. 
\end{acks}

\fi

\bibliographystyle{ACM-Reference-Format-Journals}
\bibliography{paper}
%
%
%
%
%
%
%
%

\end{document}